\documentclass[12pt]{article}
\usepackage{amsmath}
\usepackage{graphicx,psfrag,epsf}
\usepackage{enumerate}
\usepackage{color}
\usepackage{natbib}
\usepackage{url} 

\newcommand{\blind}{1}
\newcommand{\indepe}{\mathop{\perp\!\!\!\perp}}

\begin{document}

\def\spacingset#1{\renewcommand{\baselinestretch}%
{#1}\small\normalsize} \spacingset{1}


\if1\blind
{
  \title{\bf 
Within-Person Variability Score-Based Causal Inference: A Two-Step Estimation for Joint Effects of Time-Varying Treatments
} 
\author{Satoshi Usami\hspace{.2cm}\\
    Department of Education, University of Tokyo\\
    usami\_s@p.u-tokyo.ac.jp
    }
  \maketitle
} \fi

\if0\blind
{
  \bigskip
  \begin{center}
    {\Large\bf Within-Person Variability Score-Based Causal Inference: A Two-Step Estimation for Joint Effects of Time-Varying Treatments
}
\end{center}
} \fi
\vspace{-10mm}
\begin{abstract}
Behavioral science researchers have shown strong interest in 
disaggregating within-person relations from between-person differences (stable traits) using longitudinal data. In this paper, 
we propose a method of within-person variability score-based causal inference for estimating joint effects of time-varying continuous treatments
by effectively controlling for stable traits.
After explaining the assumed data-generating process and providing formal definitions of stable trait factors, within-person variability scores,
\textcolor{black}{and joint effects of time-varying treatments at the within-person level},
we introduce the proposed method, which consists of a two-step analysis.
Within-person variability scores for each person, which are disaggregated from stable traits of that person, are
first calculated using weights based on a best linear correlation preserving predictor through structural equation modeling \textcolor{black}{(SEM)}. 
Causal parameters are then estimated via a potential outcome approach, either marginal structural models (MSMs) or structural nested mean models (SNMMs),
using calculated within-person variability scores. 
\textcolor{black}{Unlike the approach that relies entirely on SEM, t}he present method does not assume linearity for observed time-varying confounders at the within-person level.
We emphasize the use of SNMMs with G-estimation because of its property of being doubly robust to 
model misspecifications in how observed time-varying confounders are \textcolor{black}{functionally related with} treatments/predictors and outcomes at the within-person level.
Through simulation, \textcolor{black}{we show that} the proposed method can recover causal parameters well and that causal estimates might be severely biased if one does not properly account for stable traits.
An empirical application \textcolor{black}{using} data regarding sleep habits and mental health status from the Tokyo Teen Cohort study \textcolor{black}{is also provided.}
\end{abstract} 
\noindent%
\vfill

\newpage
\spacingset{1.45} 
\section{INTRODUCTION}
\label{sec:intro}
Estimating the causal effects of (a sequence of) time-varying treatments/predictors on outcomes is a challenging issue in longitudinal observational studies because
researchers must account for time-varying and time-invariant confounders. For this analytic purpose,
potential outcome approaches such as marginal structural models (MSMs; Robins, 1999; Robins, Hern\'{a}n, \& Brumback, 2000) have been widely used in
epidemiology. Although actual applications have been relatively infrequent,
structural nested models (SNMs; Robins, 1989, 1992) with G-estimation are in principle more suitable \textcolor{black}{and robust} for handling violation of the usual assumptions of
no unobserved confounders and sequential ignorability (Robins, 1999; Robins \& Hern\'{a}n, 2009; Vansteelandt \& Joffe, 2014).

Parallel with such methodological development, behavioral science researchers have shown
interest in inferring within-person relations in longitudinally observed variables, namely, how changes in one variable influence another for the same person. 
Investigations based on within-person relations might produce conclusions opposite to those based on between-person relations.
For example, a person is more likely to have a heart attack during exercise (within-person relation), despite people who exercise more having a lower risk of heart attack (between-person relation; Curran \& Bauer, 2011).

Statistical inference for disaggregating within- and between-person (or within- and between-group) relations has been a concern in behavioral sciences for more than half a century. However,
recent methodological development and extensive discussion (Cole, Martin, \& Steiger, 2005; Hamaker, 2012; Hamaker, Kuiper, \& Grasman, 2015; Hoffman, 2014;
Usami, Murayama, \& Hamaker, 2019) have rapidly increased interest in this topic. In the psychometrics literature, \textcolor{black}{along with multilevel modeling (e.g., Wang \& Maxwell, 2015),} structural equation modeling (SEM)-based approaches have become one popular method for uncovering within-person relations.
Among these approaches, applications of a random-intercept cross-lagged panel model (RI-CLPM; Hamaker et~al., 2015),
which includes common factors called \emph{stable trait factors}, 
have rapidly increased, reaching more than \textcolor{black}{1250} citations on Google as of \textcolor{black}{December} 2021. This model was originally proposed
to uncover reciprocal relations among focal variables that arise at the within-person level
(i.e., simultaneous investigations for the effects of a variable $X$ on a variable $Y$, along with the effects of $Y$ on $X$), without
explicit inclusion of (time-varying) observed confounders $L$ (however, Mulder \& Hamaker (2021) discussed an extension that included a between-level predictor).

Despite its popularity and theoretical appeal, the 
concepts of stable traits and within-person relations in the RI-CLPM have not been fully characterized in the
causal inference literature. This might be partly because
psychometricians have used these terms \textcolor{black}{vaguely and ambiguously in statistical models},
without \textcolor{black}{clarifying the assumed data-generating process (DGP) and} providing clear mathematical definitions.
For this reason, the RI-CLPM has not been contrasted with many other methodologies used for causal inference (e.g., MSMs and SNMs).
One potential advantage of the RI-CLPM as SEM is that it can easily include and estimate measurement errors in statistical models
under parametric assumptions. However, the RI-CLPM demands linear regressions at the within-person level
that are correctly specified to link focal variables (as well as time-varying observed confounders, if included in the model). 
The linearity assumption typically imposed with respect to time-varying observed confounders in path modeling and SEM has often been 
criticized in the causal inference literature (e.g., Hong, 2015), and relaxing this assumption 
is often a key to consistently estimating the causal quantity of interest (e.g., Imai \& Kim, 2019). 

In this paper, we propose a method of \emph{within-person variability score}-based causal inference 
for estimating joint effects of time-varying continuous treatments/predictors \textcolor{black}{at the within-person level} by controlling for stable traits (i.e., between-person differences). 
The proposed method is a two-step analysis.
A within-person variability score for each person, which is disaggregated from the stable trait factor score of that person, is first calculated using weights based on a best linear correlation preserving predictor through SEM. 
Causal parameters are then estimated by MSMs or SNMs, using calculated within-person variability scores.
This approach is more flexible than the \textcolor{black}{one that relies entirely on SEM (e.g., 
the RI-CLPM that includes time-varying observed confounders)} in terms of modeling how time-varying observed confounders are \textcolor{black}{functionally related with} treatments/predictors and outcomes at the within-person level,
without imposing the linearity assumption in these relations.
We particularly emphasize the utility of SNMs with G-estimation because of its attractive property of being doubly robust to
model misspecifications in how time-varying observed confounders are \textcolor{black}{functionally related with} treatments/predictors and outcomes at the within-person level.

The proposed method can be viewed as one that synthesizes two traditions for factor analysis methods and SEM in psychometrics and a 
method of causal inference (MSMs or SNMs) in epidemiology. 
\textcolor{black}{Because causal estimands that are defined at the within-person level are less common 
in the causal inference literature (L\"{u}dtke \& Robitzsch, 2021), the proposed method offers new insights
for researchers in a broad range of disciplines who are interested in causal inference. Also,}
the idea of using within-person variability scores can be applied to many other issues that are closely relevant to causal hypotheses,
including reciprocal effects and mediation effects.

The remainder of this paper is organized as follows. Because the concepts of stable traits and within-person relations have not been fully characterized in the
causal inference literature, in Section~2 we start our discussion by introducing the two different DGPs in which time-invariant factors are included. 
\textcolor{black}{Notably, unlike in a previous study (Usami et~al., 2019), we argue that stable trait factors (assumed in the RI-CLPM as a statistical model) are merely random intercepts for persons and 
cannot be statistically characterized as time-invariant unobserved confounders.}
After providing formal definitions of stable trait factors (for between-person relations) and within-person variability scores (for within-person relations),
the definition of \textcolor{black}{joint effects of time-varying treatments at the within-person level and their} identification conditions are described in Section~3. We then introduce the proposed methodology in Section~4.
\textcolor{black}{In Section~5, we perform simulations and} show that the proposed method can recover causal parameters well, and that 
causal estimates might be severely biased if stable traits are not properly accounted for.
\textcolor{black}{Section~6 describes an empirical application of the proposed method using data from the Tokyo Teen Cohort (TTC) study (Ando et~al., 2019).}
The final section gives some concluding remarks and discusses our future research agenda.

\section{CAUSAL MODELS AND DATA-GENERATING PROCESSES}
In this section, we first explain two different DGPs and causal models in which time-invariant factors are included. In the first DGP,
we assume that time-invariant factors have both direct and indirect effects on measurements;
recent work by Gische, West, and Voelkle (2021), which provided a didactic presentation of the directed acyclic graph (DAG)-based
approach and key concepts regarding causal inference based on a cross-lagged panel design, 
assumed this process. In the second DGP, we assume that time-invariant factors have only direct effects;
this corresponds to the process that researchers (implicitly) assume in applying the RI-CLPM to infer within-person relations.
This distinction of processes is inspired by Usami et~al., (2019), who highlighted how common factors included in the different
statistical models to examine reciprocal relations have different conceptual and mathematical properties.

Below, we suppose that data are generated at fixed time points $t_0,t_1,\dots$,$t_{K}$. Let $A_{ik}$ denote a continuous treatment/predictor at 
time $t_k$ ($k=0,\dots,K-1$) for person $i$, and let $L_{ik}$ denote 
time-varying observed confounders at that time for person $i$.\footnote{\textcolor{black}{Time-invariant observed confounders $L_i$ can be included as a special case, but 
in the DGPs discussed herein, only time-varying observed confounders $L_{ik}$ are assumed for simplicity.}} 
Furthermore, $Y_{ik}$ is the outcome
at time $t_k$ ($k=0,\dots,K$) for person $i$ and is part of the time-varying confounders $L_{ik}$. 
Suppose that a time-varying confounder has three characteristics: it is independently associated with future outcomes
$Y_{ik'}$, it predicts subsequent levels of treatment as well as future confounders, and
it is affected by an earlier treatment and confounders (Vansteelandt \& Joffe, 2014). 
In this paper, \textcolor{black}{for the purpose of explanation,} we assume a single confounder that is measured concurrently with the outcome 
at each time point and is measured before the treatment/predictor level is determined for each person.
Thus, we presume that the variables are ordered as $L_0,A_0,L_1,A_1,\dots,L_{K-1},A_{K-1},L_K$.\footnote{$Y$ is not shown explicitly
here because it is part of $L$. We often omit $Y$ in expressing time-varying observed confounders in this paper.}
\subsection{Data-generating Process 1: Time-invariant Factors Have Both Direct and Indirect Effects on Measurements}
Gische et~al.\ (2021) introduced the DAG-based approach to causal inference, explaining 
how (SEM-based) statistical models can identify
the causal models. Figure~1a is a DAG that expresses linear causal relations among variables in $K=4$; 
this is similar to the one presented by Gische et~al.\ (2021) but with
time-varying observed confounders $L$ now included. Each solid single-headed arrow represents a direct causal
relation, and a dashed double-headed arrow indicates the existence of an unobserved confounder. Dashed circles are used to express latent variables.
To keep the illustration simple, here we assume (i) first-order (linear) lagged effects of variables and (ii) \textcolor{black}{that mechanisms that are not directly targeted by treatment are not altered} (modularity).
More importantly, for the purpose of illustration, we temporarily assume that the process does \emph{not} start
prior to the initial measurement ($k=-1,-2,\dots$), indicating that initial measurements are the beginning of the process.
Thus, time-invariant factors $\eta$ as random intercepts do \emph{not} have direct causal effects on the dynamics among 
variables that might be going on prior to the initial measurement. We will revisit this issue later.

In this linear causal DAG model, we suppose that time-invariant factors $\eta$ are additive to express person-specific differences 
in the mean levels of the respective variables ($Y$, $A$, and $L$), which do not change over time. 
Without loss of generality, we assume that these factors have zero means ($E(\eta^{(Y)})=E(\eta^{(A)})=E(\eta^{(L)})=0$).
If we are interested in the longitudinal change of sleep time in adolescents, as will be investigated in the later \textcolor{black}{empirical}
example, then $\eta$ reflects all time-invariant factors that might affect the level of
sleep time in an adolescent during the course of study (e.g., sex, year of birth,
constitution, genetic endowment, health, exercise habits, home environment including discipline,
engagement in club/extracurricular activities in school). The values of coefficients corresponding to the paths from time-invariant factors 
to measurements are restricted to be equal to one. 
These restrictions (a) assign a scale to the latent random intercept and (b) reflect the assumption that the structural coefficients
from the random intercepts to measurements do not change over time (Gische et~al., 2021). The bidirected dashed edges between time-invariant factors 
\begin{figure}[htbp]
\includegraphics[height=16cm,width=24cm,angle=90]{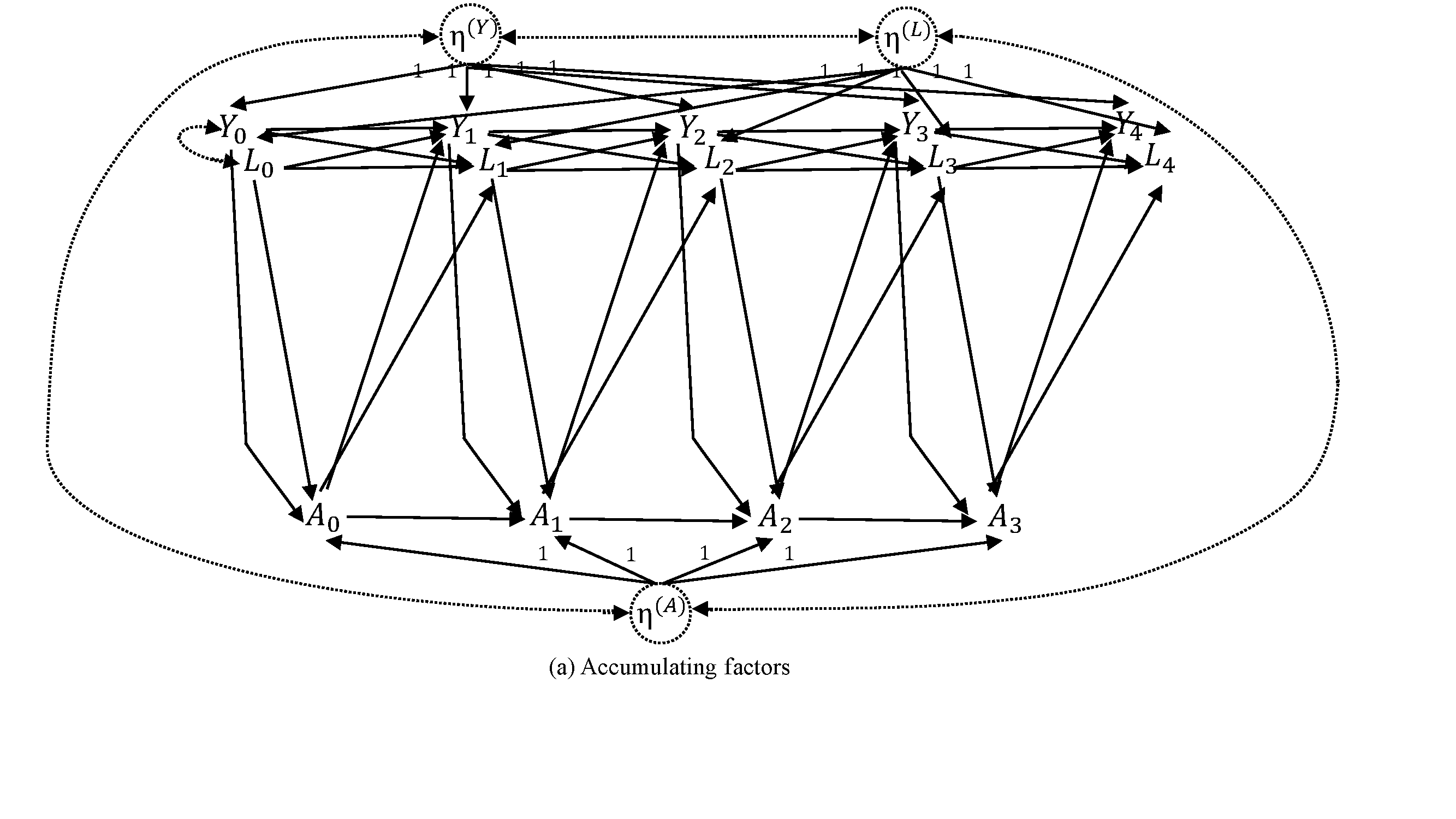}
\end{figure}
\begin{figure}[htbp]
\includegraphics[height=16cm,width=24cm,angle=90]{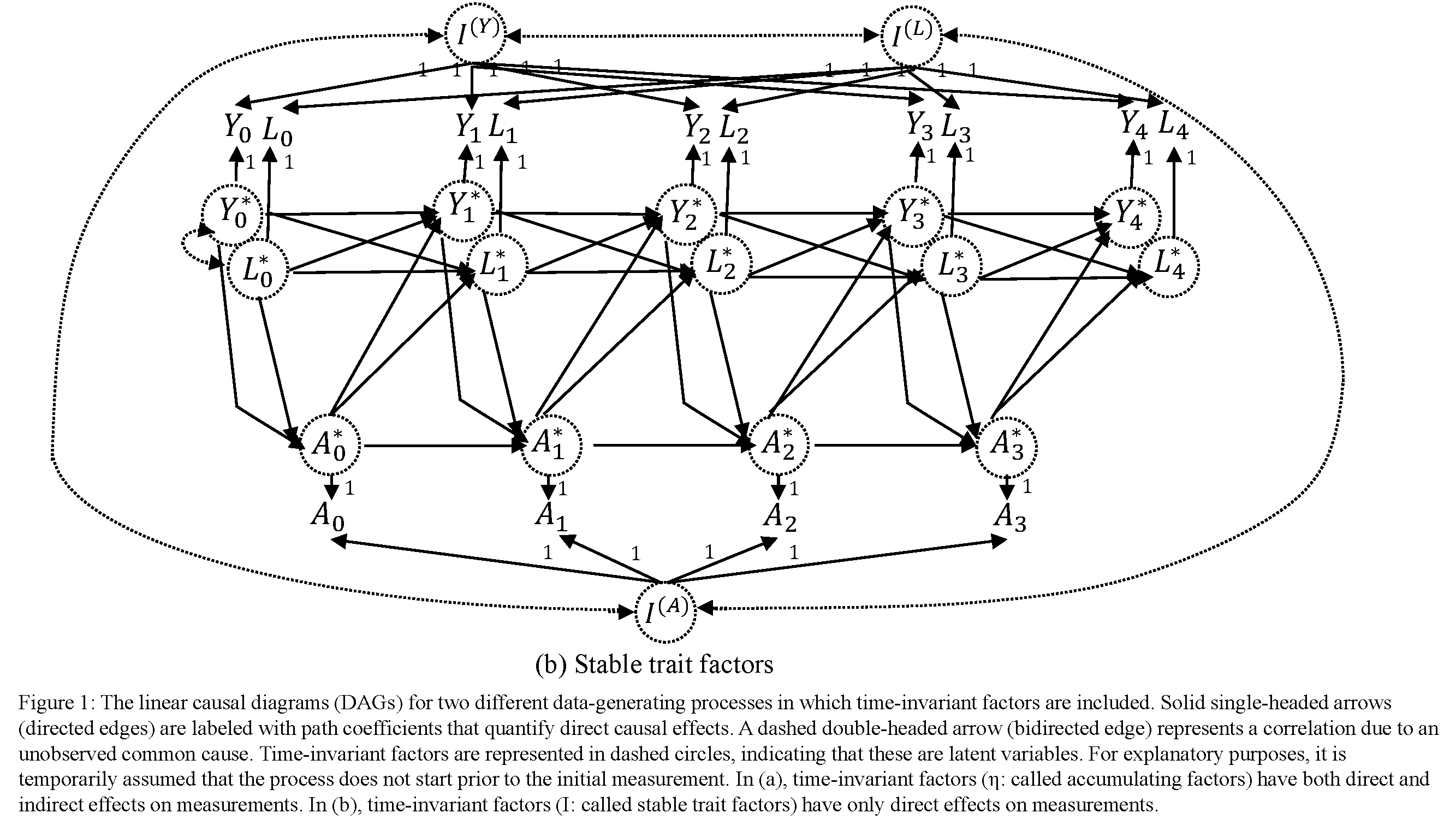}
\end{figure}
indicate that 
they might show covarying relations due to unobserved confounding. Likewise, the bidirected dashed \textcolor{black}{edge between initial measurements 
$Y_0$ and $L_0$ indicates} that they might show covarying relations due to unobserved confounding.

Under this linear causal DAG model, the DGP can be represented by the following set of linear equations ($k \geq 1$):
\begin{flalign}
Y_{ik}&=\eta^{(Y)}_{i}+\mu^{(Y)}_{k}+\alpha^{(Y)}_{k}Y_{i(k-1)}+\beta^{(Y)}_{k}A_{i(k-1)}+\gamma^{(Y)}_{k}L_{i(k-1)}+d^{(Y)}_{ik}, \notag \\
A_{ik}&=\eta^{(A)}_{i}+\mu^{(A)}_{k}+\alpha^{(A)}_{k}Y_{ik}+\beta^{(A)}_{k}A_{i(k-1)}+\gamma^{(A)}_{k}L_{ik}+d^{(A)}_{ik}, \\
L_{ik}&=\eta^{(L)}_{i}+\mu^{(L)}_{k}+\alpha^{(L)}_{k}Y_{i(k-1)}+\beta^{(L)}_{k}A_{i(k-1)}+\gamma^{(L)}_{k}L_{i(k-1)}+d^{(L)}_{ik}. \notag
\end{flalign}
Here, $\mu^{(Y)}_{k}$, $\mu^{(A)}_{k}$, and $\mu^{(L)}_{k}$ are (fixed) intercepts at time $t_k$ and are omitted in the DAG representation.
Residual terms are denoted by $d$ and are assumed to be uncorrelated with time-invariant factors; they are also usually omitted in the DAG representation.
It is also assumed that there is no unobserved common cause among these residuals (i.e., concurrent residuals are mutually uncorrelated).
\textcolor{black}{Note that the equations assume homogeneity: the coefficients $\alpha, \beta$, and $\gamma$ are fixed and constant across persons}.

As suggested by Gische et~al.\ (2021), a statistical model that captures the DAG depicted in Figure~1a (i.e., Equation~1) can be globally identified: 
all parameters can theoretically be estimated uniquely from observational data. Assuming sufficient sample size, correct model specification, and no excess multivariate
kurtosis, the SEM-based maximum likelihood (ML) method provides estimates that are asymptotically unbiased, efficient, and consistent (Bollen, 1989).
More details about causal identification and estimation in linearly parameterized causal DAG models are provided by Gische and Voelkle (\textcolor{black}{in press}).

A notable feature of this DGP is that time-invariant factors have both direct and indirect effects
on measurements. For example, $\eta^{(Y)}$ has a direct effect on $Y_3$ (i.e., $\eta^{(Y)}\rightarrow Y_3$), while $Y_3$ is also caused by $Y_2$, 
which is again caused by $\eta^{(Y)}$ (i.e., $\eta^{(Y)}\rightarrow Y_2\rightarrow Y_3$). In addition, other
time-invariant factors $\eta^{(A)}$ and $\eta^{(L)}$ also have indirect effects on $Y_3$ (e.g., $\eta^{(A)}\rightarrow A_2\rightarrow Y_3$).
These indirect effects result from the fact that time-invariant factors are modeled with lagged regressions in Figure~1a (or Equation~1), rather than being 
modeled separately.

Usami et~al.\ (2019) compared several existing statistical models to examine reciprocal relations
among variables, emphasizing that whether or not common factors are modeled with lagged regression makes 
substantial differences in the conceptual and mathematical roles of common factors.
For example, the common factors included in the latent change score model (LCS; McArdle \& Hamagami, 2001), autoregressive latent trajectory model (ALT; Bollen \& Curran, 2004), and
general cross-lagged panel model (GCLM; Zyphur et~al., 2020ab) and
individual-specific effects that are often included in longitudinal panel models of econometrics (e.g., the dynamic panel model \textcolor{black}{or random effects model}) are commonly modeled 
with lagged regressions, reflecting that they have both direct and indirect effects on measurements. This type of common factor is called an 
\emph{accumulating factor} (Usami et~al., 2019; Usami, 2021) because its effects accumulate in measurements at later time points through the lagged regression. 
However, in the RI-CLPM \textcolor{black}{(that includes time-varying observed confounders)}, which researchers are increasingly using to uncover within-person relations, common factors (i.e., stable trait factors) are 
\emph{not} modeled with lagged regression, indicating that this statistical model cannot identify parameters in the causal DAG model as depicted in Figure~1a (i.e., Equation~1).
\subsection{Data-generating Process 2: Time-invariant Factors Have Only Direct Effects on Measurements}
To clarify this point, let us consider a different (linear and first-order) DGP in which time-invariant factors are included
but have only direct effects on measurements. In Figure~1b, directed edges from time-invariant factors $I$ are 
drawn to \textcolor{black}{the corresponding} measurements. Also, directed edges from time-varying factors, which are expressed by writing the variable name with an asterisk (e.g., $Y_3^*$), are drawn to
\textcolor{black}{the corresponding} measurements. Directed edges are assumed between these time-varying factors, rather than between measurements as in Figure~1a.
Time-varying factors are also assumed to be uncorrelated with time-invariant factors. As a result, time-invariant factors $I$ have only direct effects on measurements, and
\textcolor{black}{under the linearity assumption} each measurement can be decomposed into the linear sum of time-invariant and time-varying factors that are mutually uncorrelated. 

The values of coefficients corresponding to the paths from time-varying factors to measurements are all restricted to be one,
and we assume that these time-varying factors have zero means. Under this linear causal DAG model, the DGP can be represented by the following 
linear equations \textcolor{black}{(with the assumption of homogeneity of coefficients among persons)} that have two major parts:
\begin{gather}
Y_{ik}=\mu^{(Y)}_{k}+I^{(Y)}_{i}+Y^*_{ik},\hspace{3mm}A_{ik}=\mu^{(A)}_{k}+I^{(A)}_{i}+A^*_{ik},\hspace{3mm}L_{ik}=\mu^{(L)}_{k}+I^{(L)}_{i}+L^*_{ik}
\end{gather}
for $k\geq 0$, and 
\begin{flalign}
Y^*_{ik}&=\alpha^{(Y)}_{k}Y^*_{i(k-1)}+\beta^{(Y)}_{k}A^*_{i(k-1)}+\gamma^{(Y)}_{k}L^*_{i(k-1)}+d^{(Y)}_{ik}, \notag \\
A^*_{ik}&=\alpha^{(A)}_{k}Y^*_{ik}+\beta^{(A)}_{k}A^*_{i(k-1)}+\gamma^{(A)}_{k}L^*_{ik}+d^{(A)}_{ik}, \\
L^*_{ik}&=\alpha^{(L)}_{k}Y^*_{i(k-1)}+\beta^{(L)}_{k}A^*_{i(k-1)}+\gamma^{(L)}_{k}L^*_{i(k-1)}+d^{(L)}_{ik} \notag
\end{flalign}
for $k \geq 1$. $\mu^{(Y)}_{k}$, $\mu^{(A)}_{k}$, and $\mu^{(L)}_{k}$ are the temporal group means (rather than fixed intercepts) at time point $t_k$ and 
are omitted in the DAG representation. The residual terms $d$ are assumed to be uncorrelated with both time-invariant and time-varying factors and are also omitted in the DAG representation.
As suggested from Equation~(2), under these specifications the time-varying factors $Y^*_{ik}$, $A^*_{ik}$, and $L^*_{ik}$ represent temporal deviations from the expected score for person $i$
at time point $t_k$ (i.e., $\mu^{(Y)}_{k}+I^{(Y)}_{i}$, $\mu^{(A)}_{k}+I^{(A)}_{i}$, and $\mu^{(L)}_{k}+I^{(L)}_{i}$), whereas time-invariant factors represent
stable between-person differences over time. The time series $Y^*_{ik}$, $A^*_{ik}$, and $L^*_{ik}$ can thus be interpreted as within-person variations
that are uncorrelated from time-invariant factors as stable between-person differences.

In psychology, \emph{traits} were originally considered as personality characteristics that are stable over time and in different situations.
To express such latent constructs, common factors are explicitly included in psychometric models. 
In the context of the RI-CLPM, such common factors are called \emph{stable trait factors},
and they have the same role as that of time-invariant factors $I$ in the linear causal DAG model as depicted in Figure~1b (i.e., Equation~2). 
In the RI-CLPM, the initial deviations are modeled as exogeneous variables, and their variances and covariances are estimated.
Residuals in this statistical model are usually assumed to follow a multivariate normal distribution. 
Although the original motivation for the RI-CLPM was to infer reciprocal (rather than unidirectional) relations and the model does not usually assume time-varying observed confounders $L$
and higher-order lagged effects of variables, it can be extended in a straightforward manner to investigate (joint) effects of continuous treatments/predictors $A$ on outcomes $Y$, while
including $L$ in linear regressions. Therefore, such an extended version of the RI-CLPM as a statistical model can
identify the causal parameters if the assumed DGP as in Figure~1b (i.e., Equations~2 and 3) is correct \textcolor{black}{and if $K \geq 2$ (i.e., three or more time
points; Usami et~al., 2019)}.

Usami et~al.\ (2019) explained that the conceptual and mathematical roles of common factors differ
according to whether or not they are modeled with lagged regression in the statistical model. More specifically, in models
that include accumulating factors, their influences on measurements at time $t_k$ (e.g., $\eta^{(Y)}\rightarrow Y_k$) transmit to
the future measurements (e.g., $Y_k \rightarrow Y_{k+1}$) through the lagged regression, which is also influenced by the same accumulating factors (e.g., $\eta^{(Y)}\rightarrow Y_{k+1}$);
as a result, the magnitudes of impacts from these factors change over time.
In contrast, in models that include stable trait factors (e.g., the RI-CLPM), their impacts are stable over time
because they have only direct effects. \textcolor{black}{In this way, the conceptual meaning and inferential results
for (within-person) relations among variables being modeled
differ in each statistical model according to whether researchers assume the inclusion of accumulating factors or stable trait factors; 
see Usami et~al.\ (2019, pp.~643--644) for a more detailed comparison.}
\subsection{\textcolor{black}{Implications of Comparing Different DGPs: Control for Time-invariant Unobserved Confounders and Initial Conditions}}
\subsubsection{\textcolor{black}{Control for time-invariant unobserved confounders}}
\textcolor{black}{As we have argued, the conceptual and mathematical roles differ between stable trait factors and accumulating factors.
Importantly,} the differences between these factors can also be characterized as whether or not 
they can be considered as time-invariant unobserved confounders. For example, the accumulating factors of outcomes ($\eta^{(Y)}$)
cause measurements $Y_k$ ($k \geq 1$) while also being associated with measurements of treatments/predictor at the previous time point ($A_{k-1}$). In this
sense, accumulating factors can be considered as unobserved confounders in evaluating causal effects of treatments/predictors. 
In contrast, the stable trait factors 
of outcomes ($I^{(Y)}$) also cause measurements $Y_k$ but are uncorrelated with within-person variations such as $Y^*_{k}$ and $A^*_{k-1}$.
\textcolor{black}{More specifically, when measurements $Y_k$ are unconditional, 
$I^{(Y)}$ does not confound the relations among within-person
variations (e.g., the path from $A^*_{k-1}$ to $Y^*_{k}$) 
because the path from $I^{(Y)}$ to within-person
variations $Y^*_k$ is blocked by the measurement $Y_k$, which act as colliders.}
Therefore, stable trait factors cannot be viewed as time-invariant unobserved confounders; rather, they 
\textcolor{black}{should be characterized as merely random intercepts} that are uncorrelated with predictors (i.e., within-person variations).
This view differs from that of Usami et~al.\ (2019), who explain stable trait factors as time-invariant unobserved confounders.

\textcolor{black}{
If the assumed (linear and first-order) DGP as in Figure~1b is correct and if all variables are observable, then 
controlling for only $Y^*_{k-1}$ and $L^*_{k-1}$ is sufficient to
evaluate the within-person relation between $A^*_{k-1}$ and $Y^*_{k}$. 
One could argue that controlling for stable trait factors is not required to identify causal parameters for treatment effects at the within-person level, the reason being that within-person processes (time-varying factors)
and between-person differences (stable trait factors as time-invariant factors) are mutually uncorrelated. 
However, because all these factors are actually latent variables and unobservable, we need to use measurements (as colliders) to infer treatment effects at the within-person level, 
and appropriate control of stable trait factors as latent variables is required in the statistical model. If the assumed DGP as in Figure~1b is correct, then not controlling for stable trait factors 
causes biased estimates of causal parameters for the within-person relation (e.g., Usami, Todo \& Murayama, 2019 and the later simulations).
}
\subsubsection{Initial conditions}
How to treat the initial measurements (i.e., $Y_0$, $L_0$, and $A_0$) is also important for distinguishing between the two DGPs in Figure~1. Special attention needs to be paid to the initial measurements 
because there are no incoming directed edges to these variables from variables
prior to the initial time point ($k=-1,-2,\dots$). Although we have assumed so far that the 
DGP does not start prior to the initial measurements, this assumption is not realistic in many applications,
and the initial measurements must somehow account for the past of the process 
that is not explicitly modeled (see Figure~S1 in the Online Supplemental Material for more details). Assuming a DGP similar to that in Figure~1a, Gische et~al.\ (2021, Figure~5) provided a straightforward and interpretable approach that 
freely estimates the coefficients (loading) from $\eta$ to \emph{all} initial measurements. For example, for $\eta^{(Y)}$, 
the coefficients from this factor to $Y_0$, $L_0$, and $A_0$ are freely specified rather than being fixed to either one or zero. In applying dynamic panel models in econometrics,
one usually assumes that individual-specific components (i.e., accumulating factors) are correlated with the initial measurements to account for the past of the process.

Importantly, if the second DGP (Figure~1b) is correct, then such 
special considerations are not required. This is because time-invariant factors $I$ have only direct effects on measurements
(rather than on temporal deviations as within-person variations) and no directed edges are assumed between observed variables.
In other words, past time-varying factors ($Y_{-1}^*, Y_{-2}^*,\dots$, $A_{-1}^*, A_{-2}^*,\dots$, $L_{-1}^*, L_{-2}^*,\dots$) as variations in within-person processes
cause observed variables separately from $I$ as stable between-person differences (see also Figure~S1 in the Online Supplemental Material). 
Therefore, if the assumed DGP as depicted in Figure~1b is correct, then 
the RI-CLPM, which includes time-varying observed confounders and assumes that initial variables at the within-person level ($Y_0^*$, $L_0^*$)
are exogeneous (and are mutually correlated) and that loadings from $I$ to \textcolor{black}{the corresponding} initial measurements (i.e., $I^{(Y)}\rightarrow Y_0$, $I^{(A)}\rightarrow A_0$ and $I^{(L)}\rightarrow L_0$) are all set to one, can identify causal parameters for treatment effects, 
even if the DGP actually starts prior to the initial measurements.
\subsection{Summary and Discussion}
\textcolor{black}{The critical difference between the two different DGPs in Figure~1 is whether the assumed time-invariant factors have only (stable) direct effects (i.e., stable trait factors) or
both direct and indirect effects on measurements (i.e., accumulating factors). Because of this difference,
stable trait factors as merely random intercepts cannot be viewed as time-invariant unobserved confounders, while
special considerations for initial measurements are not required if the assumed DGP includes only stable trait factors (i.e., Figure~1b).
Although researchers are increasingly using the RI-CLPM as a statistical model to uncover within-person relations, stable traits and (implicitly) assumed DGPs 
have not been fully characterized in the causal inference literature. Below,
we assume a DGP that includes stable trait factors as in Figure~1b, and we propose
a method of \emph{within-person variability score}-based causal inference for (joint) effects of
time-varying treatments at the within-person level. This approach 
is more flexible than the one that relies entirely on SEM (e.g., 
the RI-CLPM that includes time-varying observed confounders)
in terms of the linearity assumption regarding observed confounders at the within-person level.}

\textcolor{black}{A causal DAG represents a researcher's theory about the causal process and should be 
drawn based on subject-matter knowledge. However, in many cases, researchers
do not exactly know the true DGP and how time-invariant factors (if they exist) influence measurements
(e.g., linearly or nonlinearly, directly or indirectly, or both).}
Although it is ideal if one can unambiguously articulate the theoretically derived expected relations for variables,
this can be quite challenging in practical applications (Curran, 2011). If linear SEM-based statistical models are used, then 
as a data-driven approach one could compare model fit indices between two statistical models that appropriately represent the 
causal models of Equation~(1) and Equations~(2) and (3) (i.e., the RI-CLPM that includes time-varying observed confounders), 
and this would be useful for investigating the sensitivity of the conclusions.

\textcolor{black}{In the context of applying the RI-CLPM, L\"{u}dtke \& Robitzsch (2021) argued that including stable trait factors 
might be better suited for short-term studies that typically use shorter time lags between time points. 
In short-term studies, one might be more certain that there are no indirect effects from time-invariant factors (i.e., stable trait factors).
Also, if time-invariant unobserved confounders (rather than random intercepts that merely represent between-person differences as stable trait factors) are likely to be present, 
then other statistical approaches that account for such confounders might be more suitable.
However, in our opinion there are no clear criteria that delineate when and how to include (time-invariant) factors in the assumed DGP, and continued discussion that also considers empirical investigations
of each research hypothesis and sensitivity of results (e.g., the later simulations) will be required in the future.}
\section{FORMAL DEFINITIONS OF STABLE TRAIT FACTORS, WITHIN-PERSON VARIABILITY SCORES \textcolor{black}{AND JOINT EFFECTS OF TIME-VARYING TREATMENTS}}
\subsection{Definitions of Stable Trait Factors and Within-person Variability Scores}
\textcolor{black}{The terms ``(stable) traits'' and ``within-person relations'' have been used vaguely and ambiguously in statistical models,}
despite the existence of mathematical and interpretative differences among models (e.g., Usami et~al., 2019). Inspired by the discussion so far,
we provide the formal definitions of these below.

A \emph{stable trait factor} of person $i$ (say, for $Y$) is defined in this paper as 
(i) the time-invariant factor that has additive influence on measurements, and (ii) its quantity is equal to
the difference between the expected value of measurement (i.e., true score) of this person at time point $t_k$ (expressed as $\textcolor{black}{T}^{(Y)}_{ik}$) 
and the temporal group mean ($\mu^{(Y)}_{k}$), which is invariant over time:
\begin{gather}
I_{i}^{(Y)}=\textcolor{black}{T}^{(Y)}_{ik}-\mu_{k}^{(Y)}
\end{gather}
for $k=0,\dots,K$, $-\infty< {\textcolor{black}{T}}_{ik}^{(Y)}< \infty$, and $-\infty< \mu_{k}^{(Y)}< \infty$. 
Note that $E(I_{i}^{(Y)})=E({\textcolor{black}{T}}^{(Y)}_{ik}-\mu_{k}^{(Y)})=\mu_{k}^{(Y)}-\mu_{k}^{(Y)}=0$.

Next, the \emph{within-person variability score} $Y^*_{ik}$ is defined as (i) a time-varying factor that has additive influence on measurements,
and (ii) its quantity is equal to the difference between a measurement and its expected value:
\begin{gather}
Y^*_{ik}=Y_{ik}-{\textcolor{black}{T}}^{(Y)}_{ik}=Y_{ik}-(\mu_{k}^{(Y)}+I_{i}^{(Y)}),
\end{gather}
with the assumptions of $E(Y^*_{ik})=0$ and \textcolor{black}{independence between 
${T}^{(Y)}_{ik}$ and $Y^*_{ik}$}. From this formulation, stable trait factors and within-person variability scores are 
uncorrelated because 
\begin{gather}
Cov(I_{i}^{(Y)},Y^*_{ik})=E[(\textcolor{black}{T}^{(Y)}_{ik}-\mu_{k}^{(Y)})Y^*_{ik}]
=E({\textcolor{black}{T}}^{(Y)}_{ik}Y^*_{ik})-\mu_{k}^{(Y)}E(Y^*_{ik})=0.
\end{gather} 
Thus, variances of measurements at time point $t_k$ can be expressed as the 
sum of those of stable trait factor scores and within-person variability scores. This means 
that the time series for within-person variability scores have the following covariance structure:
\begin{gather}
Cov(Y^*_{ik},Y^*_{ik'})=Cov(Y_{ik},Y_{ik'})-Var(I_{i}^{(Y)}).
\end{gather}
In this paper, we use the terms \emph{within-person relation} and \emph{between-person relation} to describe
the relations between variables that are based on within-person variability scores and stable trait factor scores, respectively.
\subsection{\textcolor{black}{Definition of Joint Effects of Time-varying Treatments at the Within-person Level}} 
\textcolor{black}{Next, we explain the definition of joint (causal) effects of treatments at the within-person level} using the potential outcome approach. We
assume a similar causal DAG model to that in Figure~1b: (i) measurements are expressed by the linear sum of stable trait factors and within-person variability scores, 
and (ii) within-person variability scores are expressed by functions (with assumption of homogeneity) of those in past time. However, unlike the presentation in Section~2,
we relax some assumptions about the within-person variability score\textcolor{black}{s} to allow the following: (a) higher-order lagged effects and
interaction effects of treatments/predictors can exist at the within-person level, and (b) time-varying observed confounders
can be nonlinearly related with outcomes and treatments/predictors at the within-person level.
\textcolor{black}{The current focus is on evaluating the within-person relation between variables, that is, how 
the (joint) intervention of treatments/predictors influences future outcomes at the within-person level.}

\textcolor{black}{Below, we use overbars $\bar{Y}^*_k=\{Y^*_0,Y^*_1,\dots,Y^*_k\}$ to denote the history of $\textcolor{black}{Y^*}$ through $t_k$ and underbars
$\underline{Y}^*_k=\{Y^*_k,\dots,Y^*_K\}$ to denote the future of this variable. Let 
${Y^*_{ik}}^{{\bar{A}}^*_{i(k-1)}}$ ($k=1,\dots,K$) denote the within-person variability score for the outcome
that would take at time point $t_k$ for person $i$ were this person to receive treatment history at the within-person level
$\bar{A}^*_{i(k-1)}=\{A^*_{i0},\dots,A^*_{i(k-1)}\}$ through $t_{k-1}$. Here, $A^*_{ik}=0$ ($k=0,\dots,K-1$) indicates that the amount of treatments/predictors for person $i$
is equal to the expected score of this person at time point $t_k$ (i.e., $A_{ik}=\mu^{(A)}_{k}+I^{(A)}_{i}$).
${Y^*_{ik}}^{{\bar{A}}^*_{i(k-1)}}$ is a potential outcome, which we connect to the within-person variability score
by the consistency assumption}
\begin{gather}
Y^*_{ik}={Y^*_{ik}}^{\bar{a}^*_{i(k-1)}}
\end{gather}
\textcolor{black}{if ${\bar{A}^*_{i(k-1)}}={\bar{a}^*_{i(k-1)}}$; otherwise, ${Y^*_{ik}}^{\bar{a}^*_{i(k-1)}}$ is counterfactual.
Note that $Y^*_{ik}$ is a latent variable and unobservable, while potential outcomes for measurements (i.e., observed variables) are
assumed in the standard potential outcome approach.} 

\textcolor{black}{In potential outcome approach, causal effect refers to a contrast between potential outcomes under different treatment values. 
Therefore, for each causal effect, we can imagine a (hypothetical) randomized
experiment to quantify it (i.e., target trial; Hern$\acute{a}$n \& Robins, 2021). For example, (average)
causal effect on $Y^*_{ik}$ when a continuous treatment/predictor $A^*_{i(k-1)}$ increases one unit from the 
reference value $a^{*r}_{i(k-1)}$ at time $t_{k-1}$  
can be expressed as $E({Y^*_{ik}}^{\bar{a}^*_{i(k-2)},a^{*r}_{i(k-1)}+1}-{Y^*_{ik}}^{\bar{a}^*_{i(k-2)},a^{*r}_{i(k-1)}})=E({Y^*_{ik}}^{\bar{a}^*_{i(k-2)},a^{*r}_{i(k-1)}+1})-E({Y^*_{ik}}^{\bar{a}^*_{i(k-2)},a^{*r}_{i(k-1)}})$.}

The standard assumption of no unobserved confounders or sequential ignorability indicates that
\begin{gather}
\textcolor{black}{
{\underline{Y}^*_{ik}}^{{\bar{a}}^*_{i(k-2)},0} \indepe A^*_{i(k-1)} | \bar{L}^*_{i(k-1)}, {\bar{A}^*_{i(k-2)}}=\bar{a}^*_{i(k-2)}.
}
\end{gather}
Here, $(\bar{a}^*_{i(k-2)},0)$ is the counterfactual history, that is, the history that agrees with $\bar{a}^*_{i(k-2)}$ through time $t_{k-2}$ and is zero thereafter.
\textcolor{black}{Along with the assumed causal DAG above as well as consistency and sequential ignorability, we
impose} the stable unit treatment value assumption (SUTVA; \textcolor{black}{no unmodeled spillovers}, e.g., Hong, 2015) \textcolor{black}{and assumptions of positivity (i.e.,
the probability of receiving each level of treatment conditional on past confounders and treatments is
greater than zero) and modularity. Under these assumptions,} the average causal effect on $Y^*_{ik}$ when $A^*_{i(k-1)}$ increases one unit from the 
reference value $a^{*r}_{i(k-1)}$ at time $t_{k-1}$ 
can be expressed using the difference in \textcolor{black}{conditional means} given information on confounders and treatment history as 
\textcolor{black}{
\begin{flalign}
&E({Y^*_{ik}}^{\bar{a}^*_{i(k-2)},a^{*r}_{i(k-1)}+1})-E({Y^*_{ik}}^{\bar{a}^*_{i(k-2)},a^{*r}_{i(k-1)}})\notag \\
=&E(Y^*_{ik}| \bar{L}^*_{i(k-1)}, {\bar{A}^*_{i(k-2)}}=\bar{a}^*_{i(k-2)}, A^*_{i(k-1)}=a^{*r}_{i(k-1)}+1)\notag \\
&-E(Y^*_{ik}| \bar{L}^*_{i(k-1)}, {\bar{A}^*_{i(k-2)}}=\bar{a}^*_{i(k-2)}, A^*_{i(k-1)}=a^{*r}_{i(k-1)}).
\end{flalign}
In other words, the causal effect of treatment $A^{*r}_{i(k-1)}$ at the within-person level can be evaluated by the difference in conditional means of $Y^*_{ik}$ between persons who receive
$A^*_{i(k-1)}=a^{*r}_{i(k-1)}+1$ (i.e., treatment levels that are $a^{*r}_{i(k-1)}+1$ larger than their expected scores $\mu^{(A)}_{(k-1)}+I^{(A)}_{i}$)
and who receive $A^*_{i(k-1)}=a^{*r}_{i(k-1)}$, given information on confounders and treatment history. 
\footnote{\textcolor{black}{If we use a $do$-operator for a hypothetical experiment, which is popular for defining and quantifying the causal effect 
in the structural causal model approach (Pearl, 2009), Equation~(10) can be expressed by (conditional) interventional means (e.g., Gische et~al., 2021;
Gische \& Voelkle, in press) 
as $E(Y^*_{ik}| \bar{L}^*_{i(k-1)}, do(\bar{A}^*_{i(k-2)}=\bar{a}^*_{i(k-2)}), do(A^*_{i(k-1)}=a^{*r}_{i(k-1)}+1))
-E(Y^*_{ik}| \bar{L}^*_{i(k-1)}, do(\bar{A}^*_{i(k-2)}=\bar{a}^*_{i(k-2)}), do(A^*_{i(k-1)}=a^{*r}_{i(k-1)}))$.
The (conditional) interventional means are
numerically equivalent to conditional means (i.e., Equation~10)  
under the causal DAG such as in Figure~1b in combination with linear equations and normally distributed residual terms.
However, these quantities are conceptually different (see Gische et~al., 2021; Gische \& Voelkle, in press) and differ
under nonlinear models or non-Gaussian models. MSMs and SNMs are methods based on the potential outcome approach, and a 
$do$-operator has not been used explicitly in this context.}}}

\textcolor{black}{Similarly, the average joint (causal) effects of a sequence of treatments/predictors 
${{\bar{A}}^*_{i(k-1)}}$ on $Y^*_{ik}$ when they increase one unit from the reference values $\bar{a}^{*r}_{i(k-1)}$ can be expressed 
as
\begin{flalign}
&E({Y^*_{ik}}^{\bar{a}^{*r}_{i(k-1)}+1})-E({Y^*_{ik}}^{\bar{a}^{*r}_{i(k-1)}})\notag \\
=&E(Y^*_{ik}| \bar{L}^*_{i(k-1)}, {\bar{A}^*_{i(k-1)}}=\bar{a}^*_{i(k-1)}+1)-E(Y^*_{ik}| \bar{L}^*_{i(k-1)}, {\bar{A}^*_{i(k-1)}}=\bar{a}^*_{i(k-1)})
\end{flalign}
Figure~2 provides conceptual diagram modified from Figure 1b to account for the joint interventions ${\bar{A}}^*_{i(k-1)}={\bar{a}}^{*}_{i(k-1)}$
when DGP can be represented by linear and first-order models.}

\textcolor{black}{As a simple example, suppose $K=2$ and that the DGP can be represented
by linear and first-order models as in Equations~(2) and (3) (assuming homogeneity and no interaction 
\begin{figure}[htbp]
\includegraphics[height=16cm,width=24cm,angle=90]{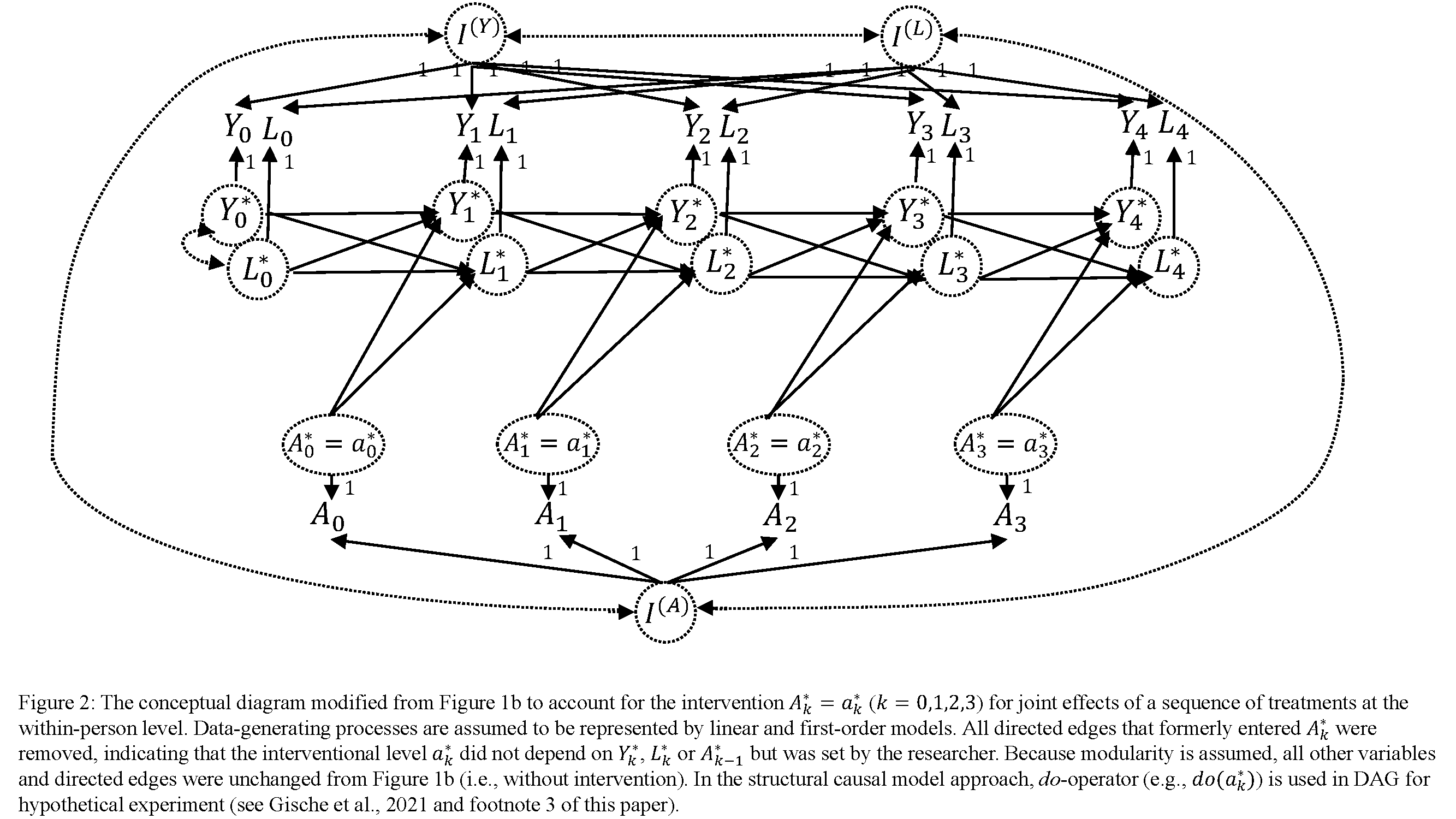}
\end{figure}
effects of treatments/predictors). Then, a conditional mean 
$E(Y^*_{i2}| \bar{L}^*_{i1}, \bar{A}^*_{i1}=\bar{a}^*_{i1})$ at $k=2$ can be expressed as the linear (weighted) sum of the terms $a^*_{i0}$ and $a^*_{i1}$:
\begin{flalign}
&E(\alpha^{(Y)}_{2}Y^*_{i1}+\beta^{(Y)}_{2}a^*_{i1}+\gamma^{(Y)}_{2}L^*_{i1}+d^{(Y)}_{i2}) \notag \\
=& \alpha^{(Y)}_{2}(\alpha^{(Y)}_{1}E(Y^*_{i0})+\beta^{(Y)}_{1}a^*_{i0}+\gamma^{(Y)}_{1}E(L^*_{i0})
)+\beta^{(Y)}_{2}a^*_{i1}+\gamma^{(Y)}_{2}(\alpha^{(L)}_{1}E(Y^*_{i0})+\beta^{(L)}_{1}a^*_{i0}+\gamma^{(L)}_{1}E(L^*_{i0})
)\notag \\
=& \underbrace{[\alpha^{(Y)}_{2}(\alpha^{(Y)}_{1}E(Y^*_{i0})+\gamma^{(Y)}_{1}E(L^*_{i0}))+
\gamma^{(Y)}_{2}(\alpha^{(L)}_{1}E(Y^*_{i0})+\gamma^{(L)}_{1}E(L^*_{i0}))]}_{\alpha_2}\notag \\
&+\underbrace{[\alpha^{(Y)}_{2}\beta^{(Y)}_{1}+\gamma^{(Y)}_{2}\beta^{(L)}_{1}]}_{\beta_{20}}a^*_{i0}+\underbrace{\beta^{(Y)}_{2}}_{\beta_{21}}a^*_{i1}\notag \\
=& \alpha_2+\beta_{20}a^*_{i0}+\beta_{21}a^*_{i1}.
\end{flalign}
From this result, joint (causal) effects
of treatments $A^*_{i0}$ and $A^*_{i1}$ when increasing one unit from the reference values 
$a^{*r}_{i0}$ and $a^{*r}_{i1}$ become $(\alpha_2+\beta_{20}(a^{*r}_{i0}+1)+\beta_{21}(a^{*r}_{i1}+1))
-(\alpha_2+\beta_{20}a^{*r}_{i0}+\beta_{21}a^{*r}_{i1})=
\beta_{20}+\beta_{21}=\alpha^{(Y)}_{2}\beta^{(Y)}_{1}+\gamma^{(Y)}_{2}\beta^{(L)}_{1}+\beta^{(Y)}_{2}.$
Note that $\beta_{20}$ (the effect of intervention $A^*_0=a^*_0$ on $Y^*_2$) can also be evaluated
by tracing the two paths $a^*_0 \rightarrow Y^*_1 \rightarrow Y^*_2$ ($=\alpha^{(Y)}_{2}\beta^{(Y)}_{1}$) and $a^*_0 \rightarrow L^*_1 \rightarrow Y^*_2$
($=\gamma^{(Y)}_{2}\beta^{(L)}_{1}$) that start at $A^*_0 (=a^*_0)$ and end at $Y^*_2$ in Figure~2. 
Likewise, $E(Y^*_{i1}| L^*_{i0}, A^*_{i0}=a^*_{i0})$ at $k=1$ can be expressed as
\begin{flalign}
E(\alpha^{(Y)}_{1}Y^*_{i0}+\beta^{(Y)}_{1}a^*_{i0}+\gamma^{(Y)}_{1}L^*_{i0})=
\underbrace{[\alpha^{(Y)}_{1}E(Y^*_{i0})+\gamma^{(Y)}_{1}E(L^*_{i0})]}_{\alpha_1}+\underbrace{\beta^{(Y)}_{1}}_{\beta_{10}}a^*_{i0}
=\alpha_1+\beta_{10}a^*_{i0},
\end{flalign}
thus the causal effect of treatment $A^*_{i0}$ when
increasing one unit from the reference values $a^{*r}_{i0}$ at the within-person level becomes $\beta_{10}=\beta^{(Y)}_{1}$,
which is equivalent to the so-called cross-lagged parameter in Equation~(3).}

\textcolor{black}{From Equations~(5) and (6) (i.e., stable trait factors are uncorrelated with within-person variability scores), 
we have the relation $E(Y^*_{ik}| \bar{L}^*_{i(k-1)}, {\bar{A}^*_{i(k-1)}}=\bar{a}^*_{i(k-1)})$\\=$E(Y_{ik}|I_{i}^{(Y)}, I_{i}^{(L)}, I_{i}^{(A)}, \bar{L}^*_{i(k-1)}, {\bar{A}^*_{i(k-1)}}=\bar{a}^*_{i(k-1)})-(\mu_{k}^{(Y)}+I_{i}^{(Y)})$.
Because $\mu_{k}^{(Y)}+I_{i}^{(Y)}$ is the term
that is not associated with treatments at the within-person level, it can be shown that
\begin{flalign}
&E(Y^*_{ik}| \bar{L}^*_{i(k-1)}, {\bar{A}^*_{i(k-1)}}=\bar{a}^*_{i(k-1)}+1)-E(Y^*_{ik}| \bar{L}^*_{i(k-1)}, {\bar{A}^*_{i(k-1)}}=\bar{a}^*_{i(k-1)})\notag \\
=&E(Y_{ik}|I_{i}^{(Y)}, I_{i}^{(L)}, I_{i}^{(A)}, \bar{L}^*_{i(k-1)}, {\bar{A}^*_{i(k-1)}}=\bar{a}^*_{i(k-1)}+1)-E(Y_{ik}|I_{i}^{(Y)}, I_{i}^{(L)}, I_{i}^{(A)}, \bar{L}^*_{i(k-1)}, {\bar{A}^*_{i(k-1)}}=\bar{a}^*_{i(k-1)}).
\end{flalign}
The right side of the equation can be interpreted as person-specific joint (causal) effects 
in the sense that it accounts for stable traits of persons ($I_i^{(Y)}$).
Therefore, joint (causal) effects of $A^*$ on $Y^*$  (i.e., Equation~11; at the within-person level) can be interpreted as
person-specific joint (causal) effects on $Y$ under the assumed linear causal DAG such as in Figure~1b.}

\textcolor{black}{
Although the current focus is the (joint) intervention of treatments/predictors at the within-person level (i.e., within-person variability scores),
we can consider how the intervention of stable traits of treatments/predictors influences outcomes. 
Let ${Y_{ik}}^{I_i^{(A)},{\bar{A}}^*_{i(k-1)}}$ ($k=1,\dots,K$) denote the potential outcome
that would take at time point $t_k$ for person $i$ were this person to have stable traits of treatments/predictors $I_i^{(A)}$
and to receive treatment history at the within-person level
$\bar{A}^*_{i(k-1)}=\{A^*_{i0},\dots,A^*_{i(k-1)}\}$ through $t_{k-1}$. The (person-specific) causal effect on $Y_{ik}$ when $I_i^{(A)}$ increases one unit from the 
reference value $z^{(A)r}_i$ given information on confounders and treatments at the within-person level as well as
stable traits of other variables (i.e., $I^{(Y)}$ and $I^{(L)}$) can be expressed as 
\begin{flalign}
&E({Y_{ik}}^{z^{(A)r}_i+1,{\bar{A}}^*_{i(k-1)}})-E({Y_{ik}}^{z^{(A)r}_i,{\bar{A}}^*_{i(k-1)}})\notag \\
=&E(Y_{ik}| I_{i}^{(Y)}, I_{i}^{(L)}, I_{i}^{(A)}=z^{(A)r}_i+1, \bar{L}^*_{i(k-1)}, \bar{A}^*_{i(k-1)})-E(Y_{ik}| I_{i}^{(Y)}, I_{i}^{(L)}, I_{i}^{(A)}=z^{(A)r}_i, \bar{L}^*_{i(k-1)}, \bar{A}^*_{i(k-1)}).
\end{flalign}
Because intervention of $I_i^{(A)}$ only influences $A_{ik}$ (which act as colliders), not $Y_{ik}$ (or $I_i^{(Y)}$, or $Y^*_{ik}$, or $A^*_{ik}$),
conditional means in Equation~(15) are not influenced by this intervention (e.g., Figure~S2 in the Online Supplemental Material for a conceptual diagram). Therefore, given information of $I^{(Y)}$,
this (person-specific) causal effect is equal to zero. This indicates that magnitudes of expected scores $\mu^{(A)}+I^{(A)}_{i}$ (i.e., $I^{(A)}_{i}$)
are not an issue. Rather, given information of $I^{(Y)}_i$, magnitudes of within-person variations as deviations from expected scores
are the keys to assess causal effects of treatments in the currently assumed DGP.}
\subsection{\textcolor{black}{Identification Conditions for Causal Parameters}} 
\textcolor{black}{So far, we have assumed a causal DAG model that is similar to that shown in Figure~1b.
In the proposed method, the assumptions for identifying parameters for joint (causal) effects can now be summarized as} 
(i) measurements ($K \geq 2$) are expressed by the linear sum of stable trait factors and within-person variability scores that are mutually uncorrelated, 
(ii) within-person variability scores are expressed by functions (with assumption of homogeneity) of those in past time,
(iii) consistency, (iv) sequential ignorability, (v) SUTVA, \textcolor{black}{(vi) positivity, (vii) modularity, and (viii) multivariate normality (if MLE is used in
the first step).}

\textcolor{black}{Regarding the second assumption, if $K \geq 2$ and the DGP can be represented
by linear (and first-order) equations such as in Equations~(2) and (3) (assuming homogeneity and no interaction effects of treatments/predictors),
then in this special case, the RI-CLPM (that includes time-varying observed confounders) as a statistical model can identify parameters for joint (causal) effects.
However}, the linearity assumption that is typically imposed for time-varying observed confounders and 
outcomes (and treatments/predictors) in path modeling and SEM (including the RI-CLPM) has often been 
criticized in the causal inference literature (e.g., Hong, 2015), and relaxing this assumption 
is often key to consistently estimating the causal quantity of interest (e.g., Imai \& Kim, 2019).
In addition, ensuring a correct specification in terms of the linearity
is very challenging in that many equations must be diagnosed in longitudinal designs. 

\textcolor{black}{As we will see, the proposed method still requires 
specifications of the structure for within-person variability scores
in each variable ($Y$, $A$, and $L$ in the first step) as well as parametric models for treatments and outcomes at
the within-person level (in the second step).
However, the assumption of linearity is not required for these parametric models in MSMs and SNMs,
and in MSMs one does not need to model the relation between outcomes and time-varying observed confounders (at the within-person level)
because it is the means of potential outcomes that are marginalized over these confounders that are of concern.
Notably, SNMs with G-estimation have the property of being doubly robust to 
model misspecifications in how time-varying observed confounders are functionally related with treatments and outcomes (at the within-person level).}
\section{PROPOSED METHODOLOGY}
We are now ready to introduce a method of within-person variability score-based causal inference 
for estimating joint effects of time-varying continuous treatments, \textcolor{black}{assuming that the above conditions for identification
are satisfied}. The proposed method consists of a two-step analysis.
First, within-person variability scores are calculated using weights through SEM that models only the measurement parts \textcolor{black}{that include stable trait factors}. 
Then, causal parameters are estimated by MSMs or SNMs, using the scores calculated in the first step.
This approach is more flexible than the \textcolor{black}{one that relies entirely on SEM (e.g., the RI-CLPM that includes time-varying observed confounders) in terms of} modeling how time-varying observed confounders are \textcolor{black}{functionally related with} treatments/predictors and outcomes at the within-person level,
without imposing the linearity assumption in these relations.
Before explaining the proposed methodology, we briefly discuss the motivation for adopting a two-step method, rather than
simultaneously estimating stable trait factors (or within-person variations) and causal parameters.

In general, partial misspecification in measurements and/or structural models is known to cause large biases in estimates of model parameters. In the present context, 
when a simultaneous estimation procedure such as the RI-CLPM is used, misspecification in the structural models at the within-person level
may greatly affect parameter estimates in the measurement model ((co)variances of stable factors and within-person variability scores), and vice versa.

To avoid such confounding in interpreting the estimation results, in the SEM context Anderson and Gerbing
(1988) proposed a two-step procedure that first confirms the measurement model
with a saturated model, so that structural relations have no impact on the measurement model. Then, using an appropriate measurement model, the substantive structural
relations model of interest is added (Hoshino \& Bentler, 2013). 
Applications of similar multistep estimation procedures can be seen
for diverse classes of latent variable models (Bakk \& Kuha, 2017; Croon, 2002; Skrondal \& Laake, 2001; Vermunt, 2010). 

Another potential advantage of two-step estimation is its feasibility.
MSMs and SNMs usually do not assume common factors, and the optimization procedure for these models is different from that in SEM. For this reason, 
fully customized programming is required if performing simultaneous estimation.
However, in two-step estimation, parameters in measurement models can be estimated in the first step through
various software packages for SEM, including Amos, SAS PROC CALIS, R packages (sem, lavaan, OpenMx), LISREL, EQS, and Mplus.
MSMs and SNMs can be straightforwardly applied just by using calculated within-person variability scores instead of measurements.

Two-step estimation is also advantageous because it poses less risk of improper solutions. This problem is encountered relatively often when applying the RI-CLPM 
because of negative variance parameters and a singular approximate Hessian matrix for stable trait factor variance--covariance (e.g., Usami, Todo, \& Murayama, 2019), which 
is likely caused by misspecifications in linear regressions (i.e., the structural model). 
We will separately estimate stable trait factors for each variable ($Y$, $A$, and $L$) without influence from specified structural models, thus minimizing the risk of improper solutions.

\subsection{Step 1: Estimation of Stable Trait Factors and Prediction of Within-person Variability Scores}
The first step of our method is divided into two sub-steps: (i)~specification of the measurement models and parameter estimation and (ii)~prediction of within-person variability scores. 
\subsubsection{Specification of the measurement models and parameter estimation}
As stated earlier, we assume that measurements are expressed by the linear sum of stable trait factors and within-person variability scores that are mutually uncorrelated, as in Equation~(2).
This equation can be viewed as a factor analysis model that includes a single common factor $I$ (whose factor loadings are all one)\footnote{Although we defined stable trait factors as the (time-invariant) difference between the expected value of a given person's measurement and the temporal group mean, one could argue for another definition that allows for time-varying influences on measurements. 
If this is the case, time-varying factor loadings can be freely specified in this step (except for one fixed factor loading for identification).
However, there may be some cost in that the minimum number of time points required to identify the measurement model becomes larger than that in specifying time-invariant loadings.}
and a unique factor as temporal deviations. 
In vector notation, the causal model of Equation~(2) for outcome $Y$ becomes
\begin{gather}
Y_{i}=\mu^{(Y)}+I^{(Y)}_{i}1_{K+1}+Y^*_i,
\end{gather} 
where $\mu^{(Y)}$ is a $(K+1) \times 1$ mean vector, $E(I^{(Y)}_{i})=0$, $Var(I^{(Y)}_{i})=\phi^2_{(Y)}$, $E(Y^*_i)=0$,
and $Cov(I^{(Y)}_i,Y^*_i)=0$. We denote as $\Psi_{(Y)}$ a $(K+1) \times (K+1)$ variance--covariance matrix of within-person variability scores. 
This implies that the variance--covariance matrix of $Y$ (denoted as $\Sigma_{(Y)}$) is of the form $\Sigma_{(Y)}=\phi^2_{(Y)}1_{K+1}1^t_{K+1}+\Psi_{(Y)}$.

Unlike the standard factor analysis model, $\Psi_{(Y)}$ has a dependence structure and is not diagonal. Therefore, in
using SEM to estimate the parameters in Equation~(16), some structure---such as compound symmetry, a Toeplitz structure, or a (first-order) autoregressive (AR) structure---must be specified in $\Psi_{(Y)}$ for model identification.
When the model is correctly specified, consistent estimators for $\mu^{(Y)}$, 
$\phi^2_{(Y)}$, and $\Psi_{(Y)}$ can be obtained by MLE in SEM (J\"{o}reskog \& Lawley, 1968). 

In SEM, missing values can be easily handled by full information maximum likelihood (Enders \& Bandalos, 2001) with the assumption of 
missing at random (MAR; Rubin, 1976). If data are suspected to be missing not at random (MNAR), then appropriate 
sensitivity analyses and/or multiple imputation should be considered (Resseguier, Giorgi, \& Paoletti, 2011). 
Models that account for MNAR can be easily estimated in popular software packages for SEM (see Enders, 2011; Newsom, 2015).

Another advantage of SEM is that validity of the specified model can be diagnosed 
via multiple model fit indices, along with model comparisons using information criteria. 
In this paper, we use three current major indices (e.g., Hu \& Bentler, 1999; Kline, 2016):
(a) the comparative fit index (CFI), (b) the root mean square error of approximation (RMSEA), and (c) the standardized root mean square residual (SRMR).

Similarly, we also set measurement models for treatments/predictors $A$ and observed confounders $L$ separately in this sub-step,
then estimate parameters for mean vectors ($\mu^{(A)}$ and $\mu^{(L)}$),
stable trait factor variances ($\phi^2_{(A)}$ and $\phi^2_{(L)}$), and variance--covariance matrices of within-person variability scores $\Psi_{(A)}$ and $\Psi_{(L)}$.

\subsubsection{Predicting within-person variability scores}
Let $X_i=(Y_i,A_i,L_i)^t$ and $X^*_i=(Y^*_i,A^*_i,L^*_i)^t$ be vectors of measurements and within-person variability scores, respectively,
and let $\mu=(\mu^{(Y)},\mu^{(A)},\mu^{(L)})^t$ be a mean vector. Also let
$\Sigma$ and $\Psi$ be covariance matrices for measurements $X_i$ and within-person variability scores $X^*_i$.

We consider linear prediction of within-person variability scores ${\hat{X}}^*_i$ under the condition that $\Sigma$ and $\Psi$ are known. Consider a $(3K+1) \times (3K+1)$ weight matrix $W$ that provides 
within-person variability scores from measurements as
\begin{gather}
{\hat{X}}^*_i=W^t(X_i-\mu),
\end{gather}
satisfying the relation
\begin{gather}
E({\hat{X}}^*_i{\hat{X}}^{*t}_i)=W^tE[(X_i-\mu)(X_i-\mu)^t]W=W^t\Sigma W=\Psi.  
\end{gather}

Unlike standard applications of factor analysis, we are interested in predicting within-person variability (unique factor) scores, rather than
stable trait factor (common factor) scores. However,
the current problem of determining weights $W$ shares the similar motivation of predicting factor scores.
In the factor analysis literature, a predictor that preserves the covariance structure of 
common factors has been developed as a linear correlation preserving predictor (Anderson \& Rubin, 1956; Green, 1969; ten~Berge, Krijinen, Wansbeek, \& Shapiro, 1999).

With this point in mind, $W$ that can provide the best linear predictor of ${\hat{X}^*}_i$
minimizing the risk function, defined as the trace of a residual covariance matrix 
(i.e., mean squared error MSE(${\hat{X}}^*_i$)=$E[({\hat{X}^*}_i-X^*_i)^t({\hat{X}^*}_i-X^*_i)]$), which also satisfies 
the relation in Equation~(18), can be obtained by utilizing singular value decomposition as
\begin{gather}
W^t=\Psi^{1/2}{(\Psi^{3/2}\Sigma^{-1}\Psi^{3/2})}^{-1/2}\Psi^{3/2}\Sigma^{-1}.
\end{gather}
Here, for a positive (semi)definite matrix $C$, we denote as $C^{1/2}$ 
the positive (semi)definite matrix such that its square equals $C$. Matrices $C^{-1/2}$
and $C^{3/2}$ are the inverse (if it exists) and the third power of $C^{1/2}$, respectively.
A derivation of $W$ is provided in the Online Supplemental Material. 

We use the sample means $\bar{X}$ and covariance matrix $S$ of $X$ as estimators of $\mu$ and $\Sigma$.
As implied from the relation in Equation~(\textcolor{black}{7}), we use estimated stable trait factor variances to estimate $\Psi$ as
\begin{gather}
\hat{\Psi} = S-\hat{\Phi}^+,
\end{gather}
where $\hat{\Phi}^+$ consists of estimated stable trait factor (co)variances. In the
simple case where the initial measurement of $Y$ ($Y_0$) is missing and
the number of measurements equals $K$ for each variable, $\hat{\Phi}^+$ becomes
\begin{gather}
\hat{\Phi}^+=\hat{\Phi} \otimes 1_K1^t_K=\left(
\begin{array}{ccc}
\hat{\phi}^2_{(Y)} & \hat{\phi}_{(Y,A)} & \hat{\phi}_{(Y,L)} \\
\hat{\phi}_{(Y,A)} & \hat{\phi}^2_{(A)} & \hat{\phi}_{(A,L)} \\
\hat{\phi}_{(Y,L)} & \hat{\phi}_{(A,L)} & \hat{\phi}^2_{(L)}
\end{array}
\right) \otimes 1_K1^t_K,
\end{gather}
where $\hat{\Phi}$ is an estimator of a $3 \times 3$ stable trait factor covariance matrix $\Phi$.
Because stable trait factor covariances are not estimated in the previous sub-step,
we use covariances between calculated linear correlation preserving predictors for variables. 
For example, this predictor for $Y$ can be expressed as
\begin{gather}
\hat{I}_{i}^{(Y)}=\frac{\hat{\phi}_{(Y)}}{\sqrt{1_{K+1}^t{\hat{\Sigma}^{-1}_{(Y)}}1_{K+1}}}1_{K+1}^t{\hat{\Sigma}^{-1}_{(Y)}}(Y_i-\bar{Y}).
\end{gather}
$\hat{I}_{i}^{(A)}$ and $\hat{I}_{i}^{(L)}$ can be calculated in the same manner, whereby we obtain
$\hat{\phi}_{(Y,A)}=Cov(\hat{I}_{i}^{(Y)},\hat{I}_{i}^{(A)})$, $\hat{\phi}_{(Y,L)}=Cov(\hat{I}_{i}^{(Y)},\hat{I}_{i}^{(L)})$,
and $\hat{\phi}_{(A,L)}=Cov(\hat{I}_{i}^{(A)},\hat{I}_{i}^{(L)})$.
Predictors $\hat{I}_i=(\hat{I}_{i}^{(Y)},\hat{I}_{i}^{(A)},\hat{I}_{i}^{(L)})^t$ satisfy 
the relation $E(\hat{I}_{i}\hat{I}_{i}^t)=\Phi$ if the model is correctly specified
in the previous sub-step. From Equations~(17) and (19)--(21), we can thus obtain ${\hat{X}}^*_i$
without specifying the structural models that connect within-person variability scores from
different variables ($Y\textcolor{black}{^*}$, $A\textcolor{black}{^*}$, and $L\textcolor{black}{^*}$), successfully maintaining independence from the next step.

\subsection{Applying MSMs and SNMMs}
The second step of the proposed method is straightforward, because we just need to apply MSMs or SNMs using calculated within-person variability scores.
Robins and co-workers developed SNMs with G-estimation (Robins, 1989; Robins, Blevins, Ritter, \& Wulfsohn, 1992)
and MSMs with
an inverse probability weight (IPW) estimator (Robins, 1999; Robins, Hern\'{a}n \& Brumback, 2000). These methods have been extended to treat clustered outcomes
(e.g., Brumback, He, Prasad, Freeman, \& Rheingans, 2014; He, Stephens-Shields, \& Joffe, 2015, 2019). However,
(joint) causal effects under the control of stable trait factors
have not been investigated in this area because inference for stable traits and within-person relations has been an issue in 
the psychometric and behavioral science literature, and these concepts have yet to be fully characterized in the causal inference literature.

MSMs are advantageous in that they can be easily understood and
fit with standard, off-the-shelf software that allows for weights (e.g., He, Stephens-Shields, \& Joffe, 2019; Vansteelandt \& Joffe, 2014).
However, it is well known that MSMs can be highly sensitive to misspecification 
of the treatment assignment model, even when there is a moderate number of time points 
(e.g., Hong, 2015; Lefebvre, Delaney, \& Platt, 2008). Imai and Ratkovic (2015) proposed a covariate balancing propensity score methodology for robust IPW estimation.

Because of the attractive property of being doubly robust in $G$-estimators, 
SNMs are a better approach for handling violation of the usual assumptions of no unmeasured confounders or sequential ignorability (Vansteelandt \& Joffe, 2014).
In addition, SNMs can allow direct modeling of the interactions and moderation effects of treatments/predictors $A$ with observed confounders $L$.
Another advantage of SNMs is that the variance of 
locally efficient IPW estimators in MSMs exceeds that of $G$-estimators in SNMs, unless $A$ and $L$ are independent.
We therefore emphasize the utility of SNMs in this paper.
Because we are now interested in evaluating the joint effects of treatments on the mean of an outcome, rather than
those on the entire distribution of the outcome, we apply structural nested mean models (SNMMs; Robins, 1994). 

Note that potential disadvantages of SNMs are their limited utility for G-estimation when applying logistic SNMs and their limited availability of off-the-shelf software.
Regarding the latter point, Wallace, Moodie, and Stephens (2017) developed an R package for G-estimation of SNMMs.

\subsubsection{MSMs using within-person variability scores}
MSMs are typically applied to evaluate \textcolor{black}{the joint effects of a sequence of treatments} on the outcome, which is measured only at the end of a fixed follow-up period ($t_K$).
For generality of discussion, as before we assume that the outcome is measured each time and that the primary interest is evaluation of effects of a sequence of past treatments on the outcome at each time point.
 
MSMs consider the marginal mean of potential outcomes \textcolor{black}{that are marginalized over the observed confounders $L$}.
In the current context, we consider potential outcomes at the within-person level, namely, $E({Y^*_{ik}}^{\bar{A}^*_{i(k-1)}})$ with 
treatment history $\bar{A}^*_{i(k-1)}=\bar{a}^*_{i(k-1)}$. $E({Y^*_{ik}}^{\bar{A}^*_{i(k-1)}})$ might take the form
\begin{gather}
E({Y^*_{ik}}^{\bar{A}^*_{i(k-1)}})=\textcolor{black}{\alpha_{k}}+\sum_{t=1}^{k}\textcolor{black}{\beta_{k(t-1)}}A^*_{i(t-1)}
\end{gather}
with $k=1,2,\dots,K.$\footnote{\textcolor{black}{Other terms such as quadratic effects (e.g., $A^{*2}_{ik}$) 
for time-varying treatments can be included in MSMs. Also, one can include
observed covariates/non-confounders to assess effect modification (Hern$\acute{a}$n \& Robins, 2021).}} \textcolor{black}{The average joint (causal) effects of $\bar{A}^*_{i(k-1)}$
on $Y^*_{ik}$ when increasing one unit from the reference values in each treatment become $\sum_{t=1}^k\beta_{k(t-1)}$}.
Parameters \textcolor{black}{$\tau=(\alpha_1,\dots,\alpha_K, \beta_{10}, \beta_{20}, \beta_{21}, \dots, \beta_{K(K-1)})^t$} can be estimated by fitting a weighted conditional model with an IPW estimator. 
One useful option for calculating weights is to use stabilized weights $w_{ik}$ for person $i$ at time point $t_k$ (Hern\'{a}n, Brumback, \& Robins, 2002) as
\begin{gather}
w_{ik}=\prod_{t=1}^{k-1}\frac{f(A^*_{it}|A^*_{i(t-1)})}{f(A^*_{it}|A^*_{i(t-1)},L^*_{it})},
\end{gather}
where $f(A^*_{it}|A^*_{i(t-1)},L^*_{it})>0$ for all $A^*_{it}$, if $f(A^*_{i(t-1)},L^*_{it}) \neq 0$ (the positivity assumption).
Parameters will be biased if the treatment assignment model $f(A^*_{it}|A^*_{i(t-1)},L^*_{it})$ is misspecified, but misspecification of $f(A^*_{it}|A^*_{i(t-1)})$ does not result in bias.
In MSMs, unlike the RI-CLPM (that includes $L$), one does not need to model the relation between outcomes and time-varying observed confounders 
at the within-person level because marginal (joint) effects of treatments/predictors are the primary focus in applying this method. 
Also, one can allow a nonlinear relation between 
treatments/predictors and confounders in the treatment assignment model $f(A^*_{it}|A^*_{i(t-1)},L^*_{it})$, although estimates
are sensitive to this model misspecification. 
 \subsubsection{SNMMs using within-person variability scores}
SNMMs simulate the sequential removal of an amount (\emph{blip}) of treatment at $t_{k-1}$ on subsequent average outcomes, after having removed the effects of all
subsequent treatments. SNMMs then model the effect of a blip in treatment at $t_{k-1}$ on the subsequent outcome means
while holding all future treatments fixed at a reference level 0 (Vansteelandt \& Joffe, 2014); in other words, the level that is equal to expected scores of a person
in the current context. 

SNMMs parameterize contrasts of ${\underline{Y}^*}_{ik}^{\bar{a}^*_{i(k-1)},0}$
and ${\underline{Y}^*}_{ik}^{\bar{a}_{i(k-2)},0}$ conditionally on treatments/predictors and
\textcolor{black}{confounder} histories through $t_{(k-1)}$ as
\begin{flalign}
&g[E({\underline{Y}^*}_{ik}^{\bar{a}^*_{i(k-1)},0}|\bar{L}^*_{i(k-1)}=\bar{l}^*_{i(k-1)}, \bar{A}^*_{i(k-1)}=\bar{a}^*_{i(k-1)})]-
 g[E({\underline{Y}^*}_{ik}^{\bar{a}^*_{i(k-2)},0}|\bar{L}^*_{i(k-1)}=\bar{l}^*_{i(k-1)}, \bar{A}^*_{i(k-1)}=\bar{a}^*_{i(k-1)})]\notag \\
&=h_{k}(\bar{l}^*_{i(k-1)},\bar{a}^*_{i(k-1)};\tau) 
\end{flalign}
for each $k=1,\dots,K$, where $g(\cdot)$ is a known link function, and $h_{k}(\bar{l}^*_{i(k-1)},\bar{a}^*_{i(k-1)};\tau)$ is a known $(K-k+1)$-dimensional function, smooth in the finite-dimensional parameter $\tau$ (Vansteelandt \& Joffe, 2014).  

In the following empirical applications using the data of $K=2$, a linear SNMM using the identity link $g(x) = x$ is given by
\begin{flalign}
&E({Y^*_{i2}}^{a_{i0},a_{i1}}-{Y^*_{i2}}^{a_{i0},0} | \bar{L}^*_{i1}=\bar{l}^*_{i1},\bar{A}^*_{i1}=\bar{a}^*_{i1})=(\textcolor{black}{\beta_{21}+\gamma_{21}} l^*_{i1})a^*_{i1}, \notag \\
&E({Y^*_{i2}}^{a_{i0},0}-{Y^*_{i2}}^{0,0} | L^*_{i0}=l^*_{i0},A^*_{i0}=a^*_{i0})=(\textcolor{black}{\beta_{20}+\gamma_{20}}l^*_{i0})a^*_{i0}, \\
&E({Y^*_{i1}}^{a_{i0},0}-{Y^*_{i1}}^{0,0} | L^*_{i0}=l^*_{i0},A^*_{i0}=a^*_{i0})=(\textcolor{black}{\beta_{10}+\gamma_{10}}l^*_{i0})a^*_{i0}.\notag 
\end{flalign}
Here, the first equation models the effect of $A^*_{i1}$ on $Y^*_{i2}$, the second models the effect of $A^*_{i0}$ on $Y^*_{i2}$, and the
third models the effect of $A^*_{i0}$ on $Y^*_{i1}$. \textcolor{black}{The (conditional) average joint effects of $A^*_{i0}$ and $A^*_{i1}$
on $Y^*_{i2}$ when increasing one unit from the reference values in each treatment become $\beta_{20}+\gamma_{20}l^*_{i0}+\beta_{21}+\gamma_{21}l^*_{i1}$.
This effect becomes $\beta_{20}+\beta_{21}$ if there are no interaction effects between confounders and treatments}.

SNMMs consider a transformation $U^*_{im}(\tau)$
of $\underline{Y}^*_{ik}$, the mean value of which is equal to the mean that would be observed
if treatment were stopped from time $t_{k-1}$ onward, in the sense that
\begin{gather}
E(U^*_{i(k-1)}(\tau)|\bar{L}^*_{i(k-1)},\bar{A}^*_{i(k-2)}=\bar{a}^*_{i(k-2)},A^*_{i(k-1)})
=E({\underline{Y}^*}_{ik}^{\bar{a}^*_{i(k-2)},0}|\bar{L}^*_{i(k-1)},\bar{A}^*_{i(k-2)}=\bar{a}^*_{i(k-2)},A^*_{i(k-1)})
\end{gather}
for $k=1,\dots,K$. Here, $U^*_{i(k-1)}(\tau)$
is a vector with components $Y^*_{im}-\sum_{l=k-1}^{m-1}h_{l,m}(\bar{L}^*_{il},\bar{A}^*_{il};\tau)$
for $m=k,\dots,K$ if $g(\cdot)$ is the identity link. For instance, in the above example of $K=2$,
\begin{flalign}
&U_{i1}^*(\tau)=Y^*_{i2}-(\textcolor{black}{\beta_{21}+\gamma_{21}}L^*_{i1})A^*_{i1}, \notag \\
&U_{i0}^*(\tau)=(Y^*_{i1}-(\textcolor{black}{\beta_{10}+\gamma_{10}}L^*_{i0})A^*_{i0}, Y^*_{i2}-(\textcolor{black}{\beta_{21}+\gamma_{21}}L^*_{i1})A^*_{i1}-(\textcolor{black}{\beta_{20}+\gamma_{20}}L^*_{i0})A^*_{i0})^t.
\end{flalign}

The assumptions of sequential ignorability (Equation~\textcolor{black}{9}) together with identity (Equation~27) 
imply that
\begin{gather}
E(U^*_{i(k-1)}(\tau^*) |\bar{L}^*_{i(k-1)},\bar{A}^*_{i(k-1)}) = E(U^*_{i(k-1)}(\tau^*) |\bar{L}^*_{i(k-1)},\bar{A}^*_{i(k-2)})
\end{gather}
for $k=1,\dots,K$. The parameters $\tau$ can therefore be estimated by solving the estimating equation
\begin{flalign}
\sum_{i=1}^N\sum_{k=1}^{K}&[d_{k-1}(\bar{L}^*_{i(k-1)},\bar{A}^*_{i(k-1)})
-E(d_{k-1}(\bar{L}^*_{i{k-1}},\bar{A}^*_{i{k-1}})|\bar{L}^*_{i(k-1)},\bar{A}^*_{i(k-2)})] \circ \notag \\
& V^{-1} \circ [U^*_{i(k-1)}(\tau)-E(U^*_{i(k-1)}(\tau)|\bar{L}^*_{i(k-1)},\bar{A}^*_{i(k-2)})]=0,
\end{flalign}
where $d_{k-1}(\bar{L}^*_{i(k-1)},\bar{A}^*_{i(k-1)})$ is an arbitrary $p \times (K-k+1)$-dimensional function, with $p$ the
dimension of $\tau$, and $V^{-1}$ is a $p \times (K-k+1)$-dimensional vector that includes the 
reciprocal of the variance of each element in $U^*_{i(k-1)}(\tau)-E(U^*_{i(k-1)}(\tau)|\bar{L}^*_{i(k-1)},\bar{A}^*_{i(k-2)})$.

This estimating equation essentially sets the sum across the time points of the conditional covariances between $U^*_{i(k-1)}(\tau)$ and the
function $d_{k-1}(\bar{L}^*_{i(k-1)},\bar{A}^*_{i(k-1)})$, \textcolor{black}{given} $\bar{L}^*_{i(k-1)}$ and $\bar{A}^*_{i(k-2)}$, are zero.
If there is homoscedasticity in $V$, then local semiparametric efficiency under
the SNMM is attained upon choosing
\begin{gather}
d_{k-1}(\bar{L}^*_{i(k-1)},\bar{A}^*_{i(k-1)})=E\left[\left.\frac{\partial U^*_{i(k-1)}(\tau^*)}{\partial \tau}\right|\bar{L}^*_{i(k-1)},\bar{A}^*_{i(k-1)}\right]
\end{gather}
(Vansteelandt \& Joffe, 2014). Solving the estimating equation~(30) requires a parametric model $\mathcal{A}$
for the treatment/predictor $A^*_{ik}$:  $f(A^*_{i(k-1)} | \bar{L}^*_{i(k-1)},\bar{A}^*_{i(k-2)};\eta)$
with $k=1,\dots,K$. It also requires a parametric model $\mathcal{B}$ for the conditional
mean of $U^*_{i(k-1)}(\tau)$, namely, 
$f(U^*_{i(k-1)}(\tau) | \bar{L}^*_{i(k-1)},\bar{A}^*_{i(k-2)};\kappa)$.
Notably, when the parameters $\eta$ and $\kappa$ are variation-independent, 
$G$-estimators that solve Equation~(30), obtained by substituting $\eta$ and $\kappa$
with consistent estimators, are doubly robust (Robins \& Rotnitzky,
2001, cited from Vansteelandt \& Joffe, 2014), meaning that \textcolor{black}{estimates of causal parameters} are consistent when either model
$\mathcal{A}$ or model $\mathcal{B}$ is correctly specified. 
In addition, unlike the RI-CLPM (that includes $L$), one can allow nonlinear effects of
time-varying observed confounders on treatments/predictors and outcomes in models $\mathcal{A}$ and $\mathcal{B}$.
\section{SIMULATION STUDIES}
\subsection{Method}
This section describes a Monte Carlo simulation for systematically investigating how effectively the proposed method using calculated within-person variability scores
can recover causal parameters, and it presents comparisons of estimation performance versus other potential (centering) methods to account for stable traits.
We consider two different scenarios: (i)
the assumed linear (and first-order) DGP of Figure~1b (i.e., causal models represented in Equations~2 and 3) is correct and other assumptions of consistency, sequential ignorability, SUTVA, 
positivity, modularity, and \textcolor{black}{multivariate normality} are all satisfied, and (ii) some assumptions are violated and the statistical model contains misspecifications. In the whole simulation, for simplicity we also assume that causal effects
are homogeneous among persons and interactions or moderation effects with observed confounders are not present.

In the first scenario, initial within-person variability scores ($Y^*_{i0}$, $A^*_{i0}$, and $L^*_{i0}$) are first
generated \textcolor{black}{so that they are normally distributed and their variances and covariances become 10 and 3, respectively}.
Then, within-person variability scores at succeeding times are sequentially generated via a first-order linear autoregressive model (i.e., Equation~3) with the stationarity assumption\footnote{In this paper, we use the term \emph{stationarity assumption} to indicate invariance of autoregressive parameters, cross-lagged parameters, and residual 
variance parameters over time, rather than indicating means and (co)variances in variables to be invariant over time.}:
\begin{flalign}
Y^*_{ik}&=0.40Y^*_{i(k-1)}+0.40A^*_{i(k-1)}+0.10L^*_{i(k-1)}+d^{(Y)}_{ik},\notag \\
A^*_{ik}&=0.20Y^*_{ik}+0.40A^*_{i(k-1)}+0.30L^*_{ik}+d^{(A)}_{ik}, \\
L^*_{ik}&=0.20Y^*_{i(k-1)}+0.20A^*_{i(k-1)}+0.50L^*_{i(k-1)}+d^{(L)}_{ik},\notag
\end{flalign}
\textcolor{black}{If $K=4$, this setting produces (see also the calculation in Equation~12)
\begin{flalign}
&E({Y^*_{i4}}^{\bar{a}^*_{i3}})=\alpha_4+\beta_{40}a^*_{i0}+\beta_{41}a^*_{i1}+\beta_{42}a^*_{i2}+\beta_{43}a^*_{i3}=\alpha_4+0.0486a^*_{i0}+0.09a^*_{i1}+0.18a^*_{i2}+0.40a^*_{i3},\notag \\
&E({Y^*_{i3}}^{\bar{a}^*_{i2}})=\alpha_3+\beta_{30}a^*_{i0}+\beta_{31}a^*_{i1}+\beta_{32}a^*_{i2}=\alpha_3+0.09a^*_{i0}+0.18a^*_{i1}+0.40a^*_{i2},\notag \\
&E({Y^*_{i2}}^{\bar{a}^*_{i1}})=\alpha_2+\beta_{20}a^*_{i0}+\beta_{21}a^*_{i1}=\alpha_2+0.18a^*_{i0}+0.40a^*_{i1},\notag \\
&E({Y^*_{i1}}^{a^*_{i0}})=\alpha_1+\beta_{10}a^*_{i0}=\alpha_1+0.40a^*_{i0},
\end{flalign}
where $\alpha_k$ $(k=1,\dots,4)$ is a constant. Because no moderation effects are assumed, estimating 10 different causal parameters $\tau_{K=4}=(\beta_{10},\beta_{20},\beta_{21},\beta_{30},\beta_{31},\beta_{32},\beta_{40},\beta_{41},\beta_{42},\beta_{43})^t$ is a common
goal between MSMs and SNMMs.} The variance of normal residual $d$ was set to 5 for each variable,
making the variance of within-person variability scores for each variable become almost 10 at each time point (the proportion of variance explained in Equation~32 becomes almost 50\%).

Independently of generating within-person variability scores, three kinds of stable trait factors ($I^{(Y)}_{i}$, $I^{(A)}_{i}$, and $I^{(L)}_{i}$) are generated by multivariate normal with a correlation of $0.3$.
Observed values are then generated using the relation of Equation~(2),
\begin{gather}
Y_{ik}=I^{(Y)}_{i}+Y^*_{ik}, \hspace{3mm}A_{ik}=I^{(A)}_{i}+A^*_{ik}, \hspace{3mm}L_{ik}=I^{(L)}_{i}+L^*_{ik},
\end{gather}
where temporal group means are set to zero at each time point (i.e., $\mu^{(Y)}_{k}=\mu^{(A)}_{k}=\mu^{(L)}_{k}=0).$

In this simulation, we systematically changed the total number of persons as $N= 200, 600$, and 1000, the number of time points as $K= 4$ and 8, and 
the size of stable trait factor variances as $\phi_{(Y)}^2=\phi_{(A)}^2=\phi_{(L)}^2= 10/9, 30/7$, and 10. 
This setting of stable trait factor variances indicates that the proportion of this variance
to that of measurements becomes around 10\%, 30\%, and 50\%, respectively, at each time point. 
\textcolor{black}{To make it easier to compare the results between the $K=4$ and $K=8$ conditions, in $K=8$ we suppose
only $A^*_{i4}$, $A^*_{i5}$, $A^*_{i6}$, and $A^*_{i7}$ are intervened, while controlling
for $A^*_{i0}$, $A^*_{i1}$, $A^*_{i2}$, and $A^*_{i3}$. This setting
produces conditional means of (potential) outcomes as functions of treatments intervened:
$\alpha_8+\beta_{84}a^*_{i4}+\beta_{85}a^*_{i5}+\beta_{86}a^*_{i6}+\beta_{87}a^*_{i7}
=\alpha_8+0.0486a^*_{i4}+0.09a^*_{i5}+0.18a^*_{i6}+0.40a^*_{i7}$ at $k=8$,
$\alpha_7+\beta_{74}a^*_{i4}+\beta_{75}a^*_{i5}+\beta_{76}a^*_{i6}
=\alpha_7+0.09a^*_{i4}+0.18a^*_{i5}+0.40a^*_{i6}$ at $k=7$,
$\alpha_6+\beta_{64}a^*_{i4}+\beta_{65}a^*_{i5}
=\alpha_6+0.18a^*_{i4}+0.40a^*_{i5}$ at $k=6$,
and $\alpha_5+\beta_{54}a^*_{i4}=\alpha_5+0.40a^*_{i4}$ at $k=5$, 
where $\alpha_k$ $(k=5,\dots,8)$ is a constant. There are a total of 10 causal parameters 
$\tau_{K=8}=(\beta_{54},\beta_{64},\beta_{65},\beta_{74},\beta_{75},\beta_{76},\beta_{84},\beta_{85},\beta_{86},\beta_{87})^t$
that are equal to those in the $K=4$ condition (i.e., $\tau_{K=4}=\tau_{K=8}$).}

By crossing these factors, we generated 200 simulation data for each combination of factors.  
For comparison, each simulation dataset was analyzed by MSMs and SNMMs using four different scores:
1)~true within-person variability scores (true factor score centering: e.g., ${Y}^*_{ik}=Y_{ik}-I^{(Y)}_i$ for $Y$), 
2)~within-person variability scores predicted by the proposed method (Equation~17), 3)~scores based on observed person-specific means 
(observed-mean centering, e.g., ${\hat{Y}}^*_{ik}=Y_{ik}-\bar{Y}_i$, where $\bar{Y}_i=\sum_{k=0}^KY_{ik}/(K+1)$), and 4)~observed scores
(no centering, e.g., ${\hat{Y}}^*_{ik}=Y_{ik}$). In the current scenario, the no-centering method totally
ignores the presence of stable traits. \textcolor{black}{On the other hand, because o}bserved means include the components of both stable traits (between-person differences) and within-person variability, 
observed-mean centering fails to perfectly disentangle stable individual differences from within-person variability.\footnote{\textcolor{black}{As a similar problem, the risk of using observed person-specific (or cluster-specific) means to express 
cluster effects is recognized as Nickell's bias and L\"{u}dtke's bias for estimates of regression coefficients in
applications of multilevel models (e.g., Asparouhov \& Muth\'{e}n, 2018; L\"{u}dtke et~al., 2008; McNeish \& Hamaker, 2020; Usami, 2017).
}} Under each simulation condition, we calculated the bias and root mean squared error (RMSE) of 10 kinds of estimates of causal parameters from MSMs and SNMMs. 

In the first step of the proposed method, to identify the measurement model (e.g., Equation~16 for $Y$) SEM that
assumes a linear AR(1) structure with time-varying autoregressive parameters and residual variances is specified
for within-person variability scores in each variable. Although a true model (i.e., AR($K$) structure) cannot be specified because of the identification problem,
we confirmed that the AR(1) structure generally provides acceptable model fits under the current parameter setting.

The results are discarded when improper solutions appear in the first step because of out-of-range parameter estimates (e.g., negative variance). In the current simulation, fewer than 0.1\% of all estimates produced such improper solutions.
We also confirmed that improper solutions were not found in the second step of applying MSMs and SNMMs.
When applying MSMs, a first-order linear regression model is specified for the treatment assignment model, namely, $f(A_t|A_{t-1},L_{t})$ (i.e., the correct specification).
For SNMMs, models $\mathcal{A}$ and $\mathcal{B}$ are also specified in an appropriate manner.

In the second scenario where model misspecifications are present, we assume various DGPs in which (a) 
measurement errors are present, (b) time-invariant factors do not influence measurements as stable trait factors, and (c) quadratic effects of time-varying observed confounders are present 
in the treatment assignment model, keeping the other conditions the same from the first scenario. More specifically, in (a),
all measurements are influenced by normally distributed measurement errors with variances of 10\% or 20\% of those of the initial measurements (=10+$\phi^2$).
In (b), the relation between outcomes and time-invariant factors ($I$) is set as $Y_{ik}=(1+0.5k/K)I^{(Y)}_{i}+0.3I^{(A)}_{i}+0.3I^{(L)}_{i}+Y^*_{ik}$,
(i.e., time-varying loadings from $I^{(Y)}$ and those from other variables ($I^{(A)}$ and $I^{(L)}$) are present),
resulting from the assumed DGP such as that in Figure~1a in which time-invariant factors have both direct and indirect effects on measurements. 
In (c), quadratic effects from time-varying observed confounders are included in the treatment assignment model 
as $A^*_{ik}=0.20Y^*_{ik}+0.08Y^{*2}_{ik}+0.40A^*_{i(k-1)}+0.30L^*_{ik}+d^{(A)}_{ik}$,
indicating that the treatment assignment model that includes only linear effects of $Y^*_{ik}$ assumed in the current MSM and SNMM is misspecified.
\textcolor{black}{Note that causal parameters for time-varying treatments ($\tau$) remain unchanged even if quadratic effects exist in the
treatment assignment model because time-varying treatments are now intervened (see Figure~2)}.

The simulation was conducted in R, using the $lavaan$ package 
(Rosseel, 2012) to estimate parameters by SEM with MLE in the first step and the $ipw$ package for MSMs in the second step. In SNMMs, we solve Equation~(30) via the Newton--Raphson method.
Simulation code is available in the Online Supplemental Material.

\subsection{Results}
Because of space limitations, Figure~\textcolor{black}{3} shows only biases of estimates of causal parameters in MSMs and SNMMs when $\phi^2= 10/9$ and 10. Because differences in the $N$ value
were minor in terms of bias, here we only show the result when $N=1000$. Results under other conditions are provided
in the Online Supplemental Material (Figures~S3 and S4).

Figure~\textcolor{black}{3} shows that true score conditions produce almost no biases in both MSMs and SNMMs.
SNMMs show smaller RMSEs compared with MSMs on average (Figure~S2). In the proposed method, estimates show biases
because of the biased estimates of stable trait factor (co)variances triggered by a model misspecification in the first step. However, the magnitude of biases 
is much smaller than in the observed-mean centering and no-centering methods. SNMMs again show smaller RMSEs than do MSMs (Figure~S2).
The observed-mean centering method shows negative biases, and their magnitude  becomes larger when $K=4$.
This result is caused by negatively biased covariances in variables 
resulting from subtracting observed means from measurements, and this impact increases as $K$ decreases.
Another critical aspect of this method is that linear dependence prevents identification of joint effects of all past treatments on $Y_K$ (in this case, $\beta_{4\textcolor{black}{0}}$, $\beta_{4\textcolor{black}{1}}$, $\beta_{4\textcolor{black}{2}}$, $\beta_{4\textcolor{black}{3}}$ in $K=4$). 
We therefore do not recommend use of observed-mean centering. The no-centering method shows serious negative 
biases when $\phi^2$ is not small,
indicating that ignoring the presence of stable traits is critical to estimating causal effects.
Magnitudes of stable trait factor variances should vary depending on the nature of variables and study period,
but in the author's experience many studies that applied the RI-CLPM have shown significant and moderate to large sizes of $\hat{\phi}^2$
(e.g., the proportion of stable trait factor variance to that of measurements is above 30\%).
The following application also demonstrates large stable trait factor variance estimates. 

As supplemental analyses, we additionally explored the performance of the methods 
\begin{figure}[htbp]
\includegraphics[height=16cm,width=24cm,angle=90]{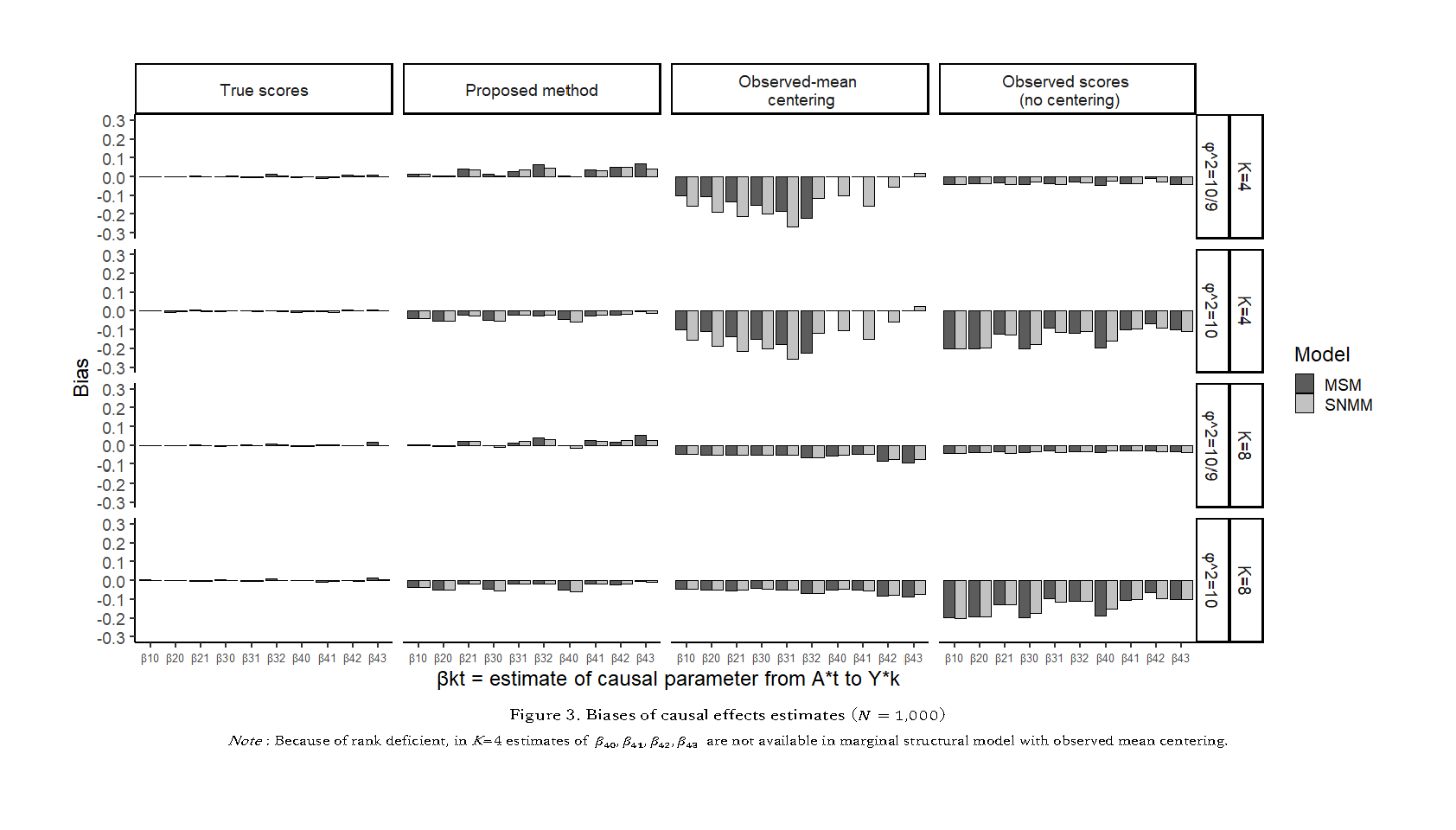}
\end{figure}
under different parameter settings,
as well as different model specification of SEM in the first step. From this, we find similar tendencies in the results (Figures~S5--S8): 
(a)~SNMMs show smaller RMSEs than do MSMs, and (b)~the proposed method shows adequate performance in terms of biases and RMSEs, and it works better than the no-centering method (especially when $\phi$ is larger) and the observed-mean centering method (especially when $K$ is smaller). 
We also investigated the performance of linear correlation preserving predictor (${\hat{I}}^{(Y)}_i$ in Equation~22) centering 
(e.g., ${\hat{Y}}^*_{ik}=Y_{ik}-{\hat{I}}^{(Y)}_i$), confirming that the proposed method worked much better than this method
on average (Figures~S5--S8).

Similar results were also observed in the second scenario, where model misspecifications are present (Figures~S9--S14).
More specifically, when measurement errors were present, the biases and RMSEs became larger in all methods (Figures~S9 and S10).
However, the proposed method still outperforms other centering methods. 
When time-invariant factors do not influence measurements as stable trait factors,
the overall results of biases and RMSEs were not largely affected (Figures~S11--S12),
regardless of the magnitude of $\phi^2$. This result is a little surprising,
considering that the specified time-varying loadings from factors ($=1+0.5k/K$ at time $t_k$) are not small
(i.e., the impact of this factor on the variance of measurement at time $t_K$ is almost twice that at time $t_0$).
This may suggest that causal parameters can be recovered relatively well even when ignoring time-varying impacts from time-invariant factors that are actually present in the first step.
However, future investigations are required in order to better clarify when estimated causal parameters 
are seriously biased under various scenarios for misspecified measurement models. When quadratic effects from time-varying observed confounders are present in the treatment assignment model
but are ignored in analyses, biases and RMSEs in MSMs become larger on average. In the proposed method,
this is salient in the RMSEs for the $K=8$ and $\phi^2=10$ conditions (Figures~S13--S14). SNMMs, which
have the property of being doubly robust in $G$-estimators, were less influenced even if these by-no-means small quadratic effects are ignored, 
and in many conditions the proposed method again outperforms other centering methods.
\section{EMPIRICAL APPLICATION}
This section describes an empirical application of the proposed method using data from the Tokyo Teen Cohort (TTC) study (Ando et~al., 2019).
We assume a similar causal DAG model to that in Figure~1b: (i) measurements are expressed by the linear sum of stable trait factors and within-person variability scores, 
(ii) within-person variability scores are expressed by functions \textcolor{black}{(with assumption of homogeneity)} of those in past time, along with
(iii) consistency, (iv) sequential ignorability, (v) SUTVA, \textcolor{black}{(vi) positivity, and (vii) modularity.}

TTC was a multidisciplinary longitudinal cohort study on the psychological and physical development of 
adolescents who were 10 years old at enrollment and lived in municipalities in the Tokyo metropolitan area (Setagaya, Mitaka, Chofu). Datasets were collected in three waves: from 2012 to 2015, from 2014 to 2017, and from 2017 to 2019 (i.e., $K=2$).
In total, 3171 children participated in the survey. See Ando et~al.\ (2019) for more-detailed information about measured variables, participant recruitment, and demographic characteristics of participants in the TTC study.

In this example, we estimate the (joint) causal effects of time-varying sleep duration ($A$) on later depressive symptoms ($Y$) in adolescents.
Several epidemiological studies have suggested a relationship between sleep habits (sleep duration, bedtime, and bedtime regularity) and mental health status (depression and anxiety)
in adolescents. For example, Matamura et~al.\ (2014) applied the CLPM to data from 314 monozygotic twins living in Japan and showed that
sleep duration had significant associations with mental health indices, 
even after controlling for genetic and shared environmental factors.
However, to the author's knowledge, no studies have investigated this relation
that accounts for stable traits in sleep duration and symptoms (i.e., at the within-person level).

The Short Mood and Feelings Questionnaire (SMFQ; Angold et~al., 1995) was used to 
measure depression in adolescents ($Y$).
The SMFQ consists of 13 items assessing depressive symptoms rated on a three-point scale (0:~\emph{not true}, 1:~\emph{sometimes true}, 2:~\emph{true})
regarding feelings and actions over the preceding two weeks. Higher SMFQ scores suggest 
more-severe symptoms. These data were measured at home by self-report questionnaires.
In this example, sleep duration in hours ($A$) was measured by the question ``How long do you usually sleep on weekdays?"
Observed confounders were body mass index (BMI; $L_B$) and bedtime ($L_A$), which was measured by the question ``When do you usually go to bed on weekdays?" Because many adolescents reported no problems for all items on the SMFQ,
the score distribution was positively skewed. In the present example, we focus on the clinical group comprising
$N=416$ adolescents (13.1\%) with SMFQ scores of 6 or higher
during the study. Katon, Russo, Richardson, McCauley, and Lozano (2008) reported 80\% sensitivity and 81\% specificity at this cut-off for diagnosis of major depression based on the Computerized Diagnostic Interview Schedule for Children (C-DISC). 
Missing data were primarily due to dropout. 
Of the 416 samples, 113 adolescents provided all three responses in the study. Descriptive statistics of sleep duration, SMFQ score, bedtime, and BMI are available in the Online Supplemental Material (Table~S1). 

In the first step, we use generalized least squares in the $lavaan$ package to estimate the model parameters
for each variable. To identify the measurement model (e.g., Equation~16 for $Y$), SEM that assumes
an AR(1) structure with time-varying autoregressive parameters and residual variances
is specified for within-person variability scores of each variable. Let $P_i$ be the total number of variables observed in adolescent $i$. 
$P_i \times P_i$ weights $W_i$ are calculated from estimated parameters under the assumption of MAR. 
Within-person variability scores $X_i^*$ are then calculated using this weight and measurements $X_{i,obs}$ for adolescent $i$ as $\hat{X}_i^*=W^t_iX_{i,obs}$. 

\textcolor{black}{Causal parameters ($\beta$ and $\gamma$)} of sleep duration at 10 and 12 years old ($A^*_0$ and $A^*_1$) on later depressive symptoms (SMFQ scores $Y^*_1$ and $Y^*_2$)
are estimated using calculated within-person variability scores by linear SNMM. In linear SNMM, blip functions and $U^*(\tau)$ are set as in Equations~(26) and (28), except that the two confounders $L_A$ and $L_B$ are present in this example. 
When applying SNMMs, models $\mathcal{A}$ and $\mathcal{B}$ are both specified using first-order linear regression models. All calculated within-person variability scores were used in the analysis under the assumption of MAR. 

We confirmed that the first step did not find improper solutions, and that current AR(1) models that assume time-varying parameters fit better than those that do not. Table~S2 summarizes the model fit indices and estimated parameters in this step.
All stable trait factor variance estimates are significant, indicating the necessity of controlling for stable traits.
Specifically, the proportions of variances in measurements attributable to estimated stable trait factors at $k=0$
are 24.5\%, 54.5\%, 48.2\%, and 74.8\% for $Y$, $A$, $L_A$, and $L_B$, respectively. 

Table~1 provides the estimation results of causal parameters, along with
estimates based on the no-centering and observed-mean centering methods for comparison.
As seen in Table~1, the proposed method reveals that \textcolor{black}{intervention of} longer sleep duration at 12 years old ($A^*_1$) has a positive effect ($\textcolor{black}{\hat{\beta}_{21}}=-2.704$, 95\%CI [-4.938,-0.470], $p<$.05)
on later depressive symptoms at 14 years old ($Y^*_2$) \textcolor{black}{at the within-person level},
but this estimate is not significant in the no-centering and observed mean score-centering methods. 
Similar positive effects of sleep duration were found in previous studies (Matamura et~al., 2014), but the present analysis newly investigates this causal hypothesis at the within-person level by controlling for stable traits of persons.
When the no-centering method is applied, the causal effect estimate of
$A^*_0$ on $Y^*_1$ is significant, showing that \textcolor{black}{intervention of} longer sleep duration at 10 years old has a \emph{negative} effect ($\textcolor{black}{\hat{\beta}_{10}}=1.693$, 95\% CI [0.405,2.981], $p<$.05) on 
later depressive symptoms at 12 years old. Considering that the magnitudes of the estimated stable trait factor variances were moderate or large 
for all variables, causal effect estimates in the no-centering method are unreliable and might be seriously biased.

In supplemental analyses, we confirmed that the major findings did not change even when using only data of adolescents who provided all three responses ($N=113$)
and a different cutoff for SMFQ (Angold et~al., 1995; Tables~S2--S5).
Again, statistical significance as well as sign and magnitude in estimates of causal parameters might change according to the choice of calculation (centering) methods for within-person variability scores,
and ignoring the presence of stable traits of persons might lead to incorrect conclusions.

\renewcommand{\arraystretch}{0.6}
\renewcommand{\doublerulesep}{1pt}
\begin{table}
\begin{center} 
{\small{
Table~1: Estimates of causal parameters of sleep duration on depression (SMFQ) ($N=416$).
\begin{tabular*}{180mm}{cccc} \hline
 & \begin{tabular}{c}Proposed method \end{tabular} & \begin{tabular}{c}Observed-mean\\ centering\end{tabular} & \begin{tabular}{c}Observed scores\\(no centering)\end{tabular}  \\ \hline
 $Sleep_1 \rightarrow SMFQ_2\hspace{4mm}(\textcolor{black}{\beta_{21}})$                & {\bf{-2.704 (1.140)}} & 0.095 (0.869) & -1.492 (1.080)  \\ 
 $(Sleep_1 \times Bedtime_1) \rightarrow SMFQ_2\hspace{4mm} (\textcolor{black}{\gamma_{21A}})$  & -0.603 (1.416) & 0.916 (1.857) &  0.336 (1.057) \\ 
 $(Sleep_1 \times BMI_1) \rightarrow SMFQ_2 \hspace{4mm}(\textcolor{black}{\gamma_{21B}})$  & -0.442 (0.638) & -0.748 (1.169) &  -0.532 (0.399) \\ 
 $Sleep_0 \rightarrow SMFQ_2 \hspace{4mm}(\textcolor{black}{\beta_{20}})$                & -0.293 (1.179) & -0.309 (1.021) &  0.185 (1.117) \\ 
 $(Sleep_0 \times Bedtime_0) \rightarrow SMFQ_2 \hspace{4mm}(\textcolor{black}{\gamma_{20A}})$  & 2.278 (1.918) & -3.012 (1.784) &  1.169 (1.792) \\ 
 $(Sleep_0 \times BMI_0) \rightarrow SMFQ_2 \hspace{4mm}(\textcolor{black}{\gamma_{20B}})$  & {\bf{-1.856 (0.780)}} & -0.251 (0.986) &  0.306 (0.287) \\ 
 $Sleep_0 \rightarrow SMFQ_1 \hspace{4mm}(\textcolor{black}{\beta_{10}})$                & 0.702 (0.686) & 0.279 (0.572) &  {\bf{1.693 (0.657)}} \\ 
 $(Sleep_0 \times Bedtime_0) \rightarrow SMFQ_1 \hspace{4mm}(\textcolor{black}{\gamma_{10A}})$  & 0.315 (1.177) & 1.037 (1.069) &  -0.918 (0.920) \\ 
 $(Sleep_0 \times BMI_0) \rightarrow SMFQ_1 \hspace{4mm}(\textcolor{black}{\gamma_{10B}})$  & 0.021 (0.473) & 0.773 (0.572) &  -0.101 (0.212) \\ \hline
\end{tabular*}
}}
\end{center}\vspace{-4.5mm}
\emph{Note:} {\footnotesize{Bold font indicates statistical significance.}}
\end{table}

\section{GENERAL DISCUSSION}
We proposed a two-step estimation method for within-person variability score-based causal inference to
estimate joint effects of time-varying (continuous) treatments/predictors by effectively controlling for stable traits. 
In the first step, a within-person variability score for each person, which is disaggregated from the stable trait factor score, is
calculated using weights based on the best linear correlation preserving predictor through SEM. 
Causal parameters are then estimated by MSMs or SNMs, using calculated within-person variability scores.
The proposed method can be viewed as one that synthesizes the
two traditions of factor analysis/SEM in psychometrics and a 
method of causal inference (MSMs or SNMs) in epidemiology.

In this paper, we began by providing formal definitions of stable trait factors (for between-person relations) and within-person variability scores (for within-person relations),
because these concepts have not been fully characterized in the causal inference literature despite the fact that they have been attracting increasing attention in psychometrics and behavioral science
(e.g., Hamaker et~al., 2015; Usami et~al., 2019). On the other hand, in epidemiology the conceptual and mathematical differences between stable trait factors and accumulating factors,
along with which kind of \textcolor{black}{time-invariant} factor is included in each statistical model, have received less attention. This paper may help bridge the gap.
We have also clarified the assumptions required to identify causal parameters for within-person variability score-based causal inference:
(i) (as depicted in Figure~1b,) measurements are expressed by the linear sum of stable trait factor scores (defined as Equation~\textcolor{black}{4}) and within-person variability scores (defined as Equation~\textcolor{black}{5}) that
are mutually uncorrelated, (ii) within-person variability scores are expressed by functions of those \textcolor{black}{(with assumption of homogeneity)} in past time, 
(iii) consistency, (iv) sequential ignorability, (v) SUTVA, \textcolor{black}{(vi) positivity, (vii) modularity, and (viii) multivariate normality (if MLE is used in the first step).}

\textcolor{black}{As for the second assumption, o}ur approach is more flexible than the RI-CLPM (that includes time-varying observed confounders), which researchers are becoming increasingly interested in for uncovering within-person relations among variables,
in that the assumption of linearity is not required with respect to time-varying observed confounders at the within-person level.
We particularly emphasize the utility of SNMs with G-estimation, because of its property of being doubly robust to the model misspecifications in how
the time-varying observed confounders are \textcolor{black}{functionally related with} treatments/predictors and outcomes, along with
flexibility in that it allows investigation of moderation effects of treatments with observed confounders. 

Through simulation and empirical application, we illustrated that ignoring 
the presence of stable traits might lead to incorrect conclusions in causal effects. We also confirmed that the proposed approach is superior to observed-mean centering\textcolor{black}{, as a conventional method to predict stable traits of persons}.
Especially when $K$ is small, observed-mean centering showed serious negative biases in estimates of causal parameters. Considering that
most research applying the RI-CLPM to uncover within-person relations used longitudinal data with two or three time points 
($K=1,2$; e.g., Usami, Todo \& Murayama, 2019), observed-mean centering cannot be recommended.

A recent study provided closed-form parametric expressions of causal effects for linear models (Gische et~al., 2021),
and Gische and Voelkle (\textcolor{black}{in press}) proposed asymptotically efficient estimators in the case of ML estimation.
It is suggested that, at least in large samples, the parametric procedure proposed by Gische and Voelkle (\textcolor{black}{in press}) may
give smaller standard errors compared with the proposed method in simulations in which data are generated by a linear model with normal residuals.
Comparing the performances of these methods and the RI-CLPM under various conditions that 
account for nonlinear relations among variables is an important topic for future studies.

One caveat for the proposed method, \textcolor{black}{which is relevant to the second assumption above,} is that in the first step, one must correctly specify the 
structure (such as the AR(1) structure) for within-person variability scores in each variable
so that the (identified) SEM can yield consistent estimates of parameters, which are required for consistent estimation of causal effects in the subsequent step.
However, in general, how to establish the correct (or even a plausible) DAG model is a major challenge (Hamaker, Mulder \& van IJzendoorn, 2020; \textcolor{black}{see also the discussion in Section~2.4}).
Relatedly, the premise that stable trait factors exist and loadings from factors are equal to those in the 
DGP might be restrictive in actual applications; therefore, we need to carefully account for the consequences of possible model misspecifications 
in the first step on the results in the second step to precisely infer within-person relations. 
The good news is that in the present simulation, we confirmed that the specified time-varying AR(1) structure works well to recover causal parameters, and its performance 
was not largely influenced even when there were model misspecifications (i.e., time-varying effects from time-invariant factors). 
However, additional large-scale simulations to further clarify the robustness of the method regarding this point are needed in future studies.

We can also use model fit indices to evaluate how well the structure specified in the first step fits 
to the data, especially when the number of time points is large. However, this procedure is not a fundamental solution.
Even if a researcher is certain that SEM that assumes stable trait factors for each variable can be specified in the first step, in general 
there is still at great risk of violating some assumptions. Notably, \textcolor{black}{relating to the first assumption above, stable trait} factors and within-person variability scores \textcolor{black}{(temporal deviations)}
at each time point might be correlated in actual applications if they share common causes (e.g., genotype; see also
McNeish \& Kelly (2019), who argued the issue of endogeneity in applying mixed-effects model). This point is closely related to the model being able to
perfectly disentangle the within- and between-person relations. Future studies should investigate how this violation impacts the estimated causal parameters.

We used two-step estimation to account for feasibility, but this issue remains
in that one still must write programming code, as that in the Online Supplemental Material. We are 
planning to develop packages for the proposed method.
Another potential limitation is that the proposed approach (as well as the RI-CLPM) demands longitudinal data with three or more time points ($K\geq 2$)
to identify the measurement model (SEM) in the first step unless strong parameter constraints are imposed.

We assumed continuous variables in this paper, but we can extend this methodology to categorical variables.
For binary and ordered categorical variables, one simple method is to use estimated
tetrachoric or polychoric correlations as input for factor analysis and then to
calculate within-person variability scores using estimated stable trait factor and unique factors (within-person variability scores) variances.
There are alternative methods for conducting factor analysis for ordinal variables (e.g., J\"{o}reskog \& Moustaki, 2001),
but whichever method is used, rescaling stable trait factor and unique factor scores variances
might be required to make calculated within-person variability scores 
interpretable. In some cases, using variances of original (categorical) data might be one option for this purpose.
However, the estimation performance for causal parameters and the influence of
biased tetrachoric or polychoric correlations caused by misspecified distributions must be investigated in future research.

Because we take an SEM approach in the first step, accounting for measurement errors,
which is closely related to violation of the consistency assumption, is feasible under the parametric assumption. Although we expect that longitudinal data with large $K$ are required for precisely estimating measurement-error variances,
we plan to investigate how the proposed method works under measurement models that include measurement errors.

This paper opens a new avenue for exploring other various research questions that are closely relevant to causal hypotheses. 
For example, use of within-person variability scores can be extended to cases in which 
one is interested in uncovering reciprocal effects (e.g., Usami et~al., 2019) and mediation effects (e.g., Goldsmith et~al., 2018; Tchetgen Tchetgen \& Shpitser, 2012),
as well as to \textcolor{black}{multilevel modeling} and hierarchical continuous time modeling (Driver \& Voelkle, 2018).
Note that there is still room for discussion on the issues of within-person relation and stable traits, 
\textcolor{black}{as well as the issue about when and how to include (time-invariant) factors in the assumed DGP (see Section~2.4)}. 
The present paper is intended to promote substantial 
discussion about the conceptual and statistical properties of the time-invariant factors (e.g., stable trait factors or accumulating factors)
included in the assumed DGP among researchers who wish to infer within-person relations and causality, 
and the hope is that the proposed method helps in exploring various causal hypotheses in longitudinal design and guiding better decision-making for researchers.

\newpage
\hspace{-6mm}{\bf{{\Large{REFERENCES}}}}\\
Anderson, J.C., \& Gerbing, D.W. (1988). Structural equation modeling in practice: A\par
review and recommended two-step approach. \textit{Psychological Bulletin, 103}, 411-423.\par
\hspace{-6mm}Anderson, T.W. \& Rubin, H. (1956). Statistical Inference in Factor Analysis. In:\par
Neyman, J., Ed., Proceedings of the 3rd Berkeley Symposium on Mathematical\par
Statistics and Probability, Vol. 5, Berkeley, 111-150.\par
\hspace{-6mm}Ando, S., Nishida, A., Yamasaki, S., Koike, S., Morimoto, Y., Hoshino, A., Kanata, S.,\par
Fujikawa, S., Endo, K., Usami, S., Furukawa, T.A., Hiraiwa-Hasegawa, M., \& Kasai,\par
K. (2019). Cohort profile: Tokyo Teen Cohort study (TTC). \textit{International Journal}\par
\textit{of Epidemiology, 48,} 1414-1414g.\par
\hspace{-6mm}Angold, A., Costello, E.J., Messer, S.C., Pickles, A., Winder, F., \& Silver, D. (1995).\par
Development of a short questionnaire for use in epidemiological studies of depression\par
in children and adolescents. \textit{International Journal of methods in Psychiatric Research,}\par
\textit{5}, 237-249.\par
\hspace{-6mm}Asparouhov, T., \& Muth\'{e}n, B. (2018). Latent variable centering of predictors and\par
mediators in multilevel and time-series models. \textit{Structural Equation Modeling: A}\par
\textit{Multidisciplinary Journal, 26,} 1-24.\par
\hspace{-6mm}Bakk, Z., \& Kuha, J. (2017). Two-step estimation of models between latent classes and\par
external variables. \textit{Psychometrika, 83}, 871-892.\par
\hspace{-6mm}Bollen, K.A. (1989). \textit{Structural equations with latent variables}. New York: Wiley.\par
\hspace{-6mm}Bollen, K.A., \& Curran, P.J. (2004). Autoregressive latent trajectory (ALT) models:\par
A synthesis of two traditions. \textit{Sociological Methods and Research, 32}, 336-383.\par
\hspace{-6mm}Brumback, B.A, He, Z., Prasad, M., Freeman, M.C., \& Rheingans, R. (2014). Using\par
structural-nested models to estimate the effect of cluster-level adherence on individual-\par
level outcomes with a three-armed cluster-randomized trial. \textit{Statistics in Medicine, 33,}\par
1490-1502.\par
\hspace{-6mm}Cole, D.A., Martin, N.C., \& Steiger, J.H. (2005). Empirical and conceptual problems with\par
longitudinal trait-state models: Introducing a trait-state-occasion model. \textit{Psychological}\par
\textit{Methods, 10}, 3-20.\par
\hspace{-6mm}Croon, M. (2002). Using predicted latent scores in general latent structure models. In\par
G. Marcoulides \& I. Moustaki (Eds.), \textit{Latent Variable and Latent Structure Modeling}\par
(pp. 195-223). Mahwah, NJ: Erlbaum.\par
\hspace{-6mm}Curran, P.J., \& Bauer, D.J. (2011). The disaggregation of within-person and between-\par
person effects in longitudinal models of change. \textit{Annual Review of Psychology, 62},\par
583-619.\par
\hspace{-6mm}Driver, C.C., \& Voelkle, M.C. (2018). Hierarchical Bayesian continuous time dynamic\par
modeling. \textit{Psychological Methods, 23}, 774-799.\par
\hspace{-6mm}Enders, C.K., \& Bandalos, D.L. (2001). The relative performance of full information\par
maximum likelihood estimation for missing data in structural equation models.\par
\textit{Structural Equation Modeling, 8}, 430-457.\par
\hspace{-6mm}Enders, C.K. (2011). Missing not at random models for latent growth curve analyses.\par
\textit{Psychological Methods, 16}, 1-16.\par
\hspace{-6mm}\textcolor{black}{Gische, C., \& Voelkle, M.C. (in press). Beyond the mean: A flexible framework for studying}\par
\textcolor{black}{causal effects using linear models. \textit{Psychometrika.}}\par
\hspace{-6mm}Gische, C., West, S.G., \& Voelkle, M.C. (2021). Forecasting causal effects of interventions\par
versus predicting future outcomes. \textit{Structural Equation Modeling, 28}, 475-492.\par
\hspace{-6mm}Goldsmith, K.A., MacKinnon, D.P., Chalder, T., White, P.D., Sharpe, M., \& Pickles, A.\par
(2018). Tutorial: The practical application of longitudinal structural equation\par
mediation models in clinical trials. \textit{Psychological Methods, 23}, 191-207.\par
\hspace{-6mm}Green, B.F. (1969). Best linear composites with a specified structure. \textit{Psychometrika, 34},\par
301-318. \par
\hspace{-6mm}Hamaker, E.L. (2012). Why researchers should think ``within-person'': A paradigmatic\par
rationale. In T.S. Conner (Ed.), \textit{Handbook of research methods for studying daily life}\par
(pp. 43-61). New York: Guilford Press.\par
\hspace{-6mm}Hamaker, E.L., Kuiper, R.M., \& Grasman, R.P.P.P. (2015). A critique of the cross-lagged\par
panel model. \textit{Psychological Methods, 20,} 102-116.\par
\hspace{-6mm}Hamaker, E.L., Mulder, J.D., \& van IJzendoorn, M.H. (2020). Description, prediction and\par
causation: Methodological challenges of studying child and adolescent development.\par
\textit{Developmental Cognitive Neuroscience, 46}, 1-14.\par
\hspace{-6mm}He, J., Stephens-Shields, A., \& Joffe, M. (2015). Structural nested mean models to\par
estimate the effects of time-varying treatments on clustered outcomes. \textit{International}\par
\textit{Journal of Biostatistics, 11}, 203-222.\par
\hspace{-6mm}He, J., Stephens-Shields, A., \& Joffe, M. (2019). Marginal structural models to estimate\par
the effects of time-varying treatments on clustered outcomes in the presence of\par
interference. \textit{Statistical Methods and Medical Research, 28,} 613-625.\par
\hspace{-6mm}Hern$\acute{a}$n, M.A., Brumback B., \& Robins J.M. (2002). Estimating the causal effect of\par
zidovudine on CD4 count with a marginal structural model for repeated measures.\par
\textit{Statistics in Medicine, 21}, 1689-1709. \par
\textcolor{black}{\hspace{-6mm}Hern$\acute{a}$n, M.A., \& Robins, J.M. (2021). \textit{Causal Inference: What If}. Boca Raton: Chapman} \par
\textcolor{black}{\& Hall/CRC.}\par
\hspace{-6mm}Hoffman, L. (2014). \textit{Longitudinal analysis: Modeling within-person fluctuation and change}.\par
New York, NY: Routledge/Taylor \& Francis.\par
\hspace{-6mm}Hong, G. (2015). \textit{Causality in a social world: Moderation, mediation and spill-over.}\par
West Sussex, UK: John Wiley \& Sons, Ltd.\par
\hspace{-6mm}Hoshino, T., \& Bentler, P.M. (2013). Bias in factor score regression and a simple solution.\par
In A.R. de Leon \& K.C. Chough (Eds.), \textit{Analysis of Mixed Data: Methods \& }\par
\textit{Applications} (pp. 43-61). Boca Raton, FL: Chapman \& Hall. \par
\hspace{-6mm}Hu, L., \& Bentler, P.M. (1999). Cutoff criteria for fit indexes in covariance structure\par
analysis: Conventional criteria versus new alternatives. \textit{Structural Equation Modeling,}\par
\textit{6}, 1-55.\par
\hspace{-6mm}Imai, K., \& Kim, S., (2019). When should we use unit fixed effects regression models for\par
causal inference with longitudinal data? \textit{American Journal of Political Science, 63},\par
467-490.\par
\hspace{-6mm}Imai, K., \& Ratkovic, M. (2015). Robust estimation of inverse probability weights for\par
marginal structural models. \textit{Journal of the American Statistical Association, 110,}\par
1013-1023.\par
\hspace{-6mm}J$\ddot{o}$reskog, K.G. (1970). A general method for analysis of covariance structures. \textit{Biometrika,}\par
\textit{57,} 239-251.\par
\hspace{-6mm}J\"{o}reskog. K.G, \& Lawley, D.N. (1968). New methods in maximum likelihood factor\par
analysis. \textit{British Journal of Mathematical and Statistical Psychology, 21}, 85-96.\par
\hspace{-6mm}J\"{o}reskog, K.G., \& Moustaki, I. (2001). Factor analysis of ordinal variables: A\par
comparison of three approaches. \textit{Multivariate Behavioral Research, 36}, 347-387.\par
\hspace{-6mm}Katon, W., Russo, J., Richardson, L., McCauley, E., \& Lozano, P. (2008). Anxiety and\par
depression screening for youth in a primary care population. \textit{Ambulatory Pediatrics,}\par
\textit{8}, 182-188.\par
\hspace{-6mm}Kline, R.B. (2016). \textit{Principles and practice of structural equation modeling (4th ed.)}.\par
Guilford Press.\par
\hspace{-6mm}Lefebvre, G., Delaney, J.A.C., \& Platt, R.W. (2008). Impact of mis-specification of the\par
treatment model on estimates from a marginal structural model. \textit{Statistics in Medicine,}\par
\textit{27}, 3629-3642.\par
\hspace{-6mm}L\"{u}dtke, O., Marsh, H.W., Robitzsch, A., Trautwein, U., Asparouhov, T., \& Muth\'{e}n, B. \par
(2008). The multilevel latent covariate model: A new, more reliable approach to group-\par
level effects in contextual studies. \textit{Psychological Methods, 13}, 203-229.\par
\textcolor{black}{\hspace{-6mm}L\"{u}dtke, O., \& Robitzsch, A. (2021). A critique of the random intercept cross-lagged}\par
\textcolor{black}{panel model. https://doi.org/10.31234/osf.io/6f85c}\par
\hspace{-6mm}Matamura, M., Tochigi, M, Usami, S., Yonehara, H., Fukushima, M., Nishida, A.,\par
Togo, F., \& Sasaki, T. (2014). Associations between sleep habits and mental health\par
status and suicidality in the longitudinal survey of monozygotic-twin adolescents.\par
\textit{Journal of Sleep Research, 23}, 290-294.\par
\hspace{-6mm}McArdle, J.J., \& Hamagami, F. (2001). Latent difference score structural models for linear\par
dynamic analyses with incomplete longitudinal data. In L. Collins \& A. Sayer (Eds.),\par
\textit{New methods for the analysis of change} (pp. 137-175). Washington, DC:\par
American Psychological Association.\par
\hspace{-6mm}McNeish, D. \& Hamaker, E.L. (2020). A primer on two-level dynamic structural equation\par
models for intensive longitudinal data in Mplus. \textit{Psychological Methods, 25}, 610-635.\par
\hspace{-6mm}Mulder, D. \& Hamaker, E.L. (\textcolor{black}{2021}). Three extensions of the random intercept cross-lagged\par
panel model, \textit{Structural Equation Modeling, 28}, \textcolor{black}{638-648}.\par
\hspace{-6mm}Muth\'{e}n, B., \& Asparouhov, T. (2002). Latent variable analysis with categorical outcomes:\par
Multiple-group and growth modeling in Mplus. Mplus Web Note: No.4.\par
\hspace{-6mm}Newsom, J.T. (2015). \textit{Longitudinal structural equation modeling: A comprehensive}\par
\textit{introduction.} New York: Routledge.\par
\hspace{-6mm}Pearl, J. (2009). \textit{Causality (2nd ed.)}. Cambridge University Press.\par
\hspace{-6mm}Resseguier, N., Giorgi, R., \& Paoletti, X. (2011). Sensitivity analysis when data are missing\par
not-at-random. \textit{Epidemiology, 22}, 282.\par
\hspace{-6mm}Robins, J.M. (1989). The analysis of randomized and non-randomized AIDS treatment\par
trials using a new approach to causal inference in longitudinal studies. In Health\par
Service Research Methodology: A Focus on AIDS (L. Sechrest, H. Freeman and A.\par
Mulley, eds.) 113-159. U.S. Public Health Service, National Center for Health Services\par
Research, Washington, DC.\par
\hspace{-6mm}Robins, J.M. (1992). Estimation of the time-dependent accelerated failure time model in\par
the presence of confounding factors. \textit{Biometrika, 79}, 321-334.\par
\hspace{-6mm}Robins, J.M. (1994). Correcting for non-compliance in randomized trials using structural\par
nested mean models. \textit{Communications in Statistics--Theory and Methods, 23,}\par
2379-2412.\par
\hspace{-6mm}Robins, J.M. (1999). Marginal structural models versus structural nested models as tools\par
for causal inference. \textit{Epidemiology, 116,} 95-134.\par
\hspace{-6mm}Robins, J.M., Blevins, D., Ritter, G. \& Wulfsohn, M. (1992). G-estimation of the effect\par
of prophylaxis therapy for pneumocystic carinii pneumonia on the survival of AIDS\par
patients. \textit{Epidemiology, 3,} 319-336.\par
\hspace{-6mm}Robins, J.M., Hern$\acute{a}$n, M.A. \& Brumback, B. (2000). Marginal structural models and\par
causal inference in epidemiology. \textit{Epidemiology, 11,} 550-560.\par
\hspace{-6mm}Robins, J.M. \& Rotnitzky, A. (2001). Comment on ``Inference for semiparametric models:\par
Some questions and an answer'', by P.J. Bickel and J. Kwon, \textit{Statistica Sinica, 11},\par
920-936.\par
\hspace{-6mm}Robins, J.M., \& Hern$\acute{a}$n, M.A. (2009). Estimation of the causal effects of time-varying\par
exposures. In G. Fitzmaurice et~al. (Eds.), Handbooks of modern statistical methods:\par
Longitudinal data analysis (pp. 553-599). Boca Raton: CRC Press.\par
\hspace{-6mm}Rosseel, Y. (2012). Lavaan: an R package for structural equation modeling. \textit{Journal of}\par
\textit{Statistical software, 48}, 1-36.\par
\hspace{-6mm}Rubin, D.B. (1976). Inference and missing data. \textit{Biometrika, 63,} 581-592.\par
\hspace{-6mm}Skrondal, A., \& Laake, P. (2001). Regression among factor scores. \textit{Psychometrika, 66},\par
563-575.\par 
\hspace{-6mm}Tchetgen Tchetgen, E.J., \& Shpitser, I. (2012). Semiparametric theory for causal\par
mediation analysis: Efficiency bounds, multiple robustness, and sensitivity analysis,\par
\textit{Annals of Statistics, 40}, 1816-1845.\par
\hspace{-6mm}ten Berge, T.M.F., Krijinen, W.P., Wansbeek, T.J., \& Shapiro, A. (1999). Some new\par
results on correlation-preserving factor scores prediction methods. \textit{Linear Algebra and}\par
\textit{its Applications, 289}, 311-318.\par
\hspace{-6mm}Usami, S., (2017).  Generalized sample size determination formulas for investigating\par
contextual effects by a three-level random intercept model. \textit{Psychometrika, 82},\par
133-157.\par
\hspace{-6mm}Usami, S. (2021). On the differences between general cross-lagged panel model and random-\par
intercept cross-lagged panel model: Interpretation of cross-lagged parameters and model\par
choice. \textit{Structural Equation Modeling, 28}, 331-344.\par
\hspace{-6mm}Usami, S., Murayama, K., \& Hamaker, E.L. (2019). A unified framework of longitudinal\par
models to examine reciprocal relations. \textit{Psychological Methods, 24}, 637-657.\par
\hspace{-6mm}Usami, S., Todo, N., \& Murayama, K. (2019). Modeling reciprocal effects in medical\par
research: Critical discussion on the current practices and potential alternative models.\par
\textit{PLOS ONE, 14(9)}: e0209133.\par
\hspace{-6mm}Vansteelandt, S., \& Joffe, M. (2014). Structural nested models and g-estimation: the\par
partially realized promise. \textit{Statistical Science, 29,} 707-731.\par
\hspace{-6mm}Vermunt, J.K. (2010). Society for political methodology latent class modeling with\par
covariates: Two improved three-step approaches. \textit{Political Analysis, 18}, 450-469.\par
\hspace{-6mm}Wallace, M.P., Moodie, E.E., \& Stephens, D.A. (2017). An R package for G-estimation of\par
structural nested mean models. \textit{Epidemiology, 28}, e18-e20.\par
\hspace{-6mm}Wang, L. \& Maxwell, S.E. (2015). On disaggregating between-person and within-person\par
effects with longitudinal data using multilevel models. \textit{Psychological Methods, 20},\par
63-83.\par
\hspace{-6mm}Zyphur, M.J., Allison, P.D., Tay, L., Voelkle, M.C., Preacher, K.J., Zhang, Z., Hamaker,\par
E.L., Shamsollahi, A., Pierides, D.C., Koval, P., \& Diener, E. (2020a). From data to\par
causes I: Building a general cross-lagged panel model (GCLM). \textit{Organizational}\par
\textit{Research Methods, 23}, 651-687.\par
\hspace{-6mm}Zyphur, M.J., Voelkle, M.C., Tay, L., Allison, P.D., Preacher, K.J., Zhang, Z., Hamaker,\par
E.L., Shamsollahi, A., Pierides, D. C., Koval, P., \& Diener, E. (2020b). From data to\par
causes II: Comparing approaches to panel data analysis. \textit{Organizational Research}\par
\textit{Methods, 23}, 688-716.
\begin{figure}[htbp]
\includegraphics[height=21cm,width=16cm,angle=0]{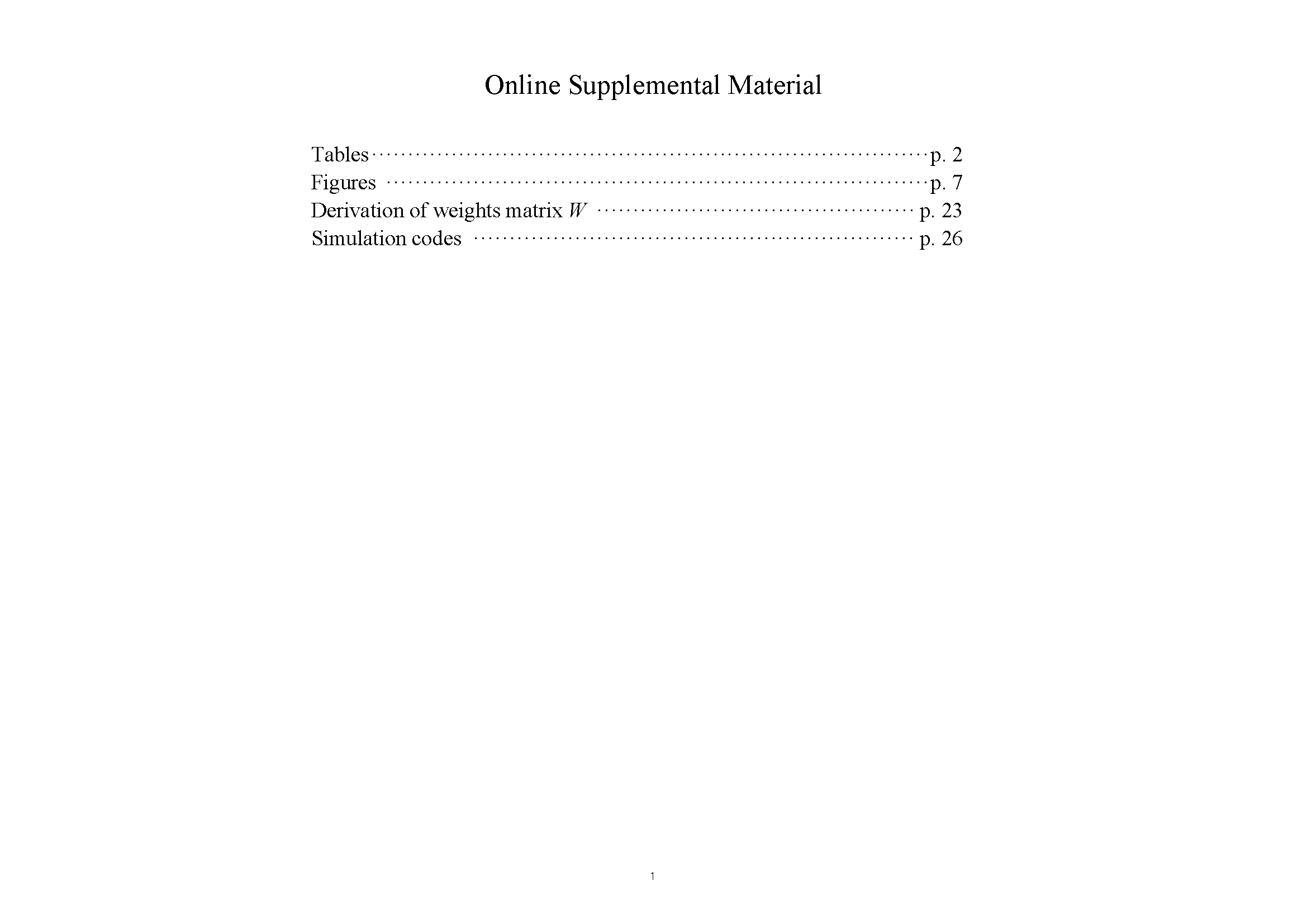}
\end{figure}
\begin{figure}[htbp]
\includegraphics[height=21cm,width=16cm,angle=0]{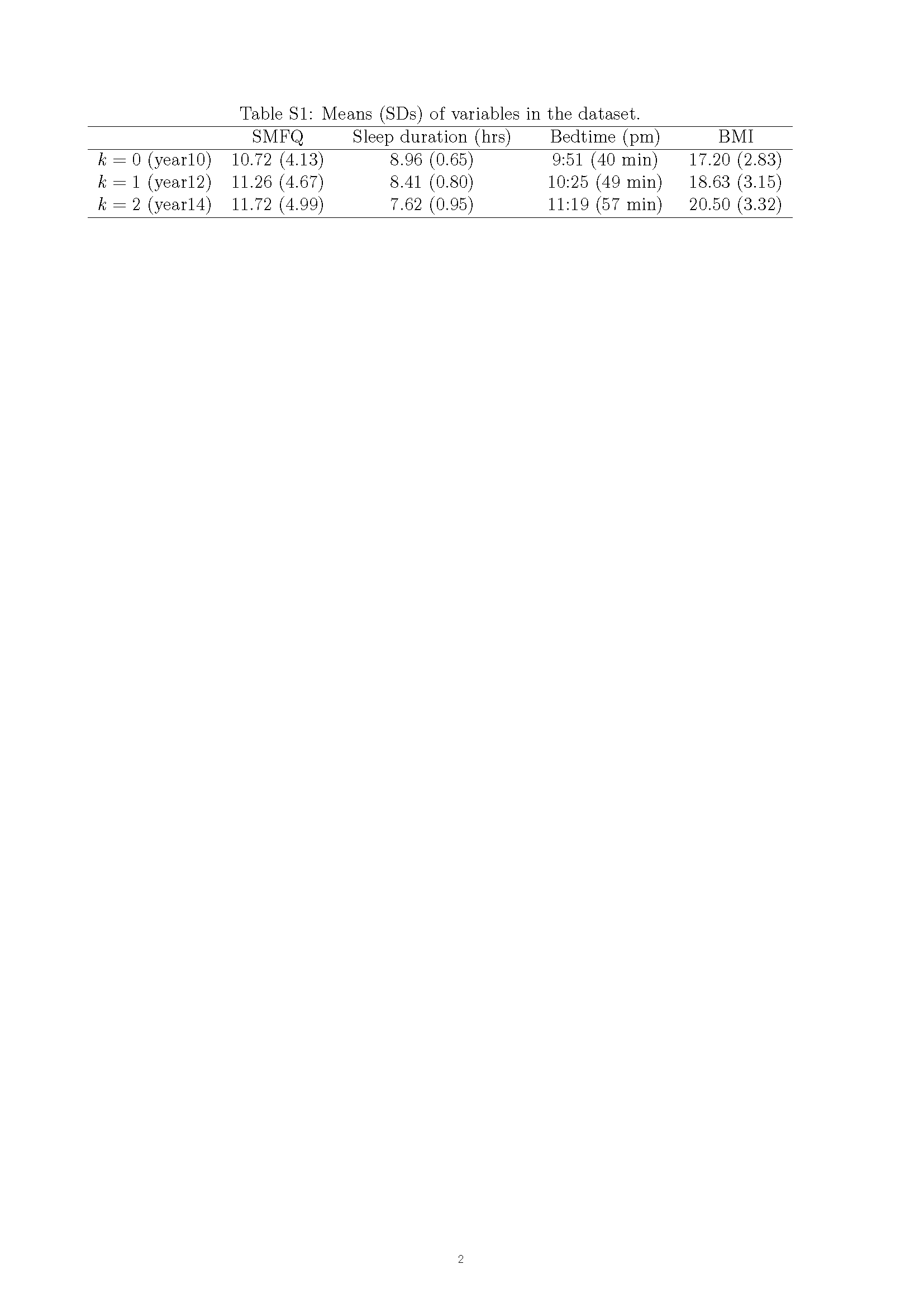}
\end{figure}
\begin{figure}[htbp]
\includegraphics[height=21cm,width=16cm,angle=0]{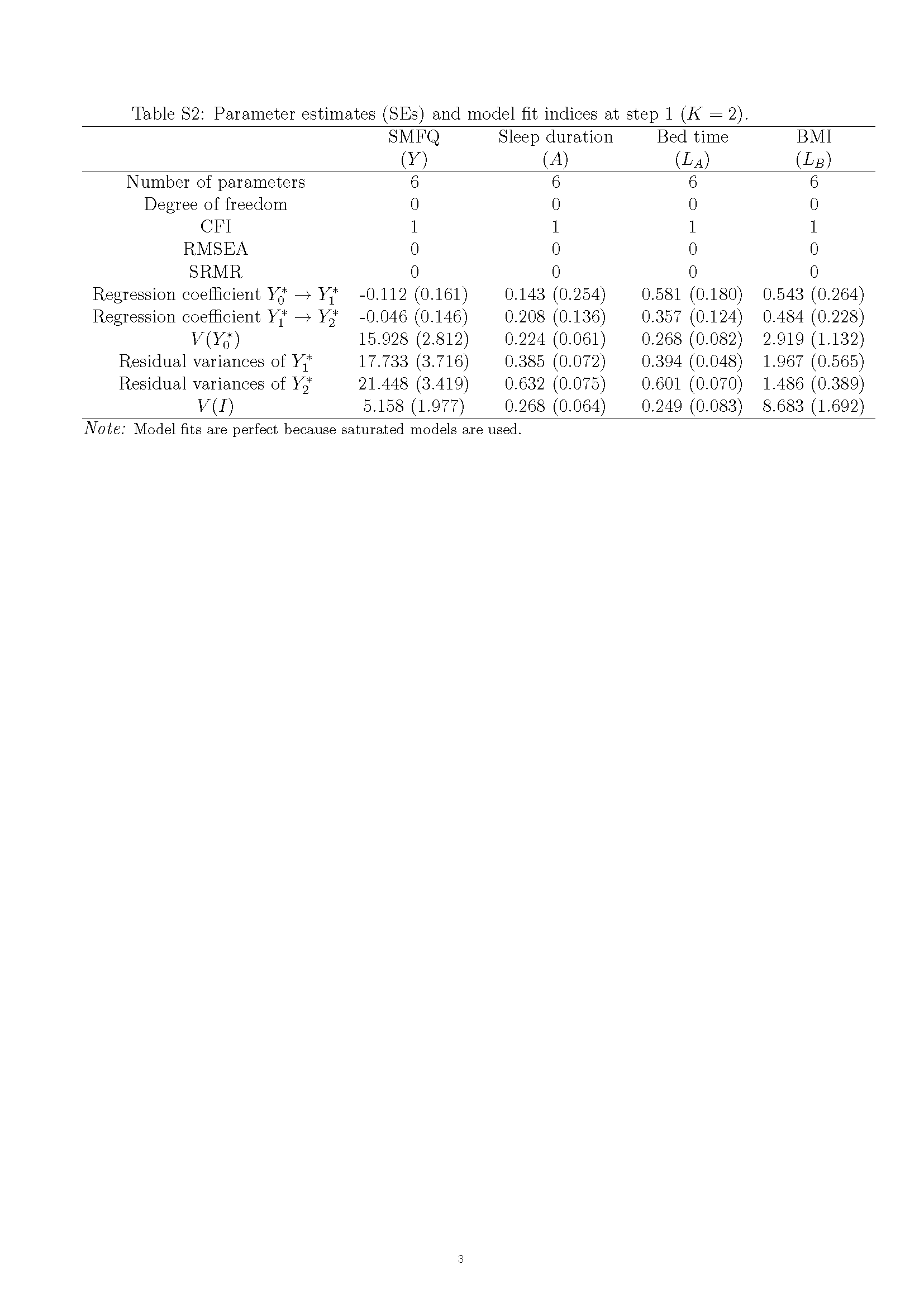}
\end{figure}
\begin{figure}[htbp]
\includegraphics[height=21cm,width=16cm,angle=0]{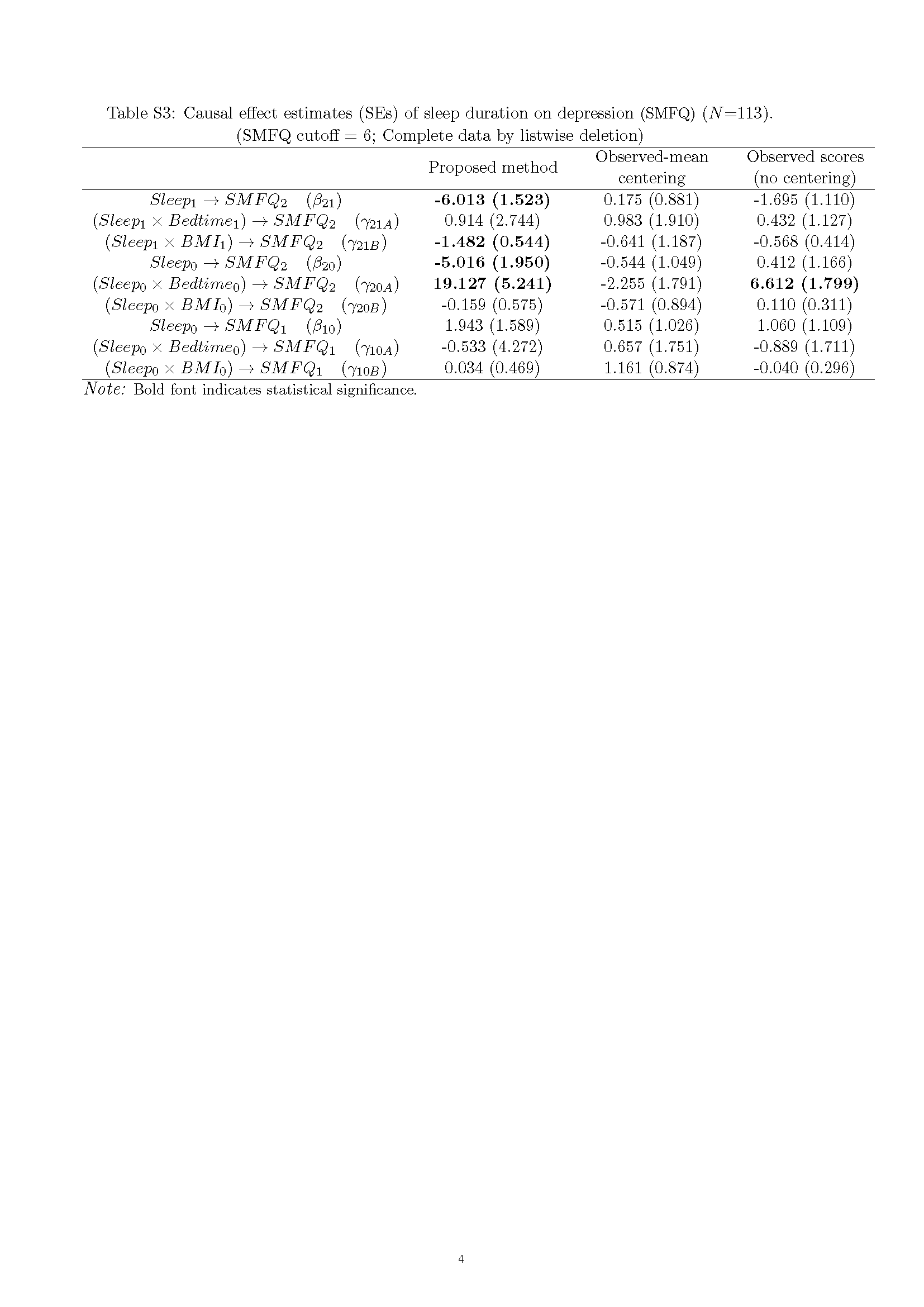}
\end{figure}
\begin{figure}[htbp]
\includegraphics[height=21cm,width=16cm,angle=0]{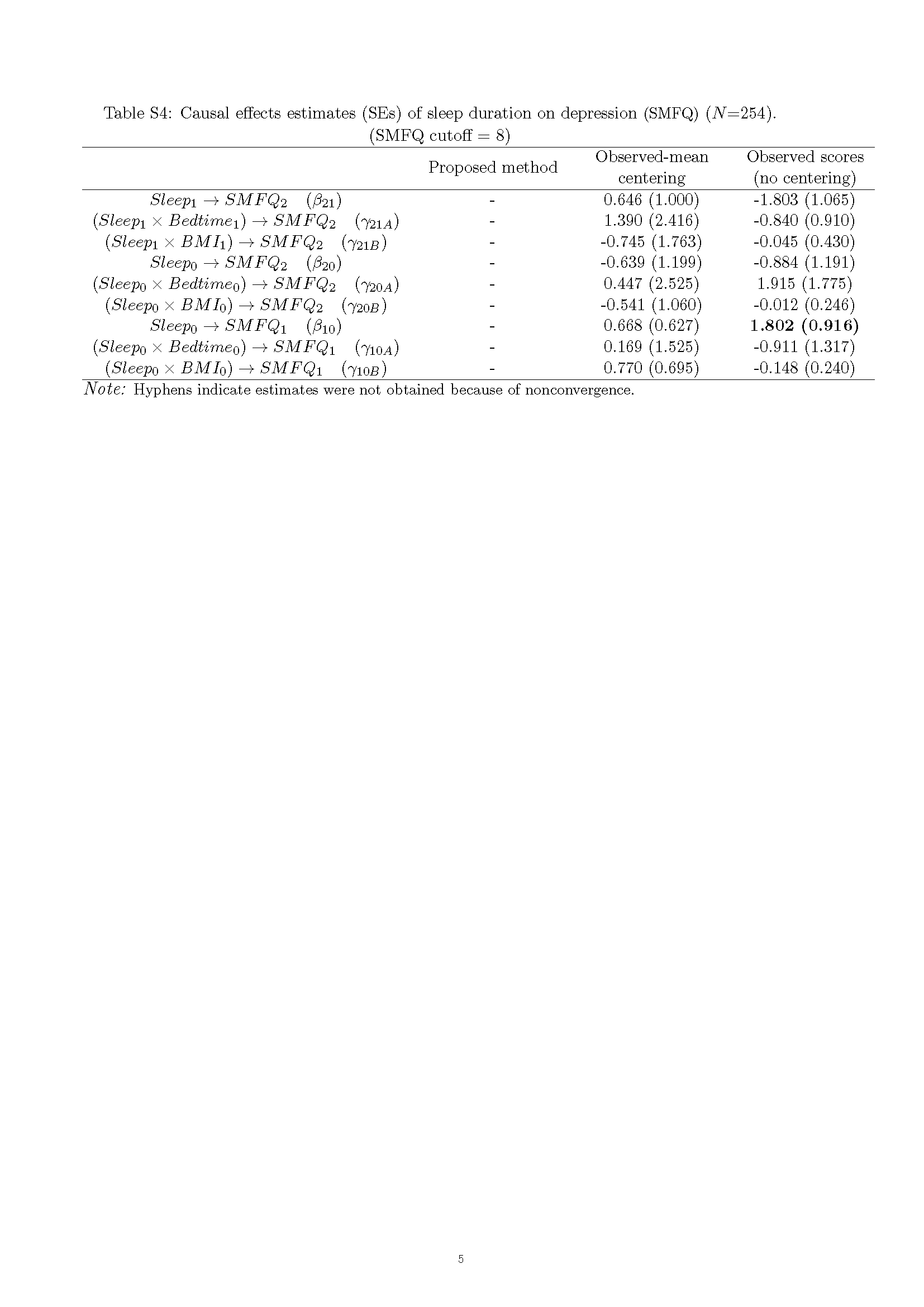}
\end{figure}
\begin{figure}[htbp]
\includegraphics[height=21cm,width=16cm,angle=0]{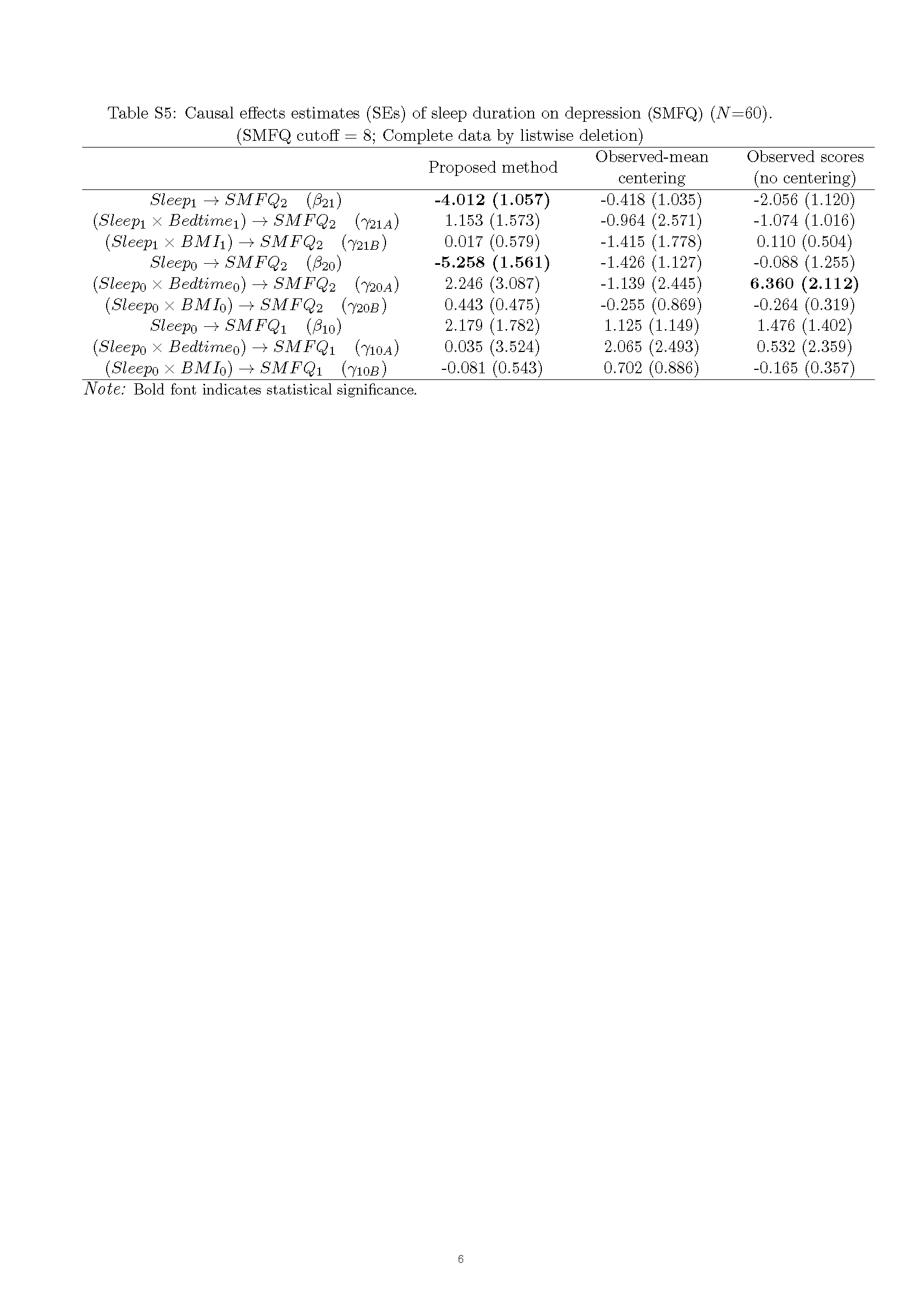}
\end{figure}
\begin{figure}[htbp]
\includegraphics[height=16cm,width=24cm,angle=90]{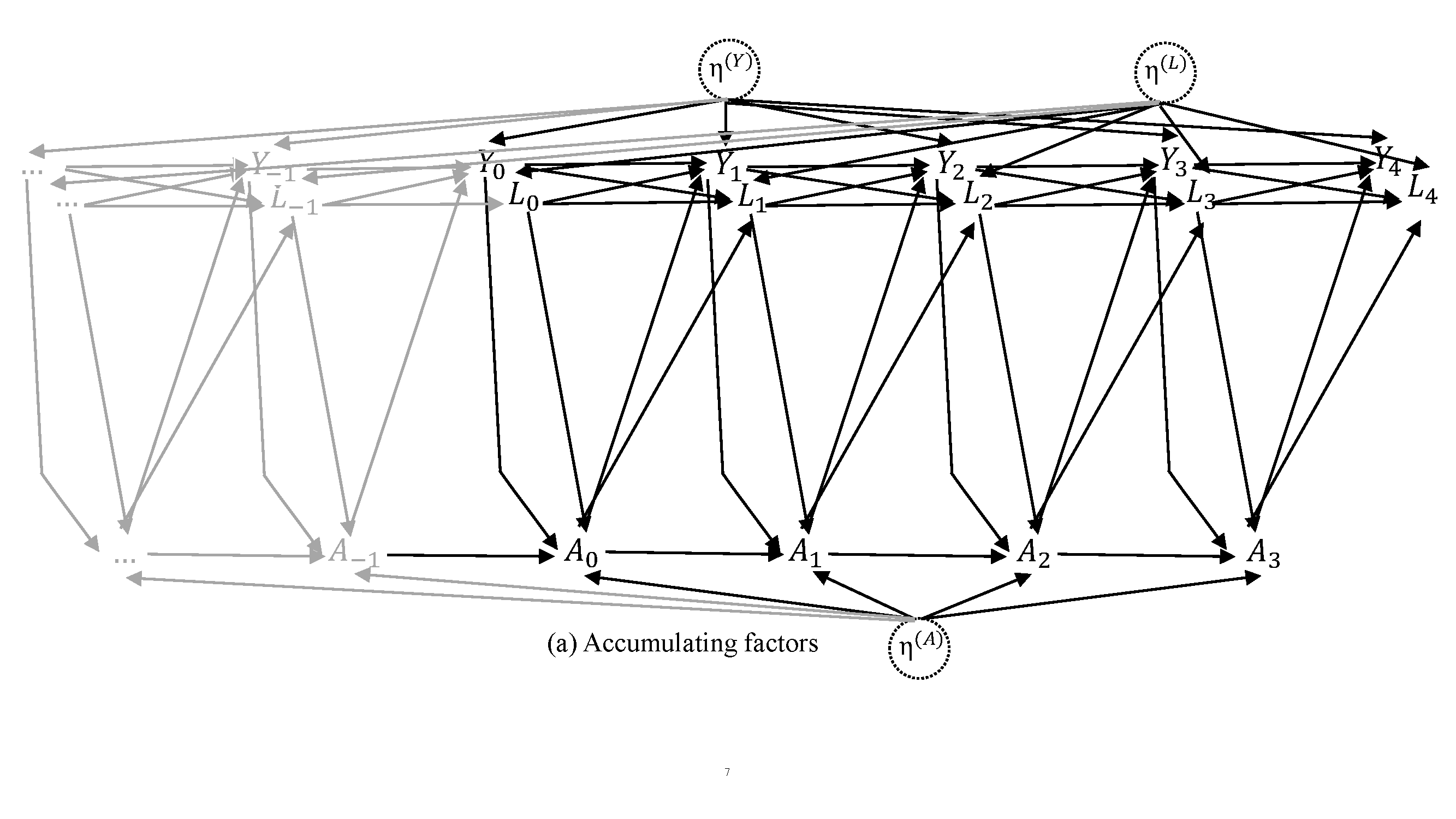}
\end{figure}
\begin{figure}[htbp]
\includegraphics[height=16cm,width=24cm,angle=90]{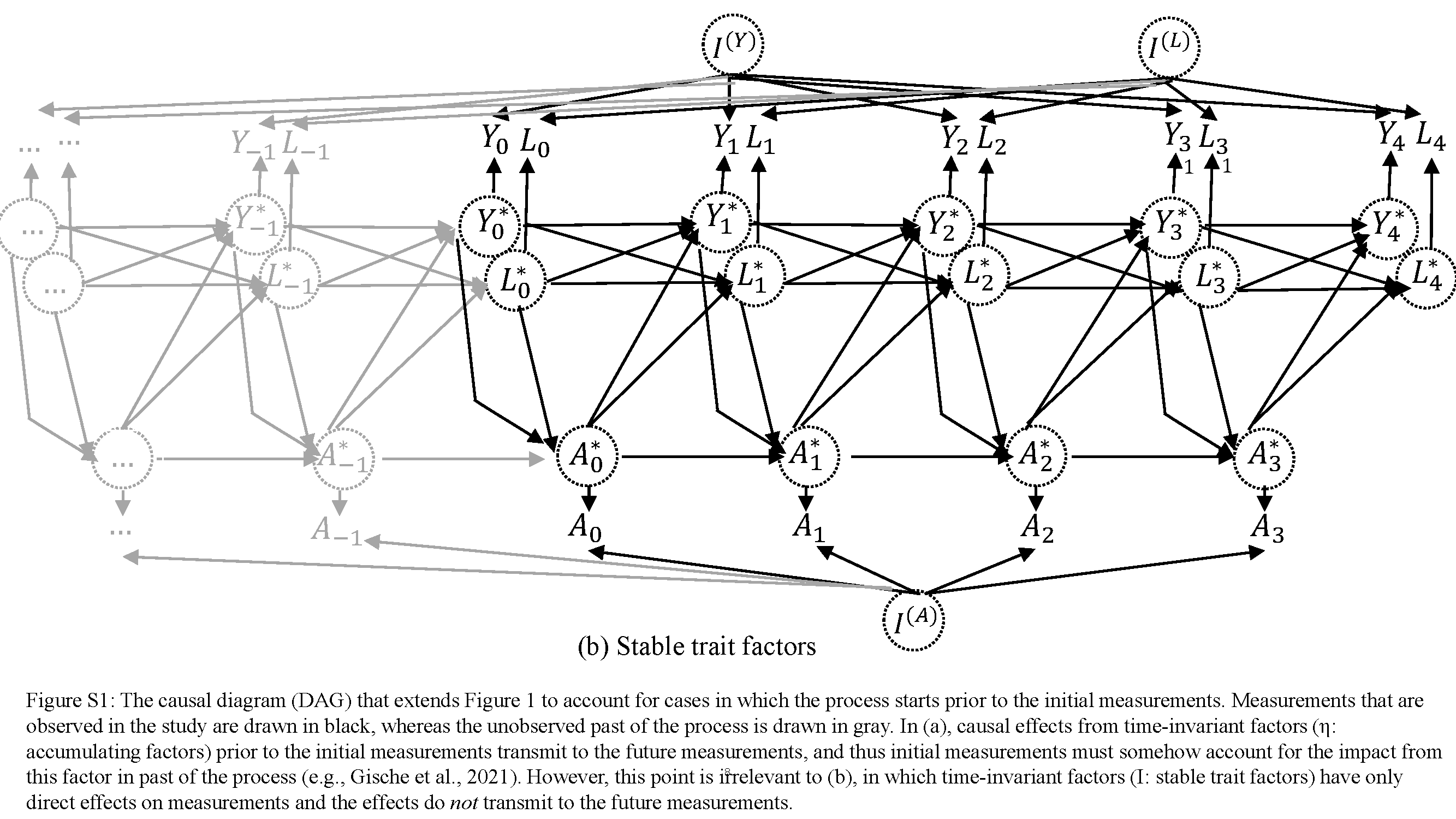}
\end{figure}
\begin{figure}[htbp]
\includegraphics[height=16cm,width=24cm,angle=90]{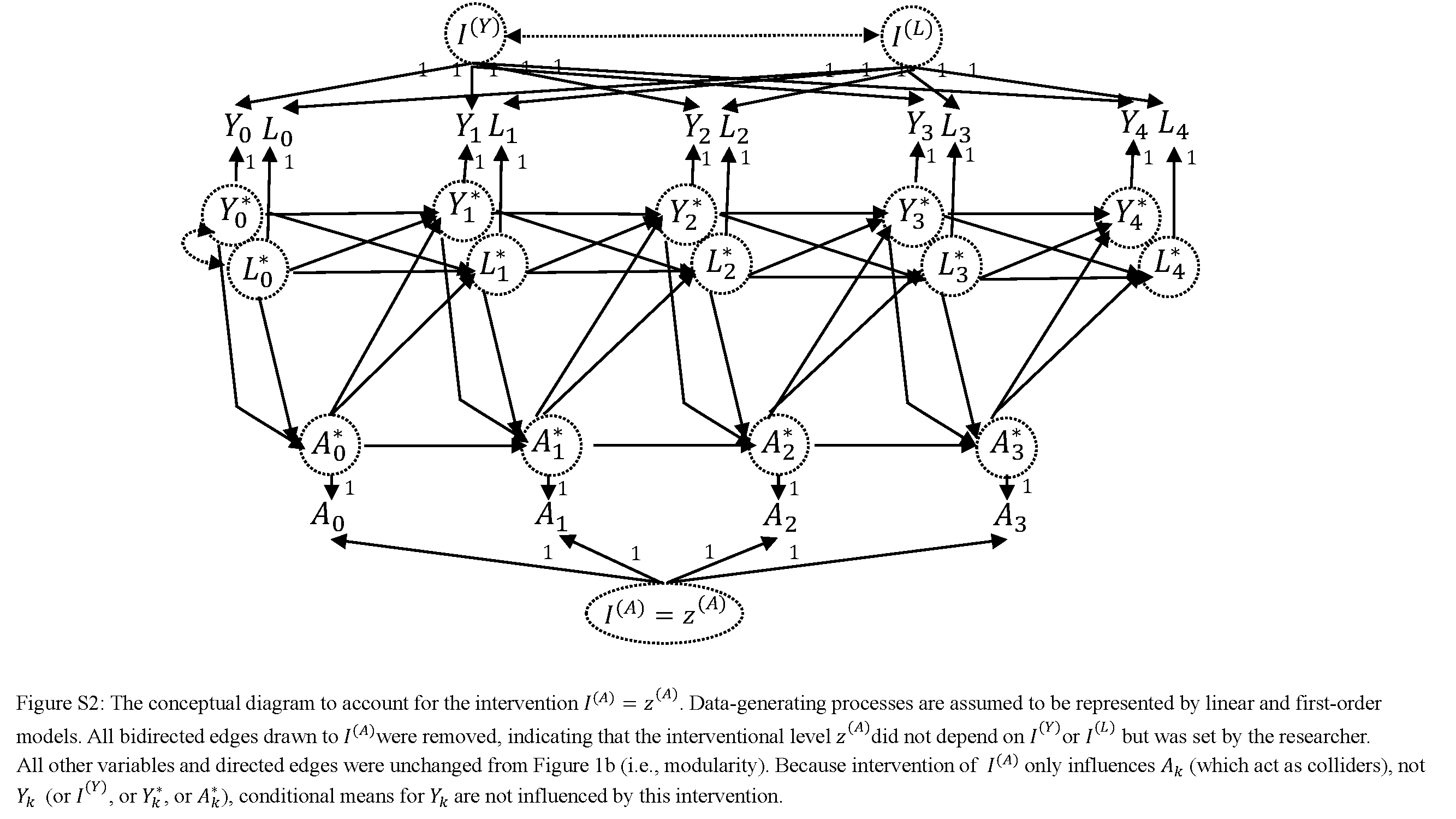}
\end{figure}
\begin{figure}[htbp]
\includegraphics[height=16cm,width=24cm,angle=90]{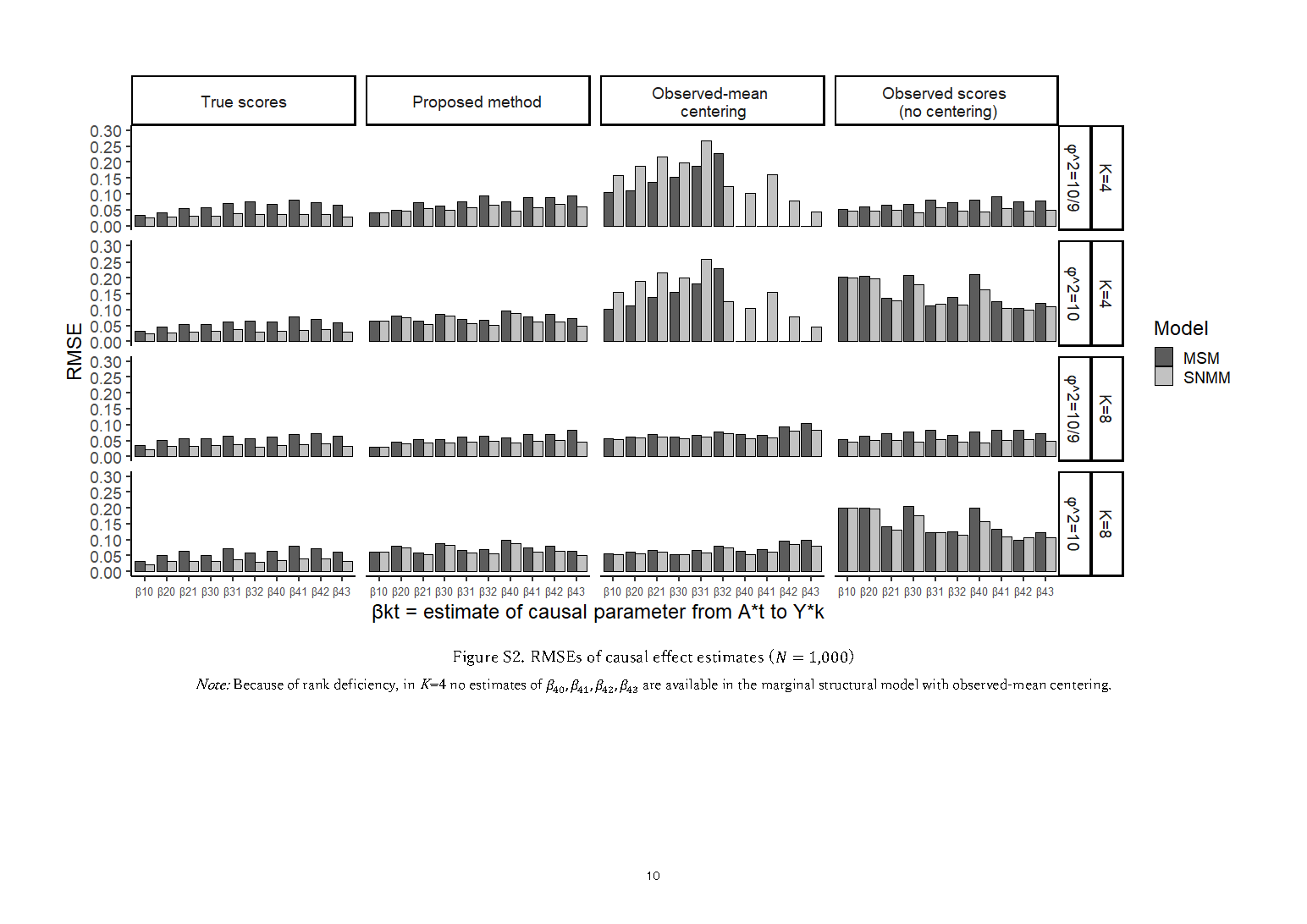}
\end{figure}
\begin{figure}[htbp]
\includegraphics[height=16cm,width=24cm,angle=90]{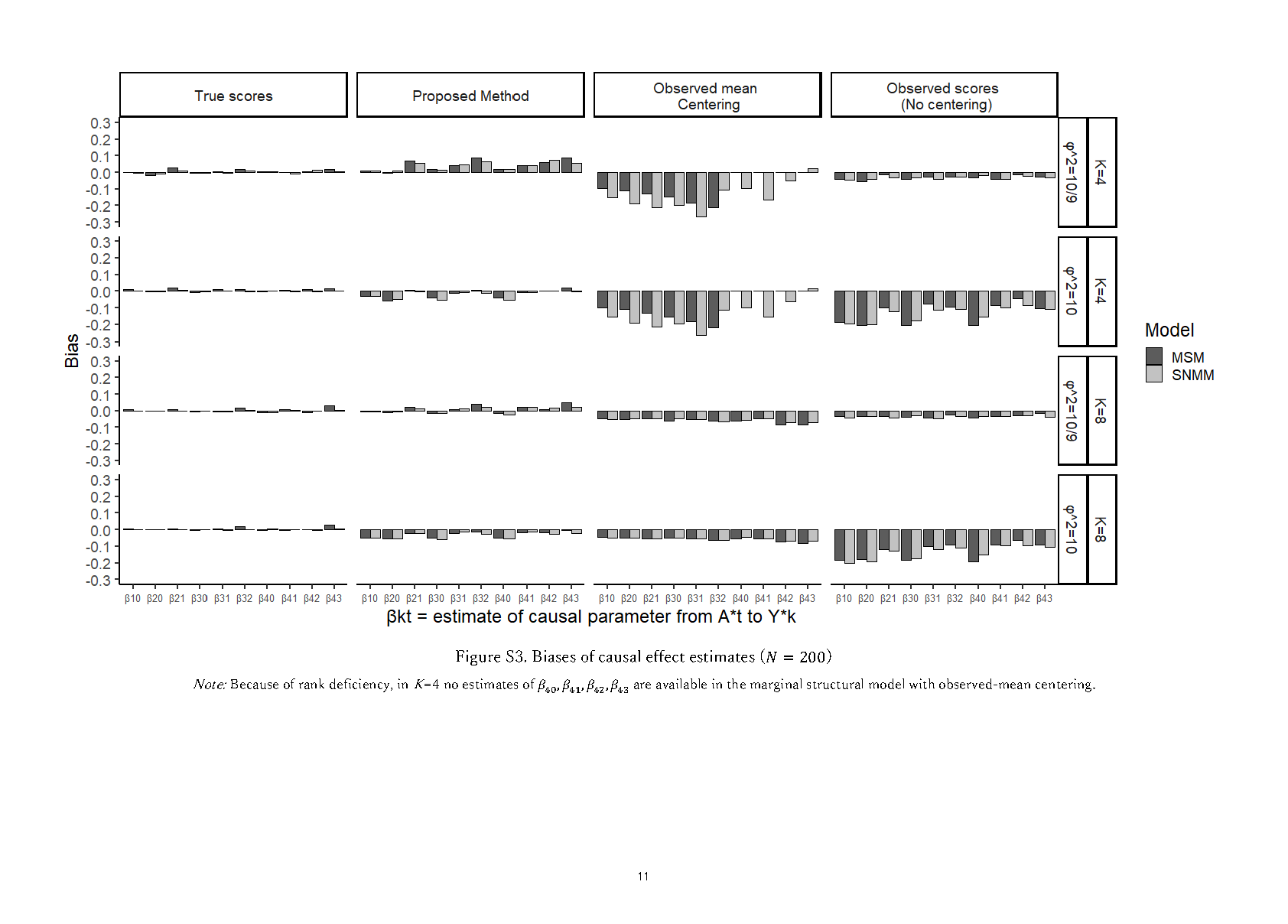}
\end{figure}
\begin{figure}[htbp]
\includegraphics[height=16cm,width=24cm,angle=90]{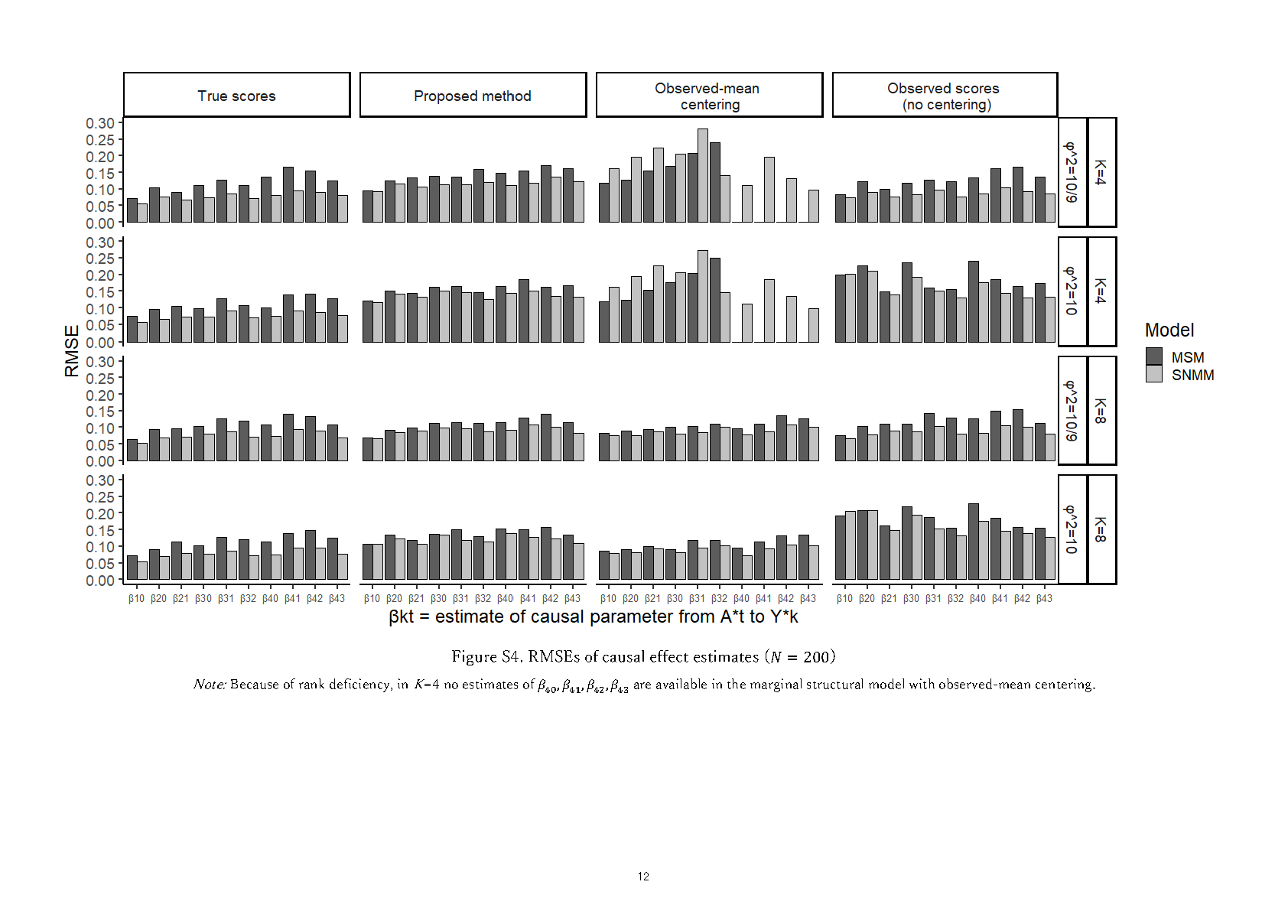}
\end{figure}
\begin{figure}[htbp]
\includegraphics[height=16cm,width=24cm,angle=90]{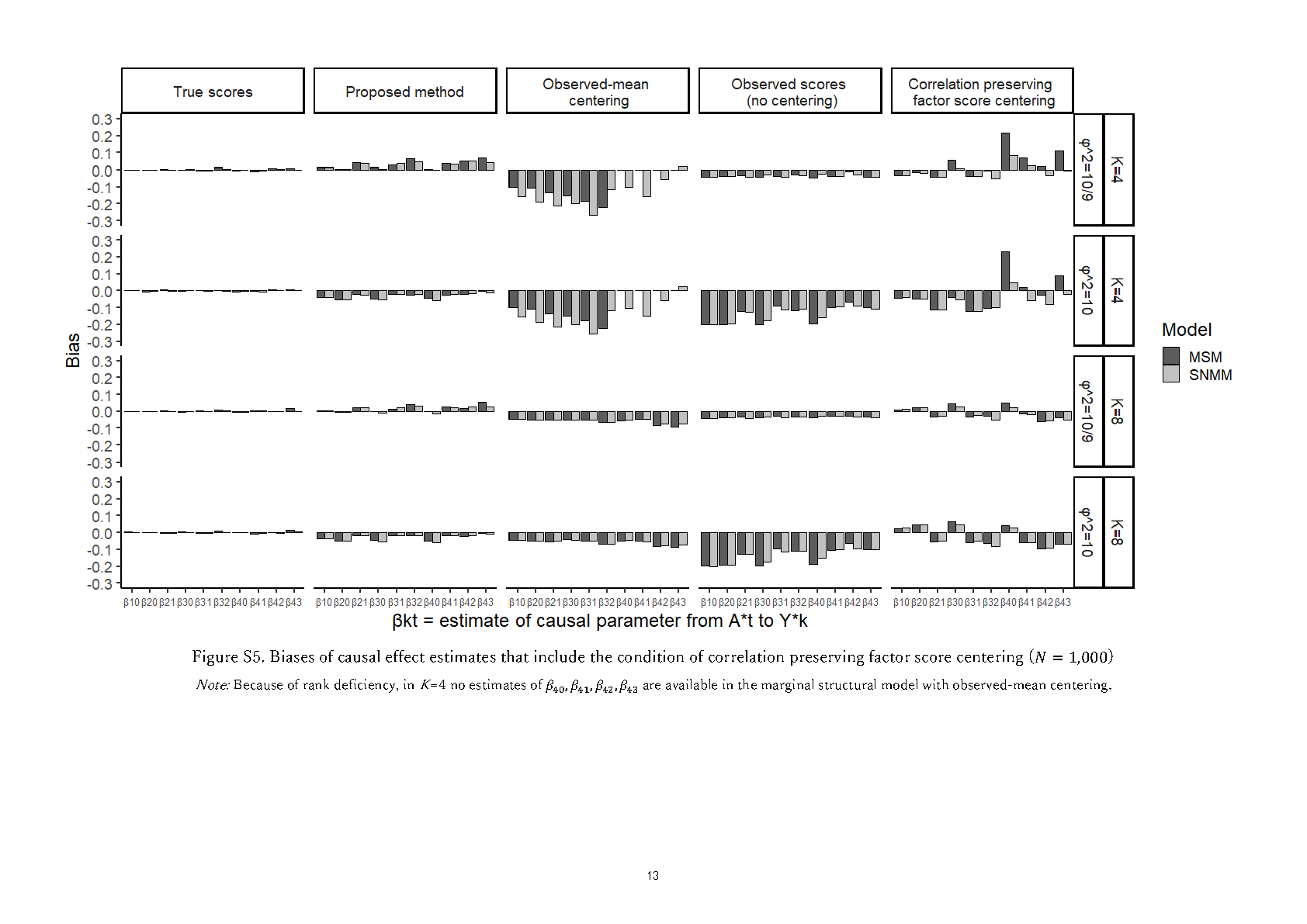}
\end{figure}
\begin{figure}[htbp]
\includegraphics[height=16cm,width=24cm,angle=90]{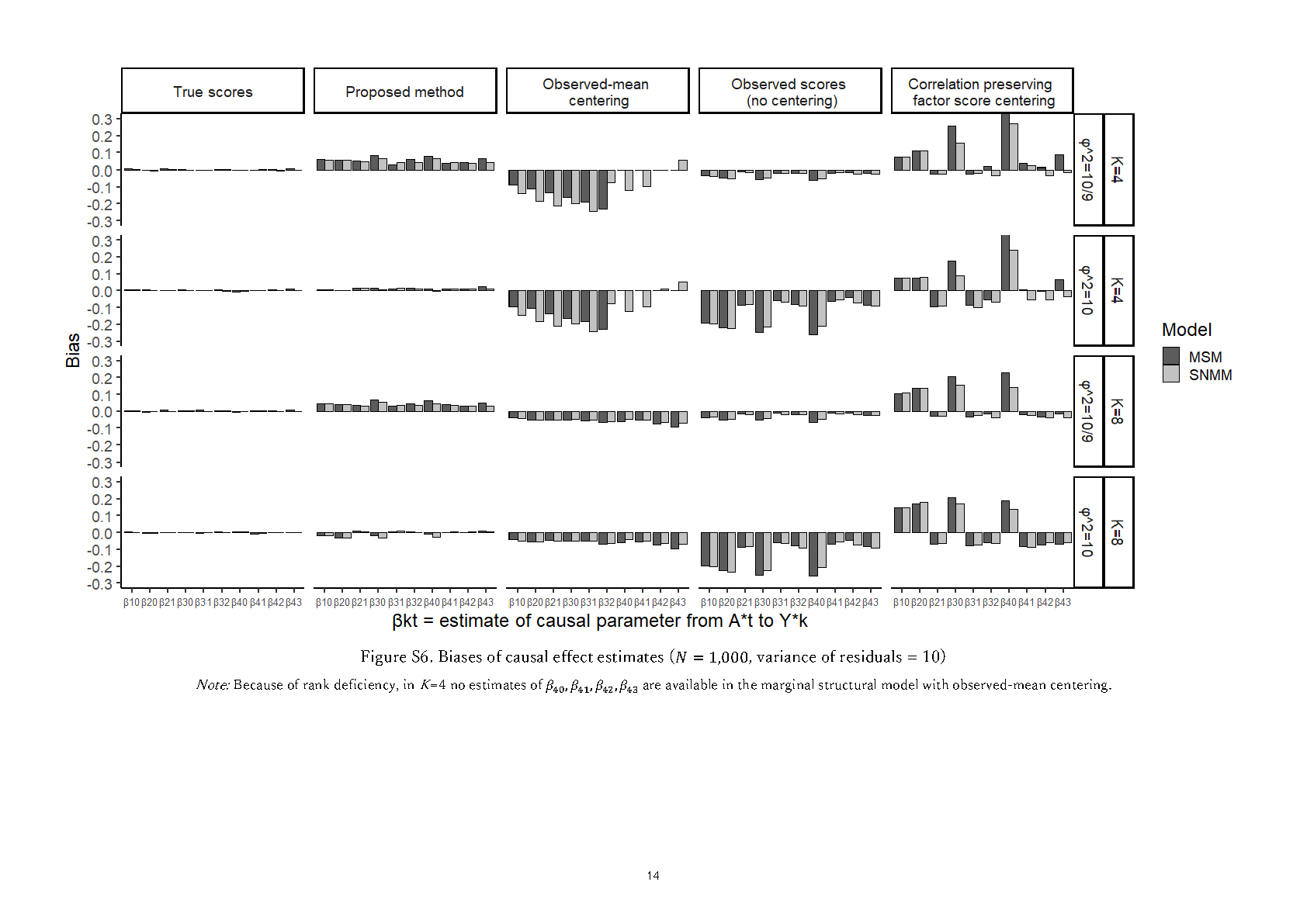}
\end{figure}
\begin{figure}[htbp]
\includegraphics[height=16cm,width=24cm,angle=90]{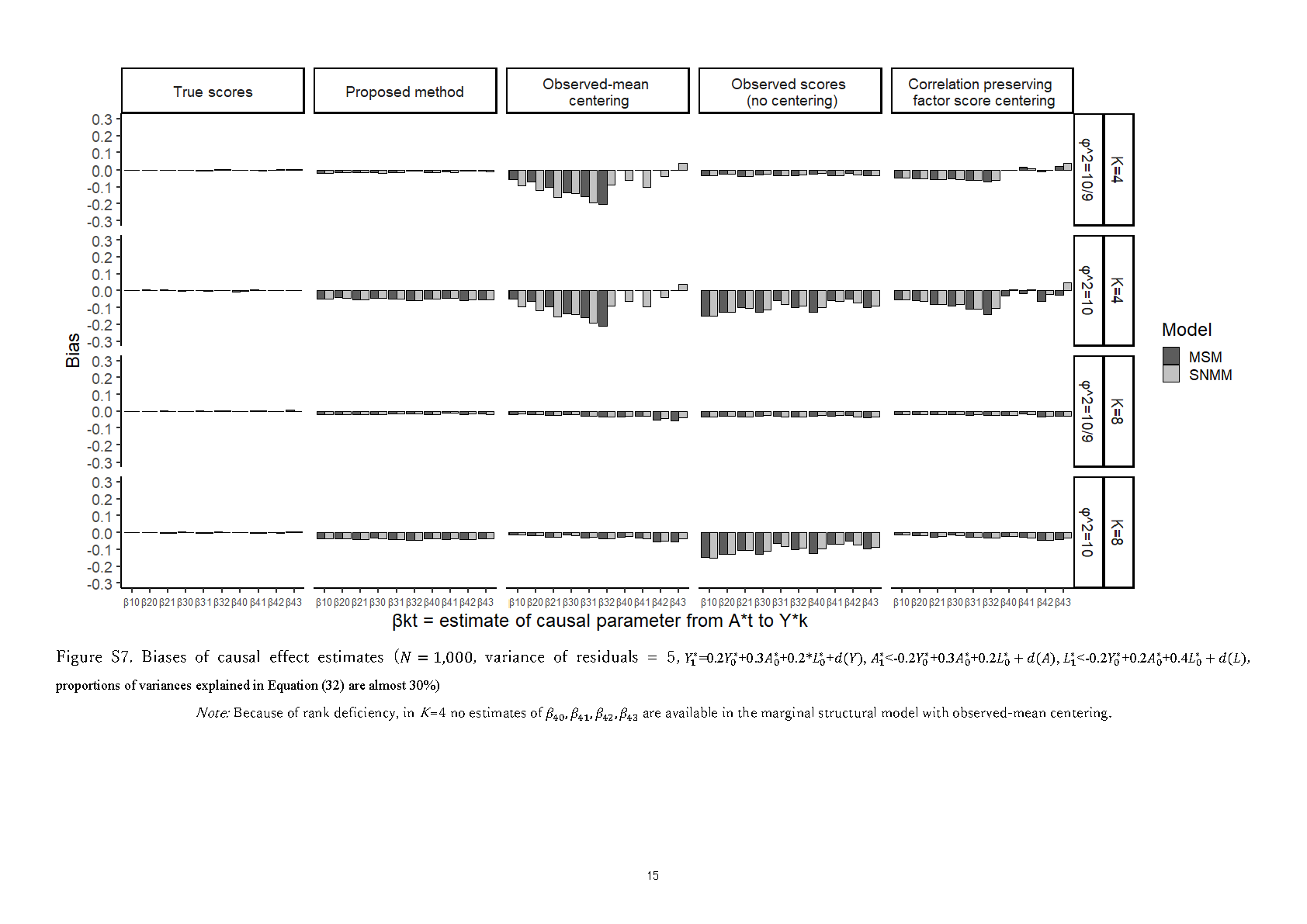}
\end{figure}
\begin{figure}[htbp]
\includegraphics[height=16cm,width=24cm,angle=90]{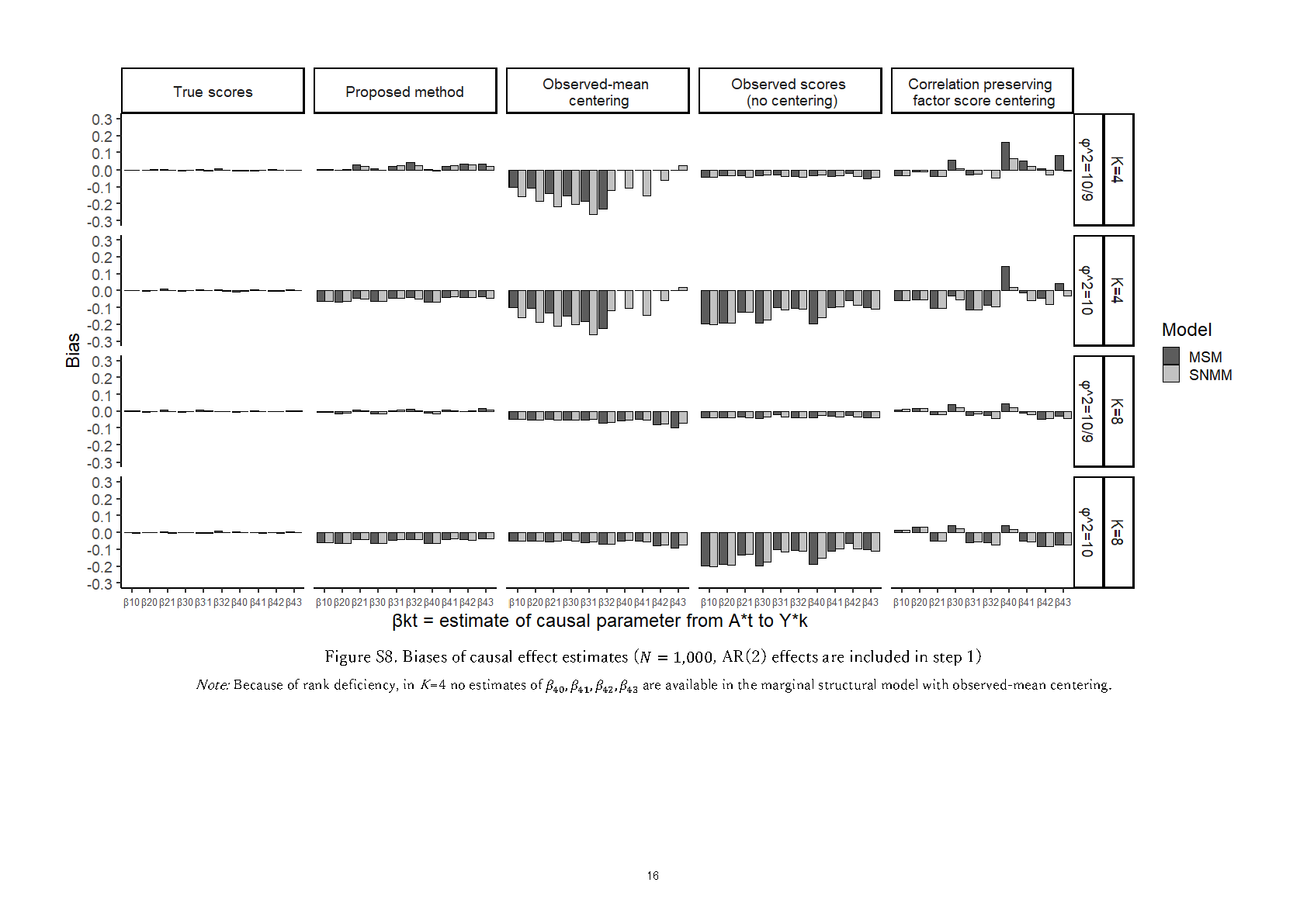}
\end{figure}
\begin{figure}[htbp]
\includegraphics[height=16cm,width=24cm,angle=90]{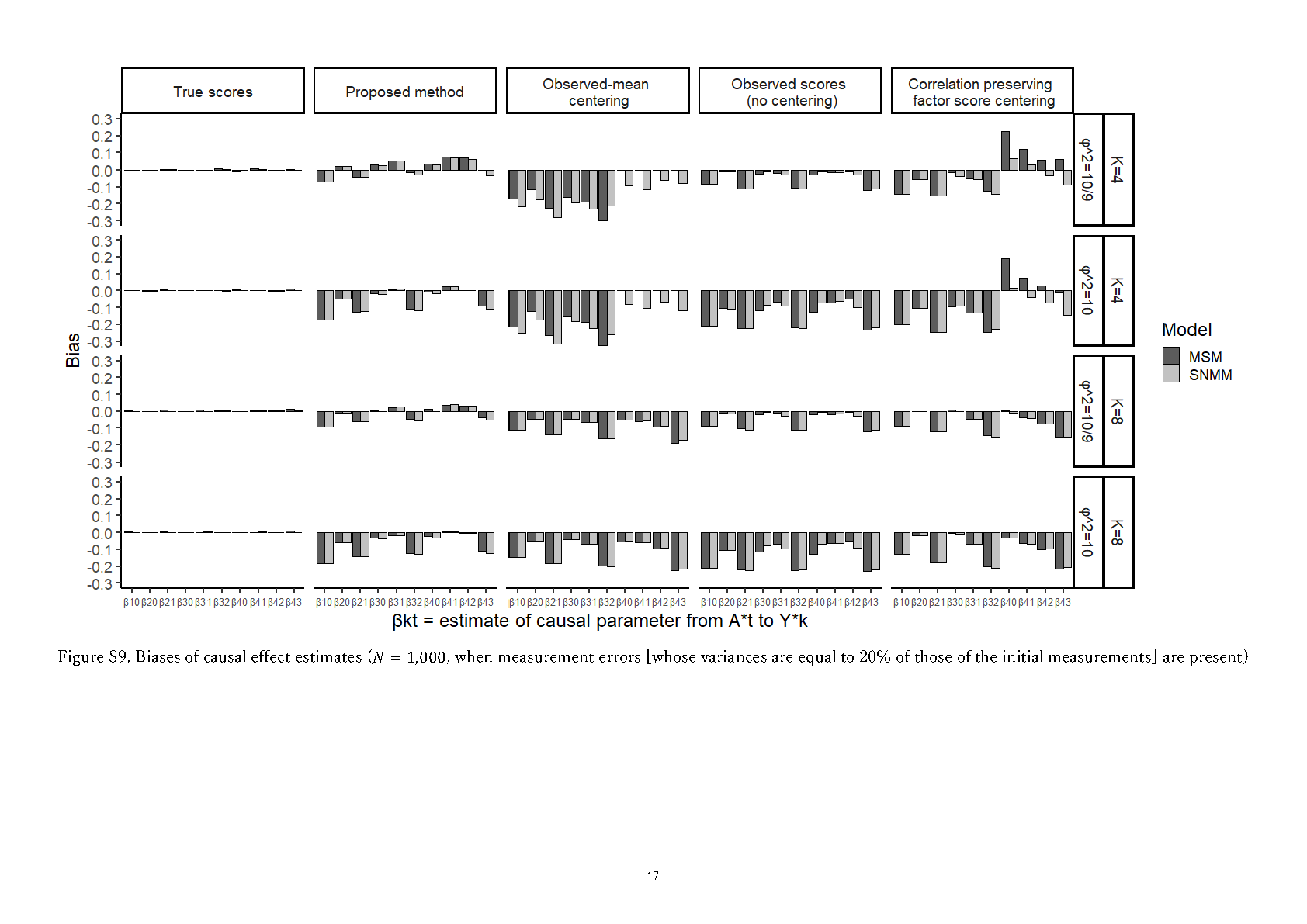}
\end{figure}
\begin{figure}[htbp]
\includegraphics[height=16cm,width=24cm,angle=90]{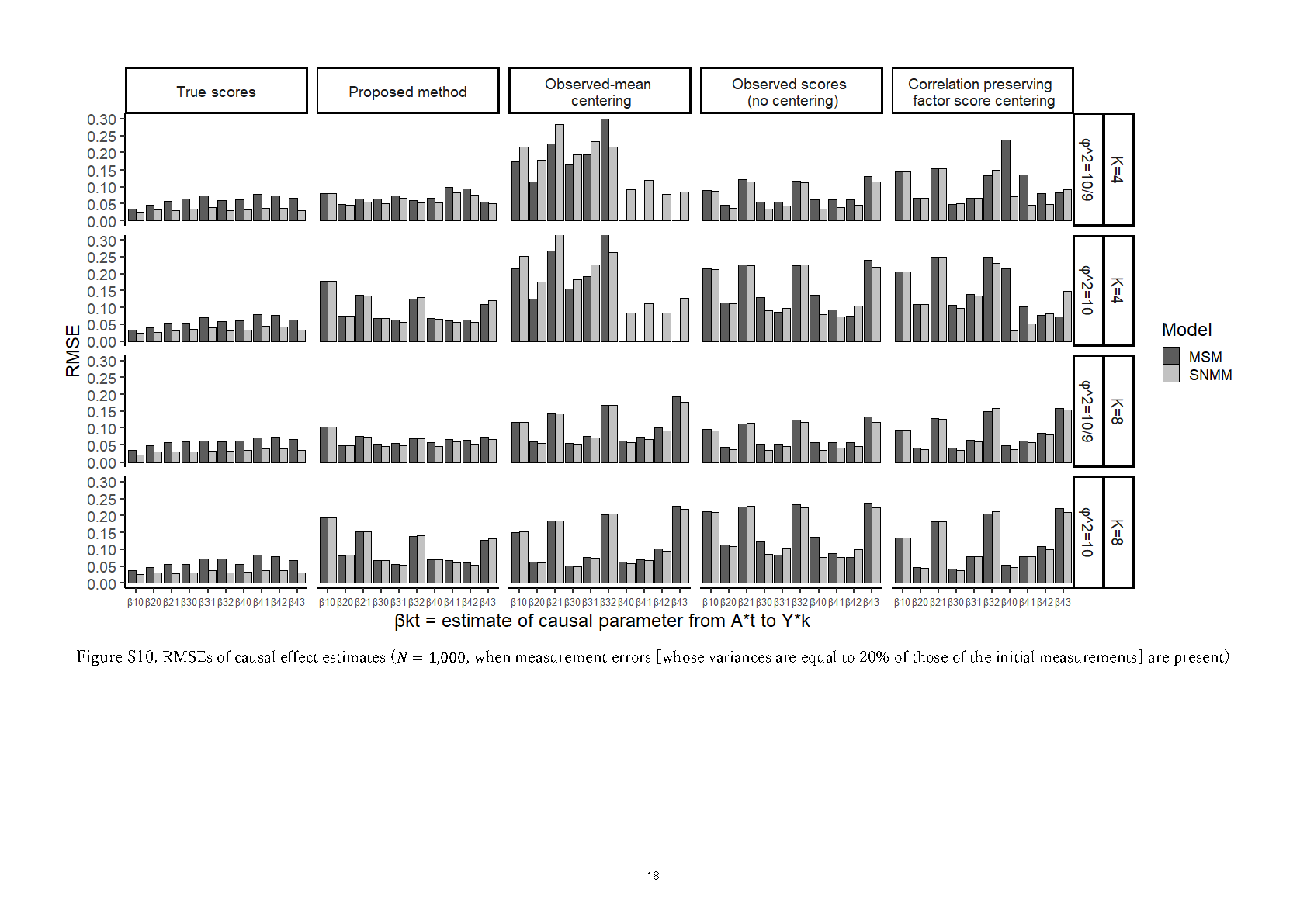}
\end{figure}
\begin{figure}[htbp]
\includegraphics[height=16cm,width=24cm,angle=90]{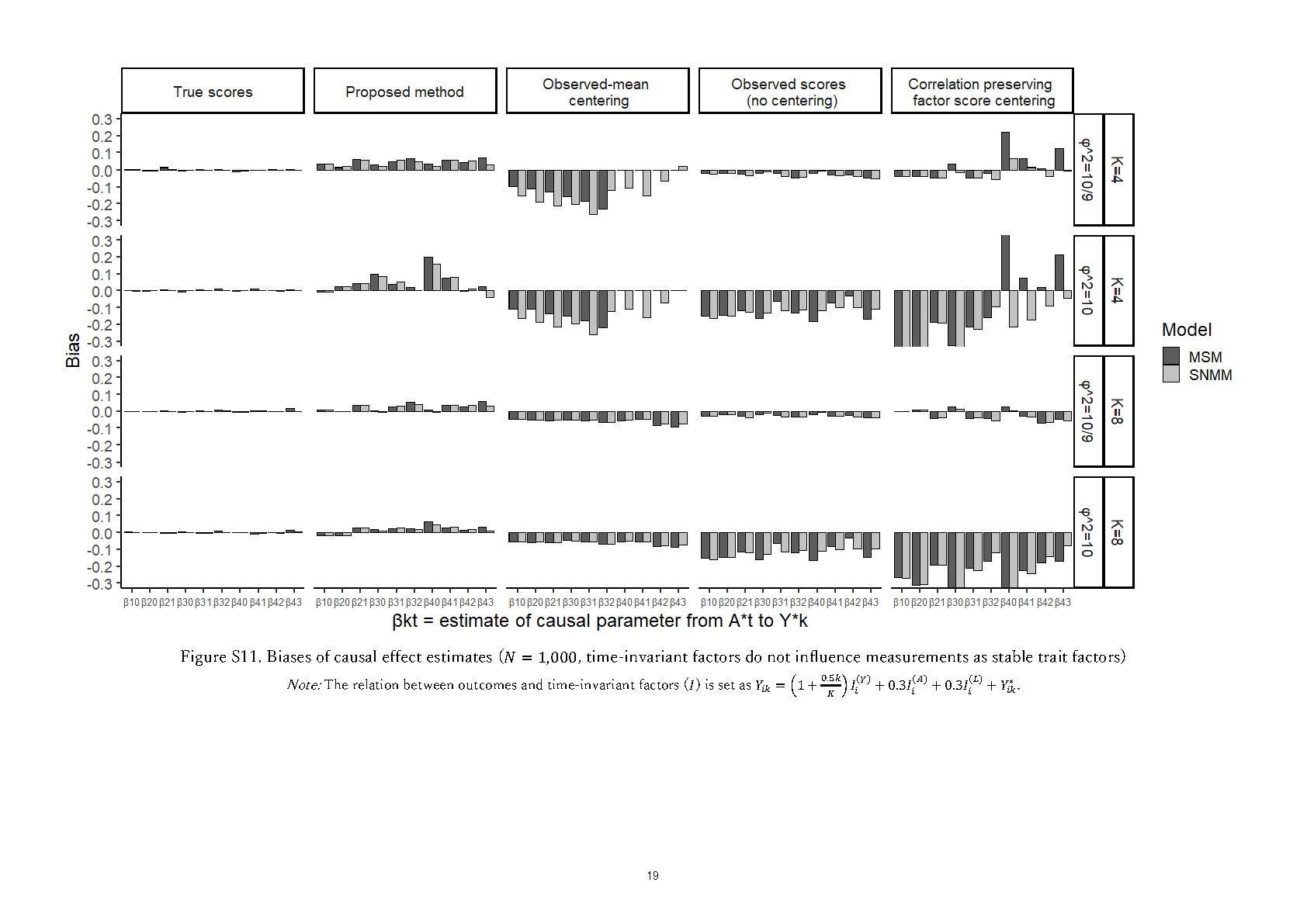}
\end{figure}
\begin{figure}[htbp]
\includegraphics[height=16cm,width=24cm,angle=90]{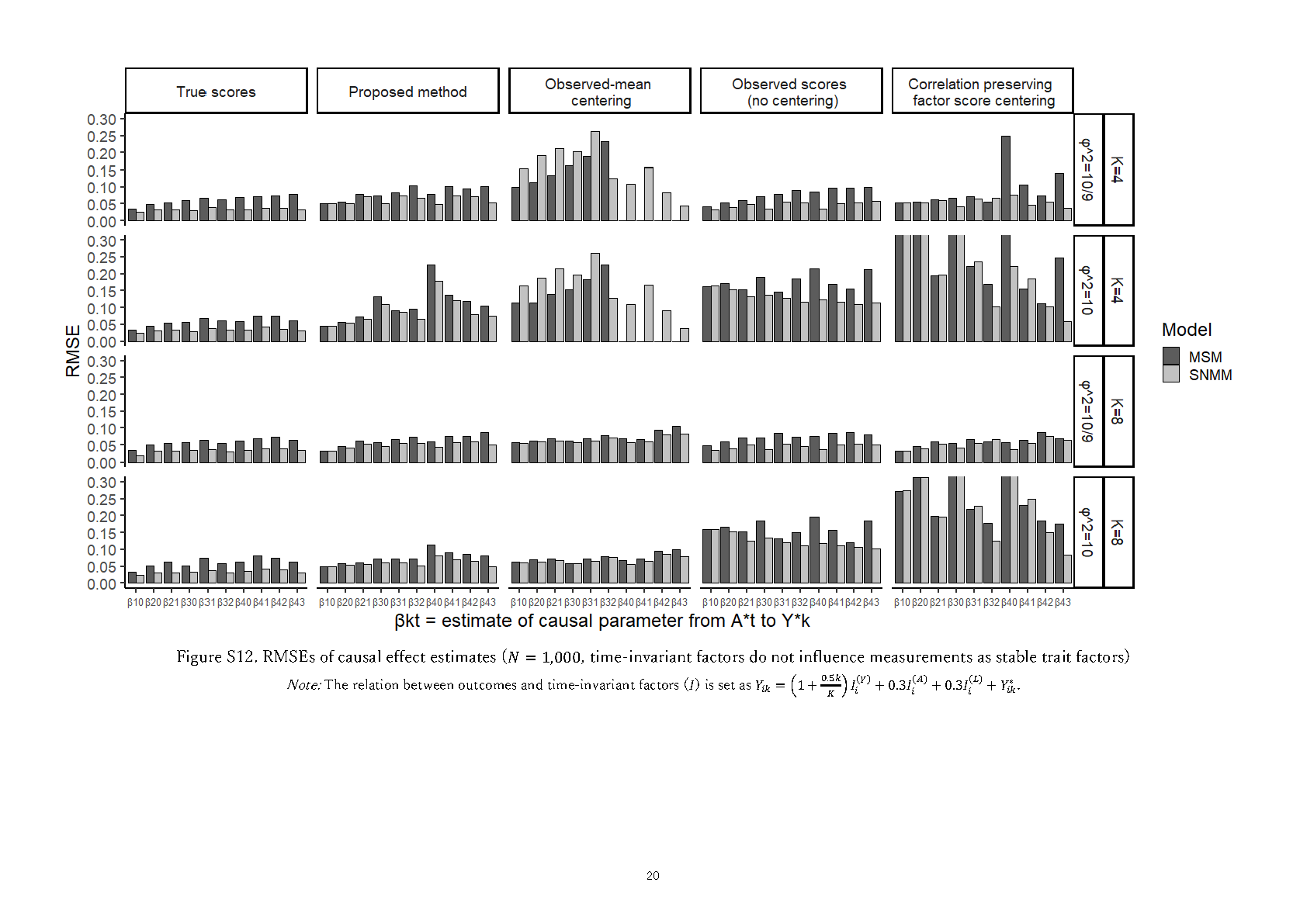}
\end{figure}
\begin{figure}[htbp]
\includegraphics[height=16cm,width=24cm,angle=90]{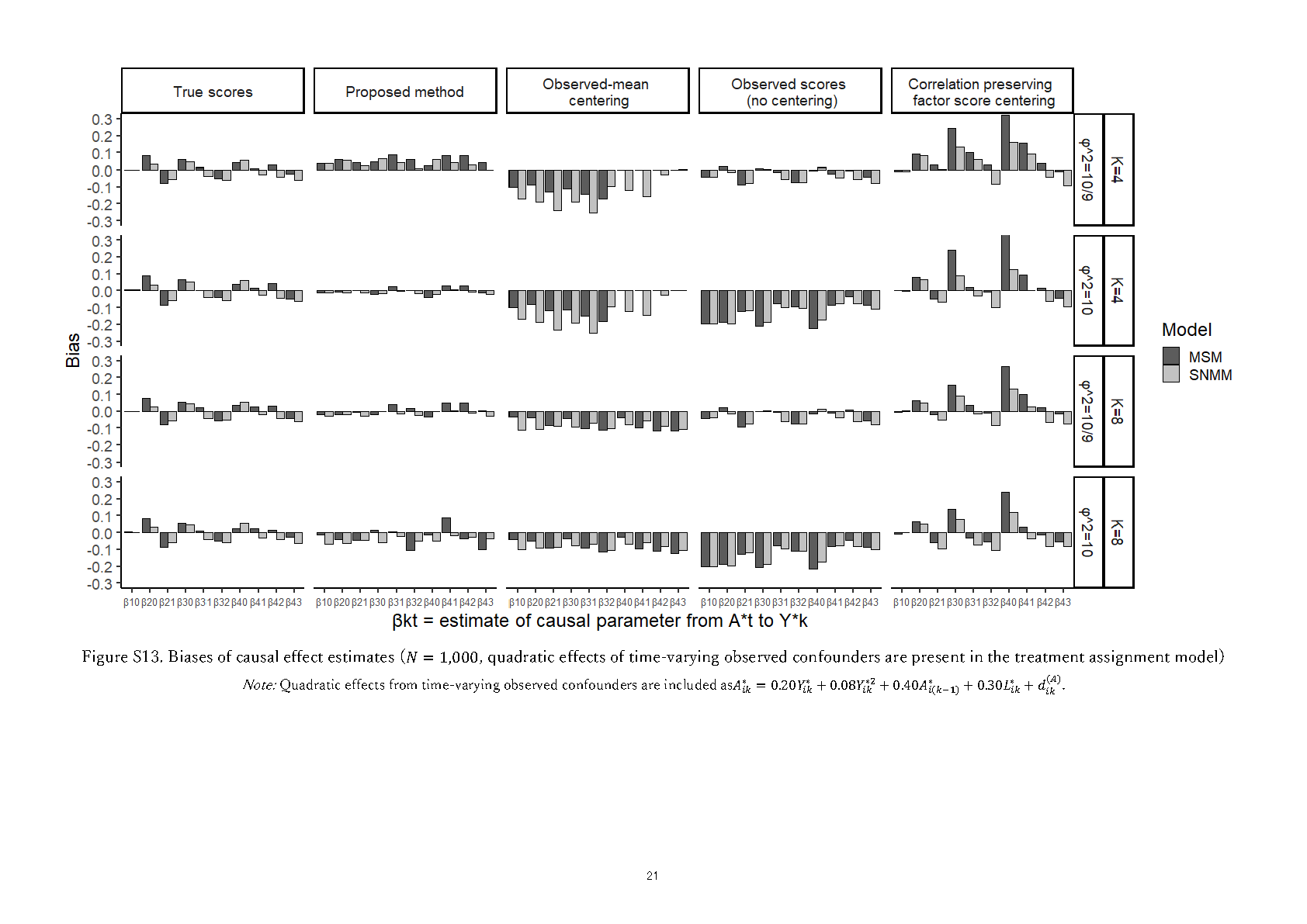}
\end{figure}
\begin{figure}[htbp]
\includegraphics[height=16cm,width=24cm,angle=90]{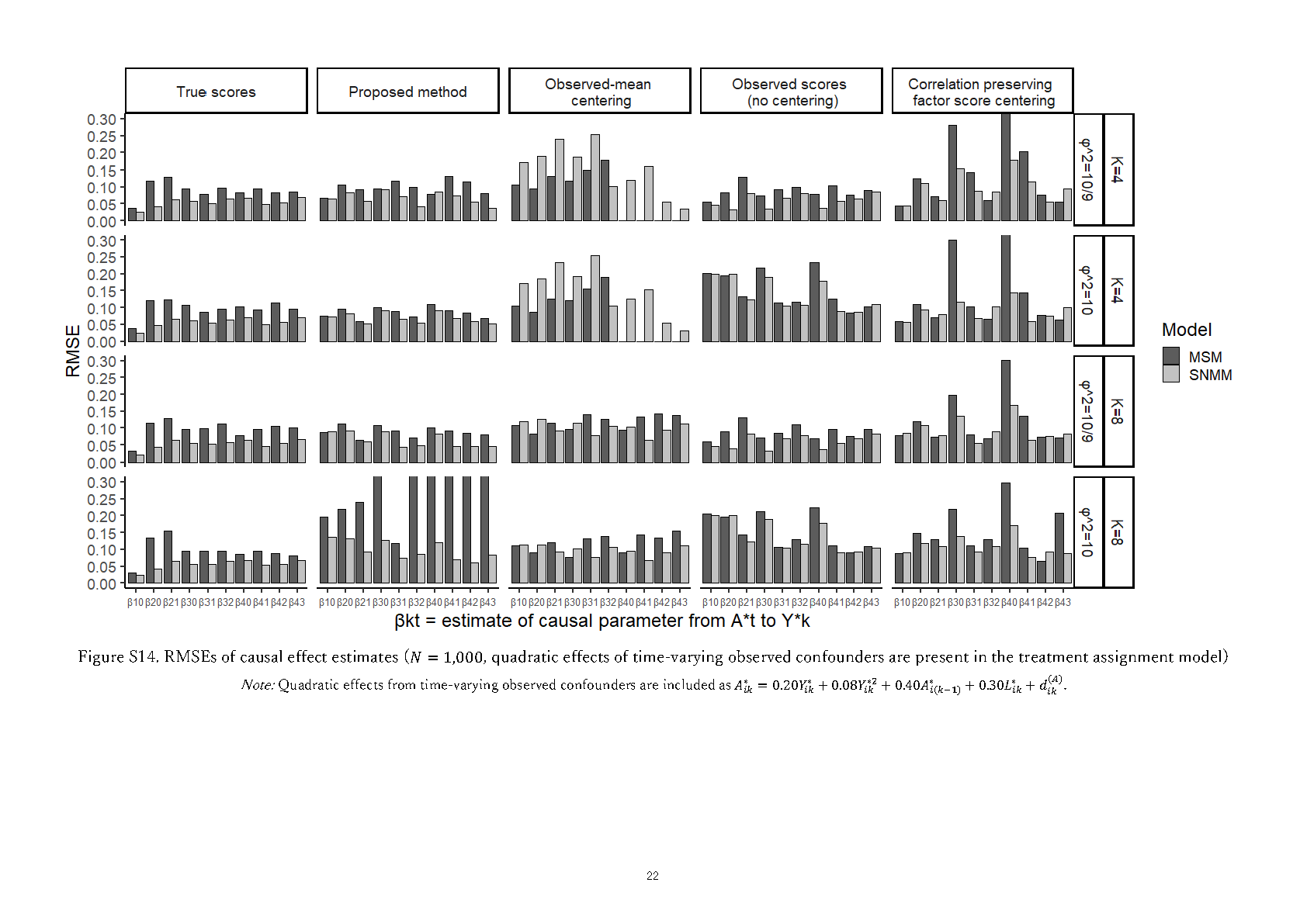}
\end{figure}
\begin{figure}[htbp]
\includegraphics[height=16cm,width=16cm,angle=0]{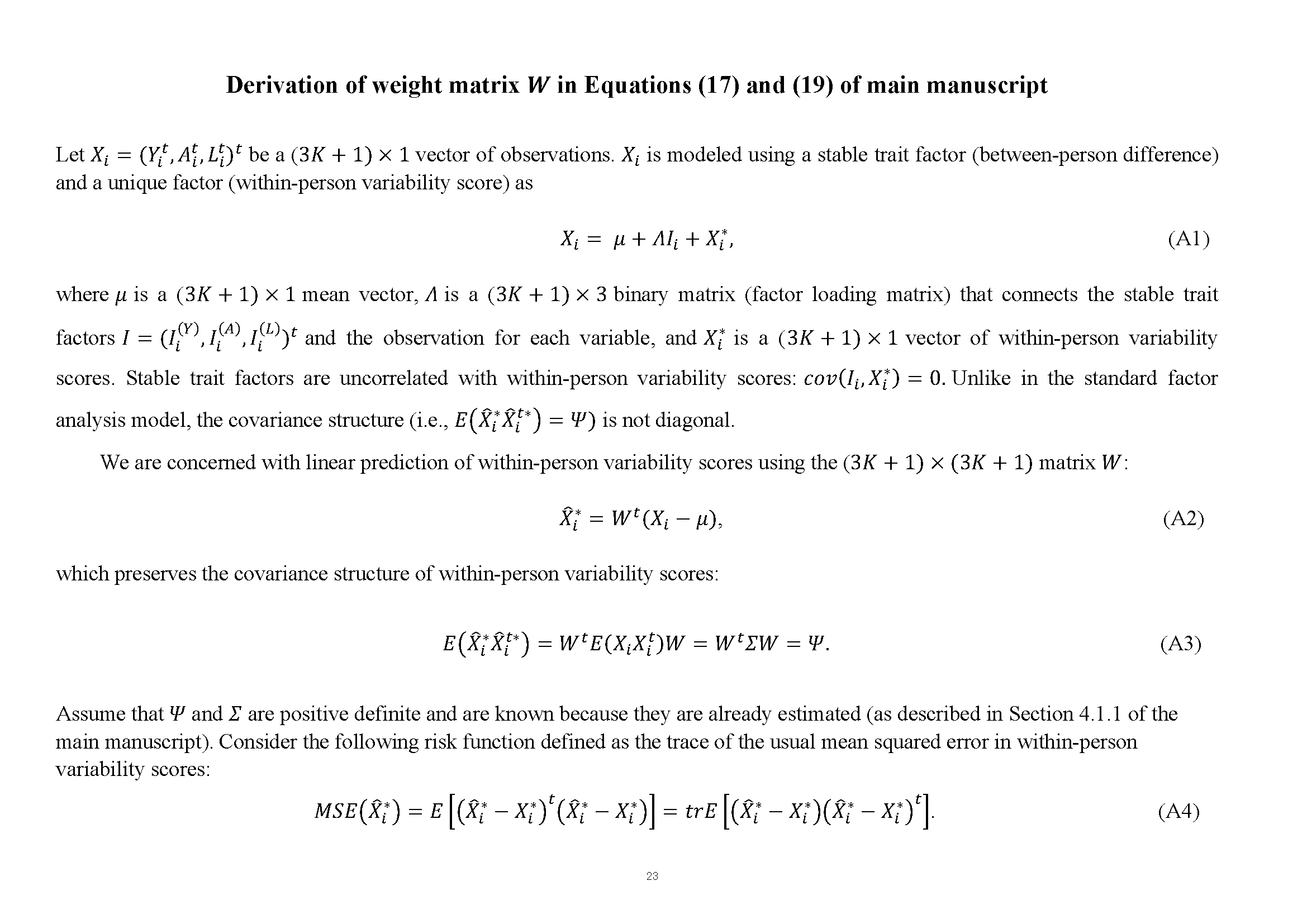}
\end{figure}
\begin{figure}[htbp]
\includegraphics[height=16cm,width=16cm,angle=0]{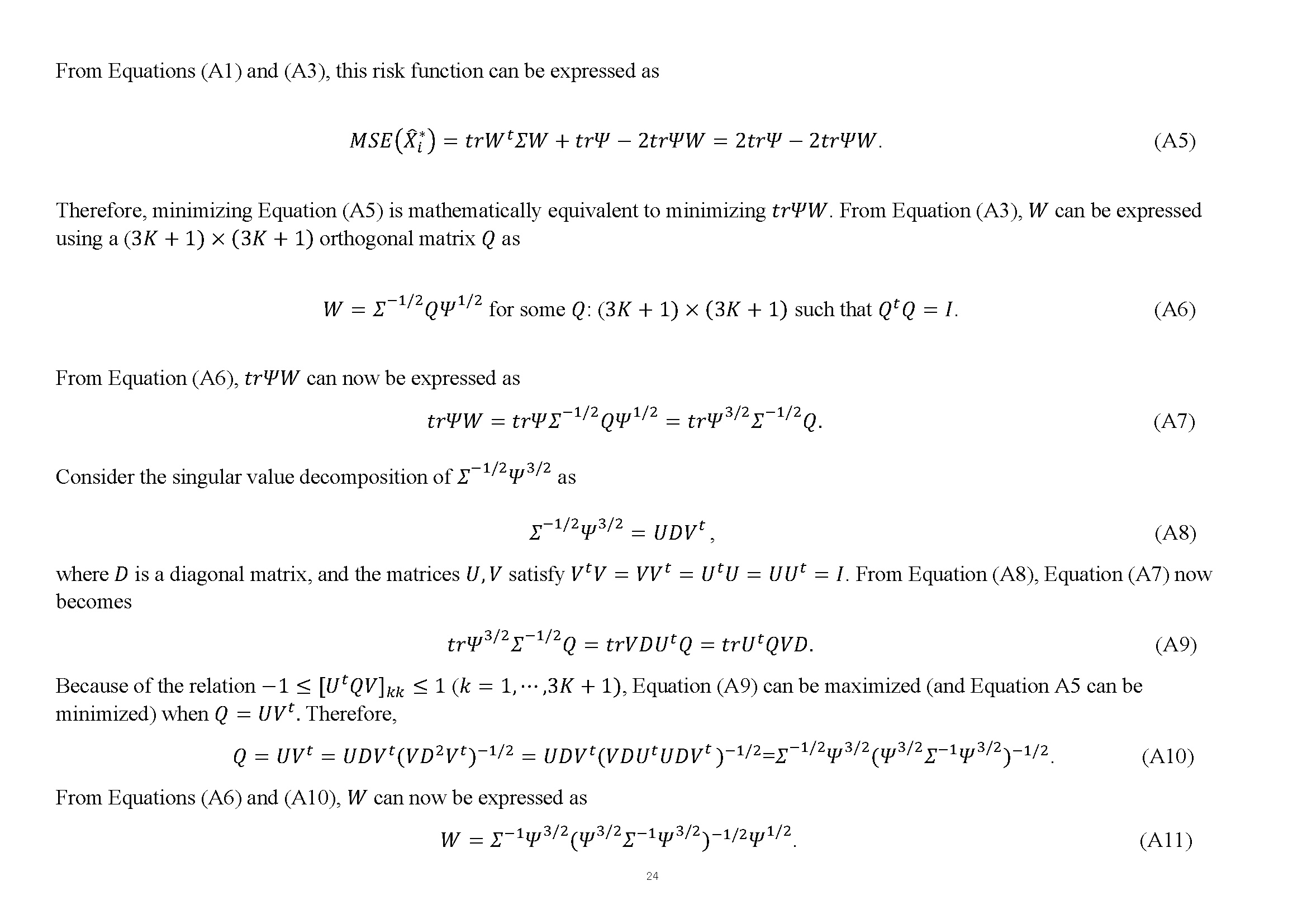}
\end{figure}
\begin{figure}[htbp]
\includegraphics[height=16cm,width=16cm,angle=0]{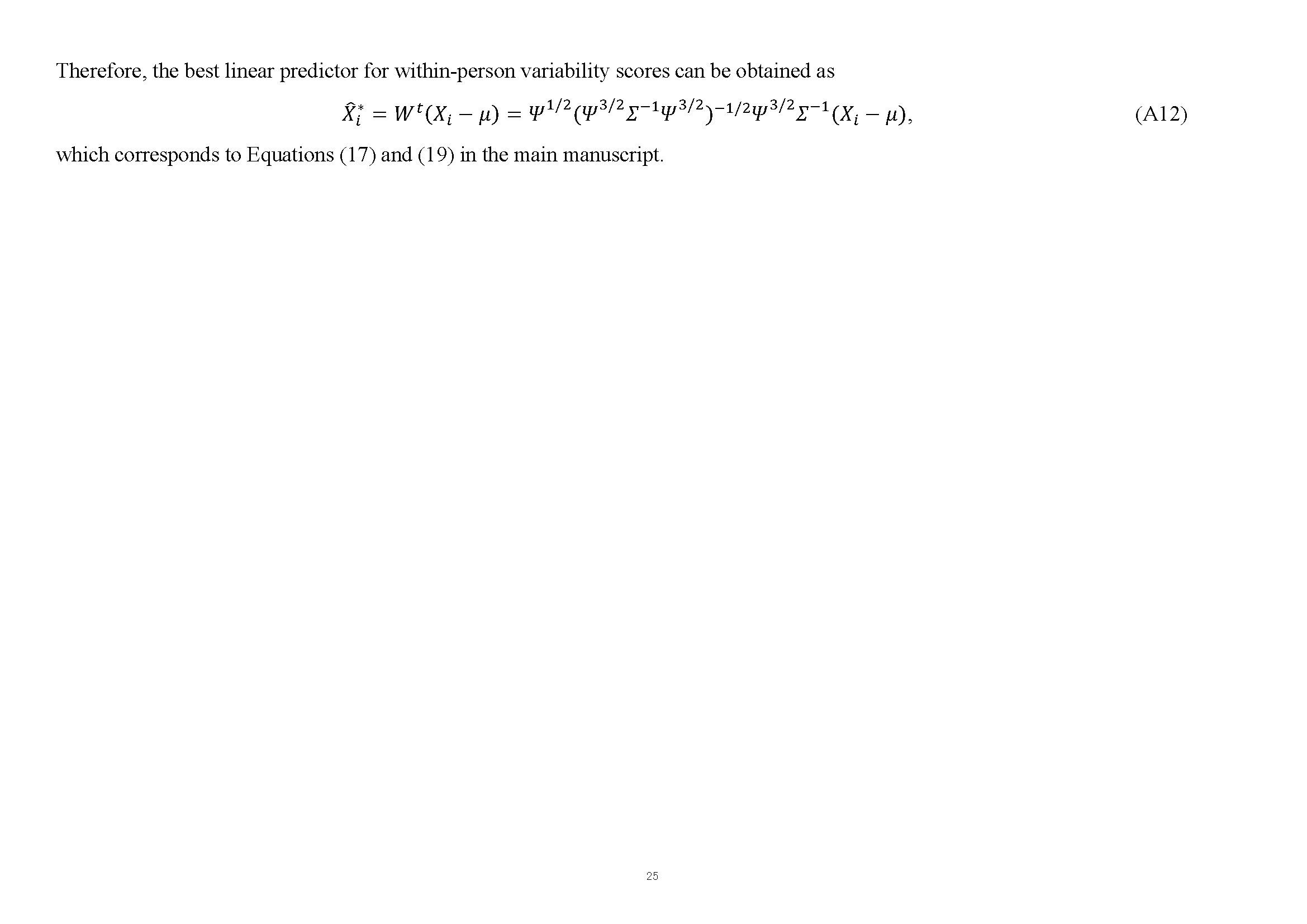}
\end{figure}
\begin{figure}[htbp]
\includegraphics[height=21cm,width=16cm,angle=0]{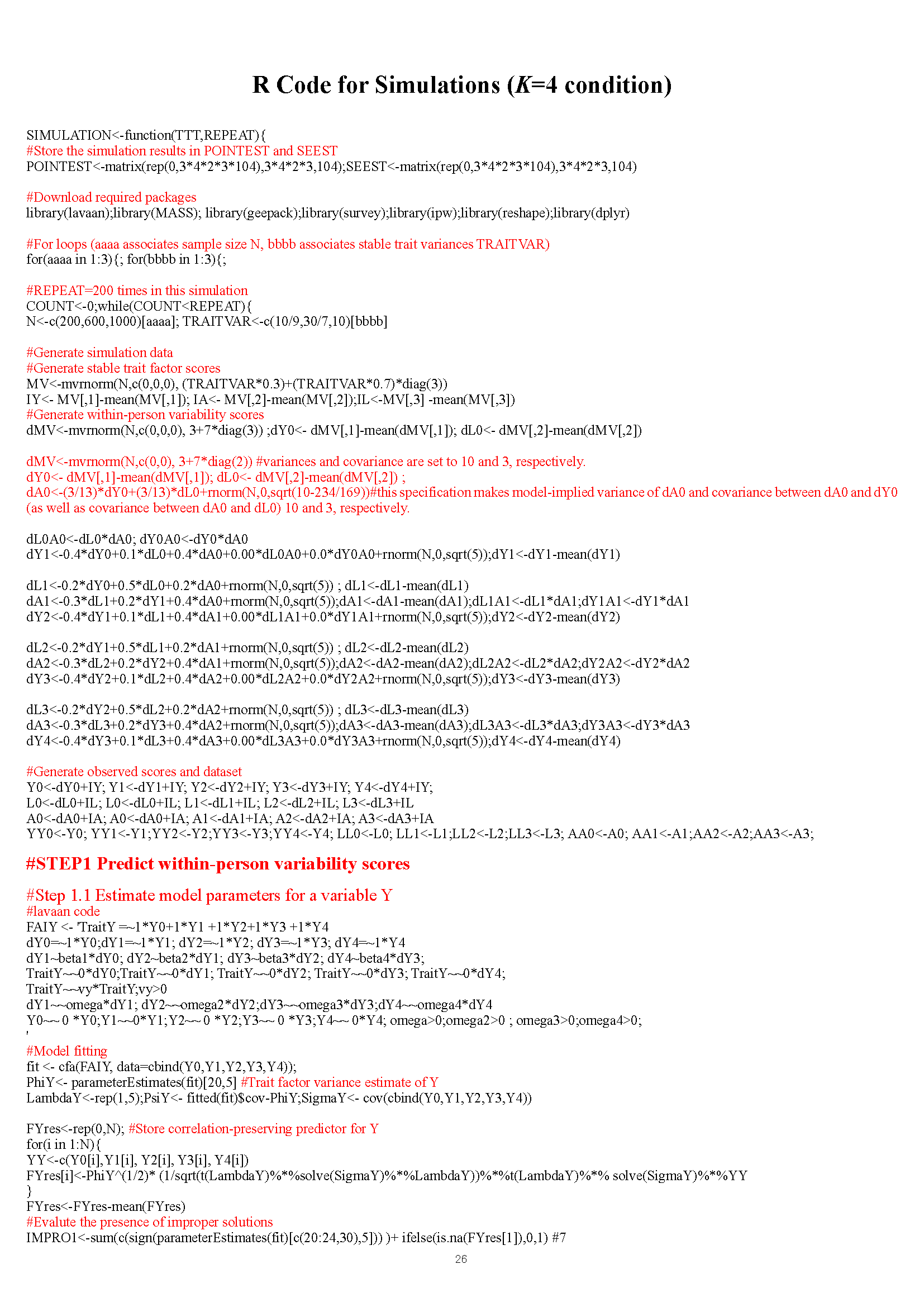}
\end{figure}
\begin{figure}[htbp]
\includegraphics[height=21cm,width=16cm,angle=0]{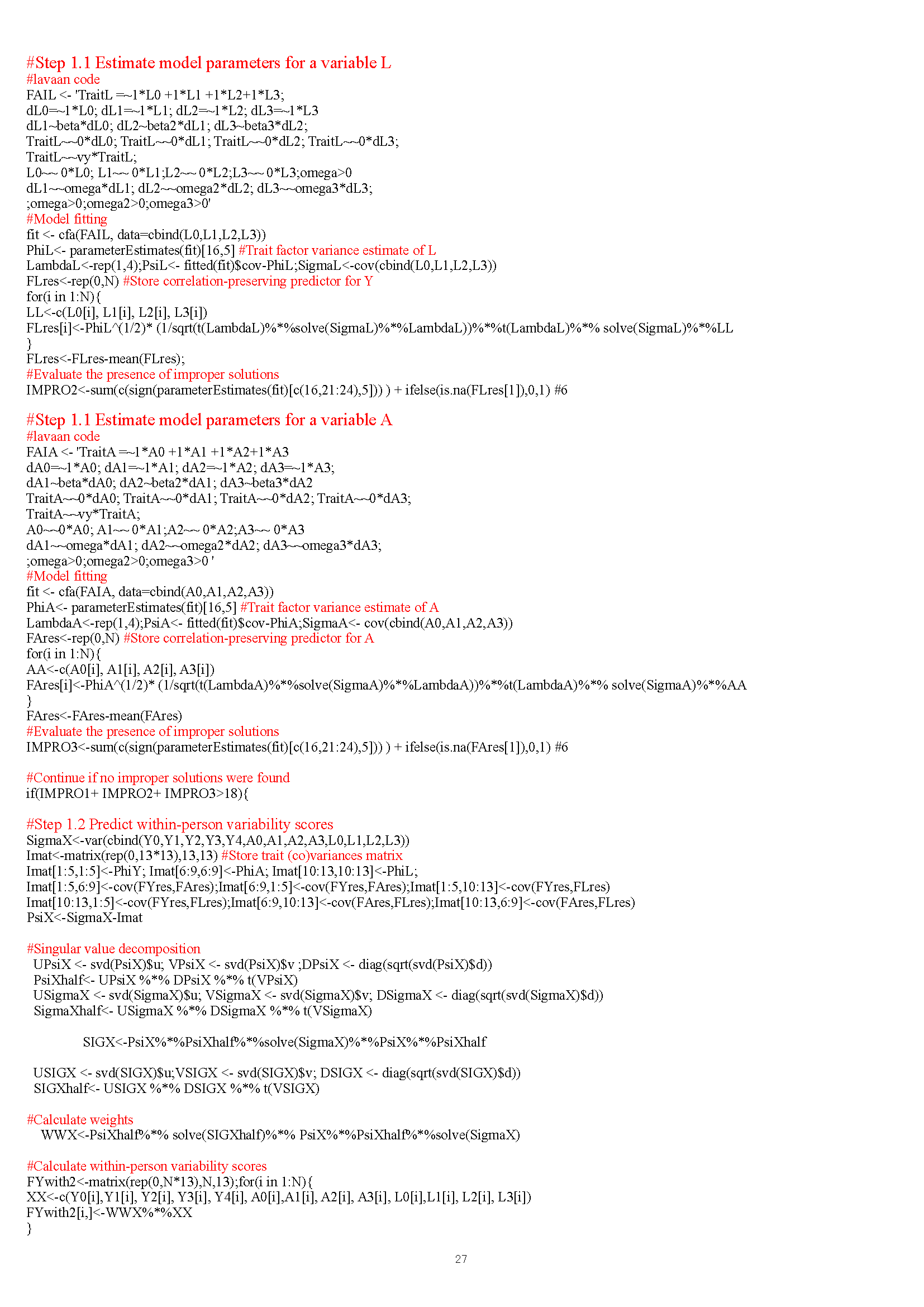}
\end{figure}
\begin{figure}[htbp]
\includegraphics[height=21cm,width=16cm,angle=0]{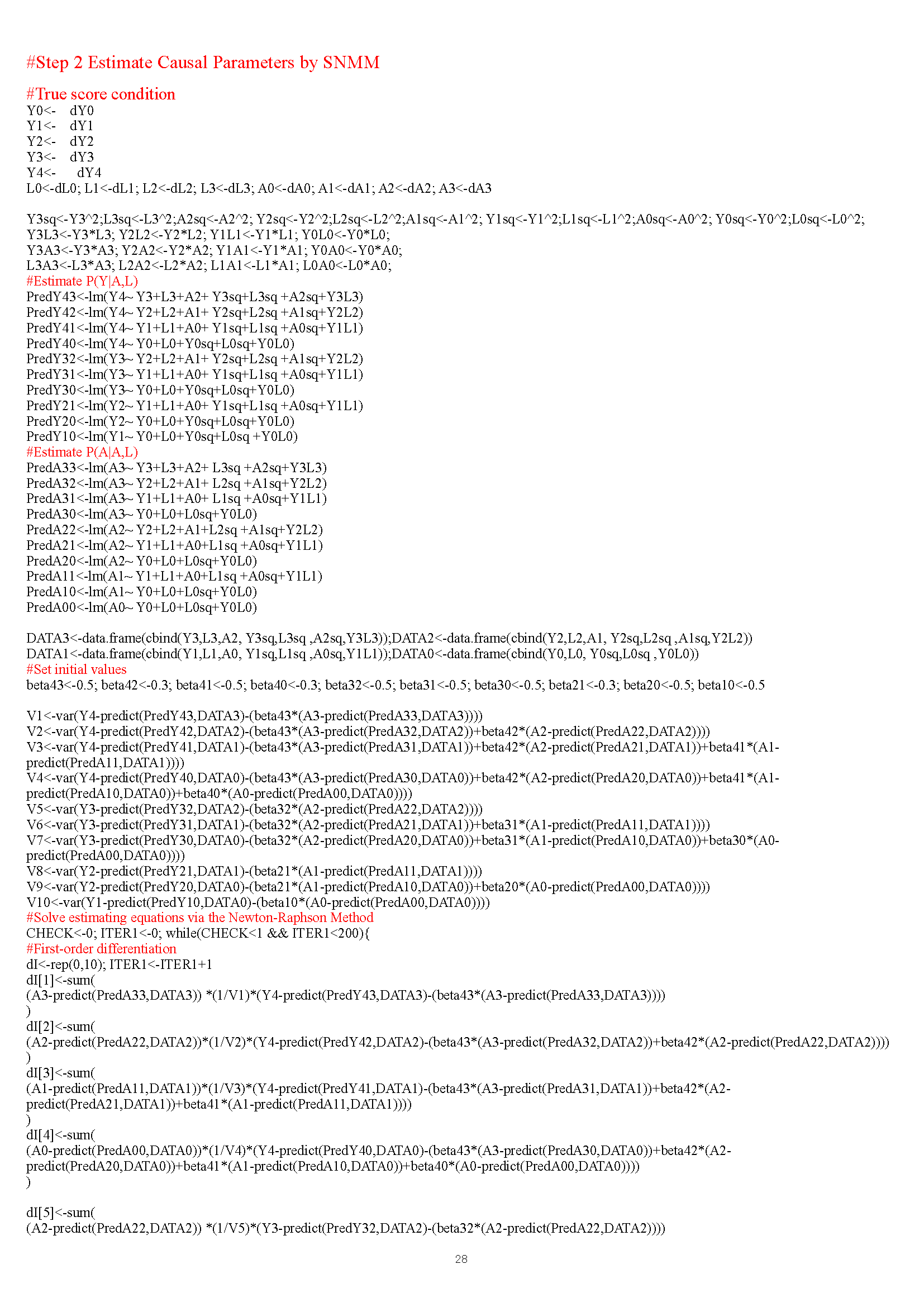}
\end{figure}
\begin{figure}[htbp]
\includegraphics[height=21cm,width=16cm,angle=0]{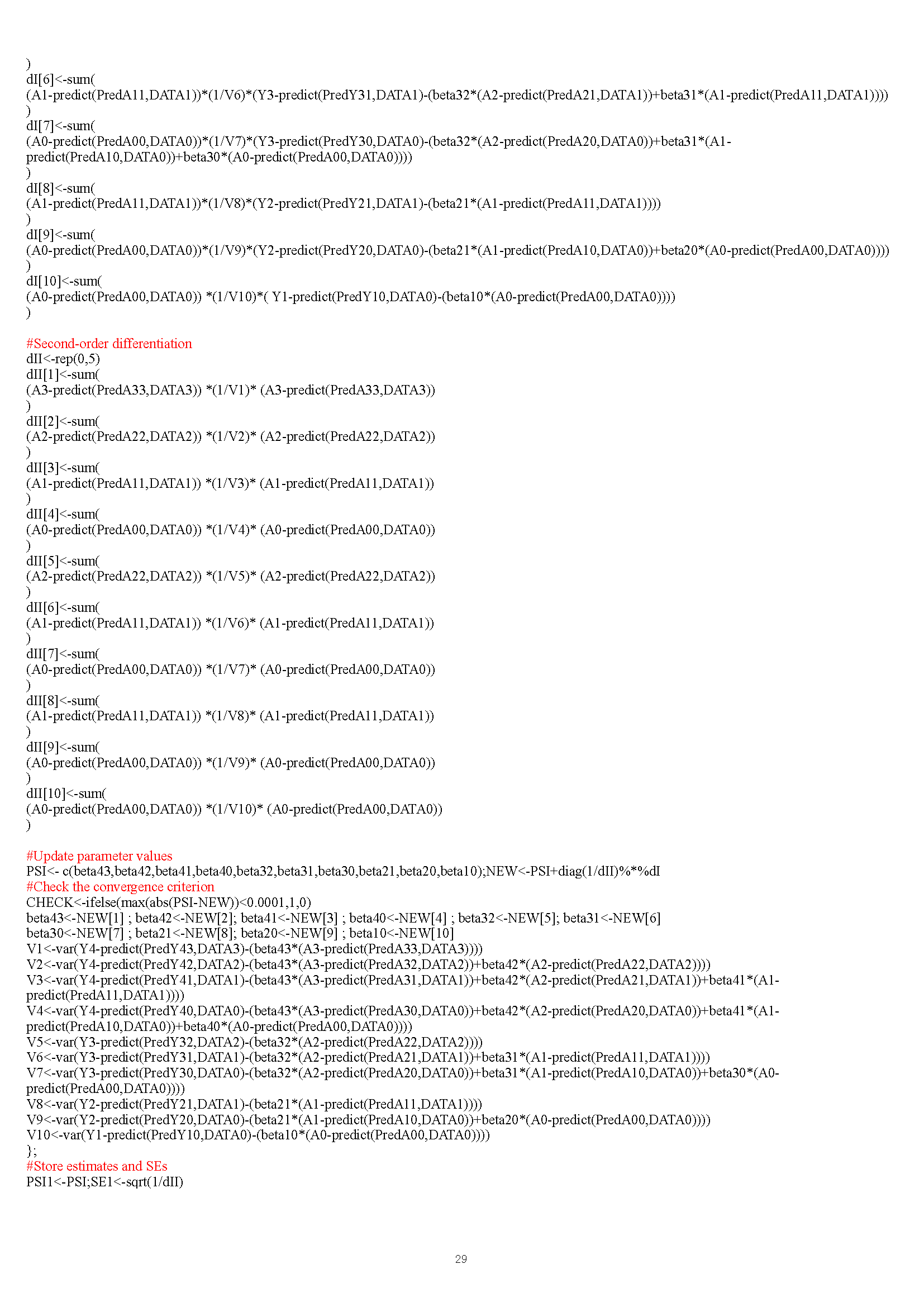}
\end{figure}
\begin{figure}[htbp]
\includegraphics[height=21cm,width=16cm,angle=0]{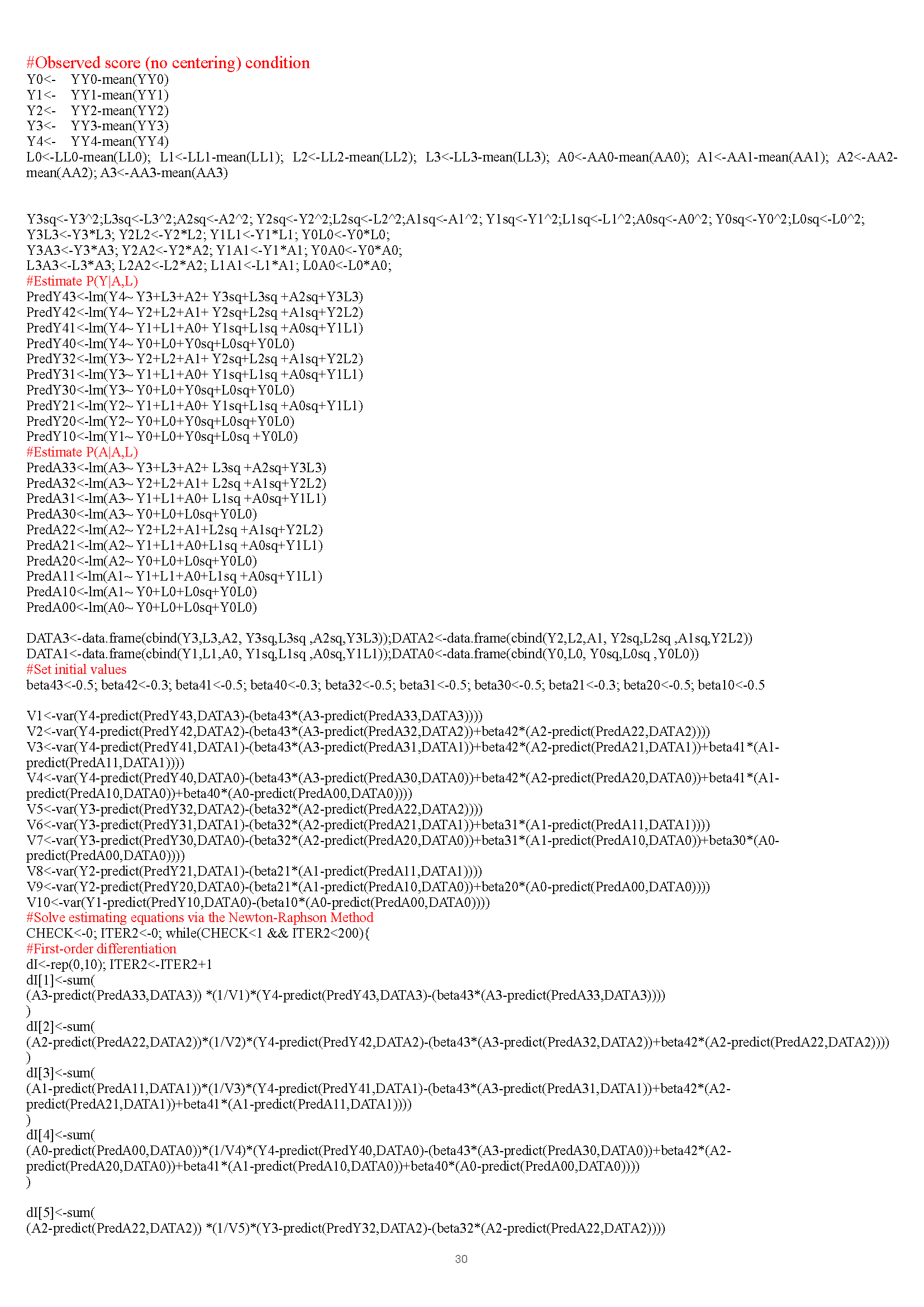}
\end{figure}
\begin{figure}[htbp]
\includegraphics[height=21cm,width=16cm,angle=0]{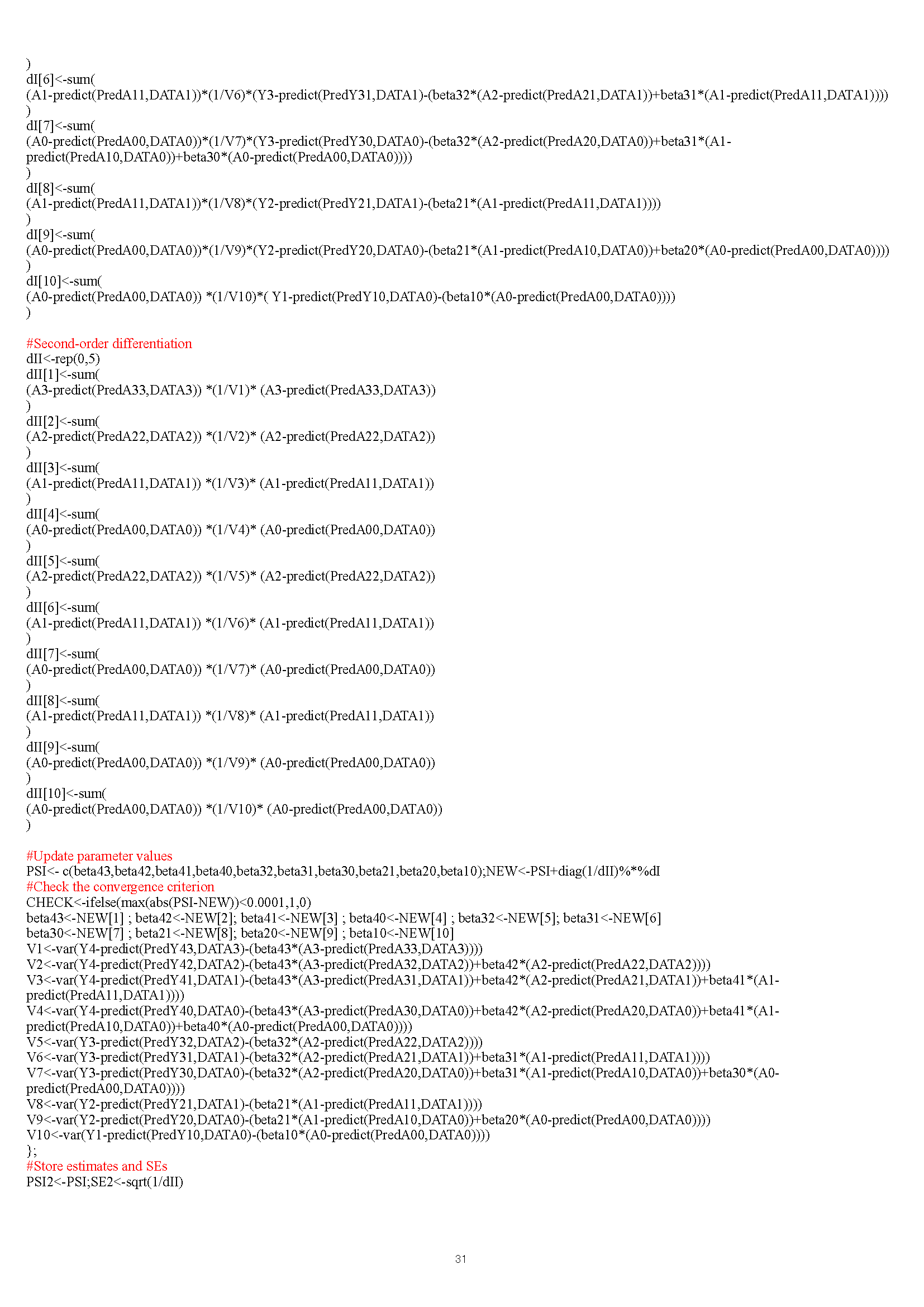}
\end{figure}
\begin{figure}[htbp]
\includegraphics[height=21cm,width=16cm,angle=0]{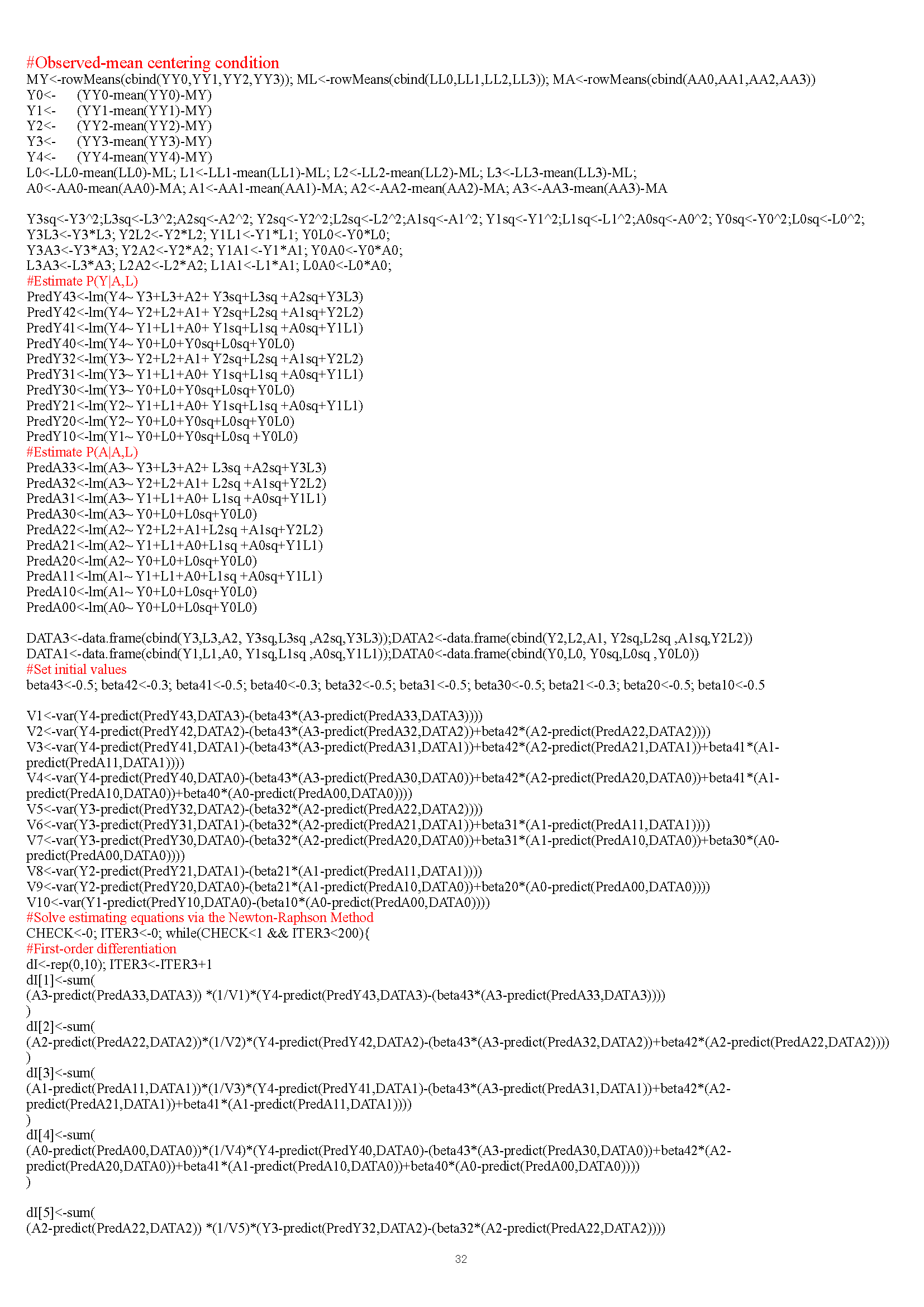}
\end{figure}
\begin{figure}[htbp]
\includegraphics[height=21cm,width=16cm,angle=0]{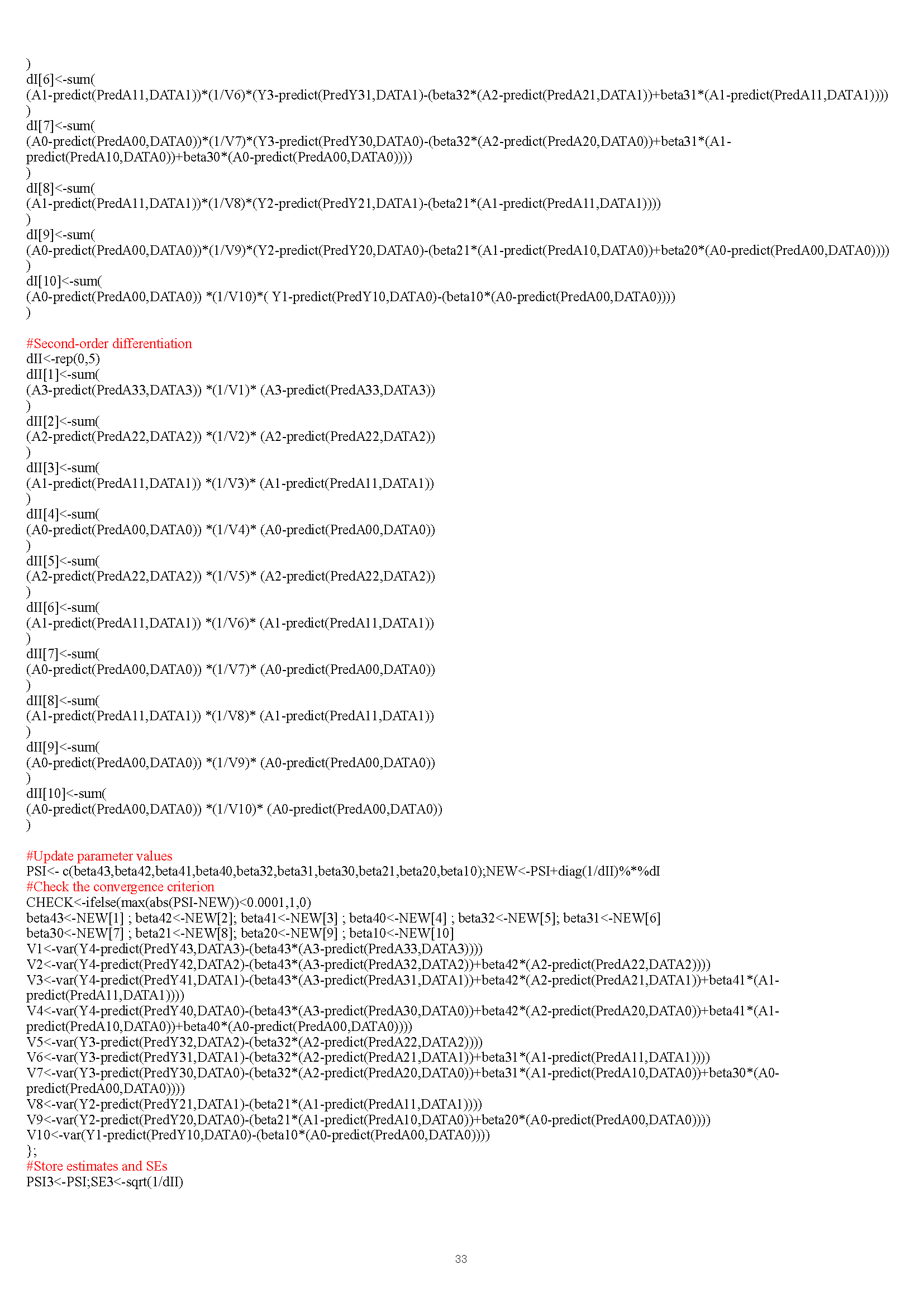}
\end{figure}
\begin{figure}[htbp]
\includegraphics[height=21cm,width=16cm,angle=0]{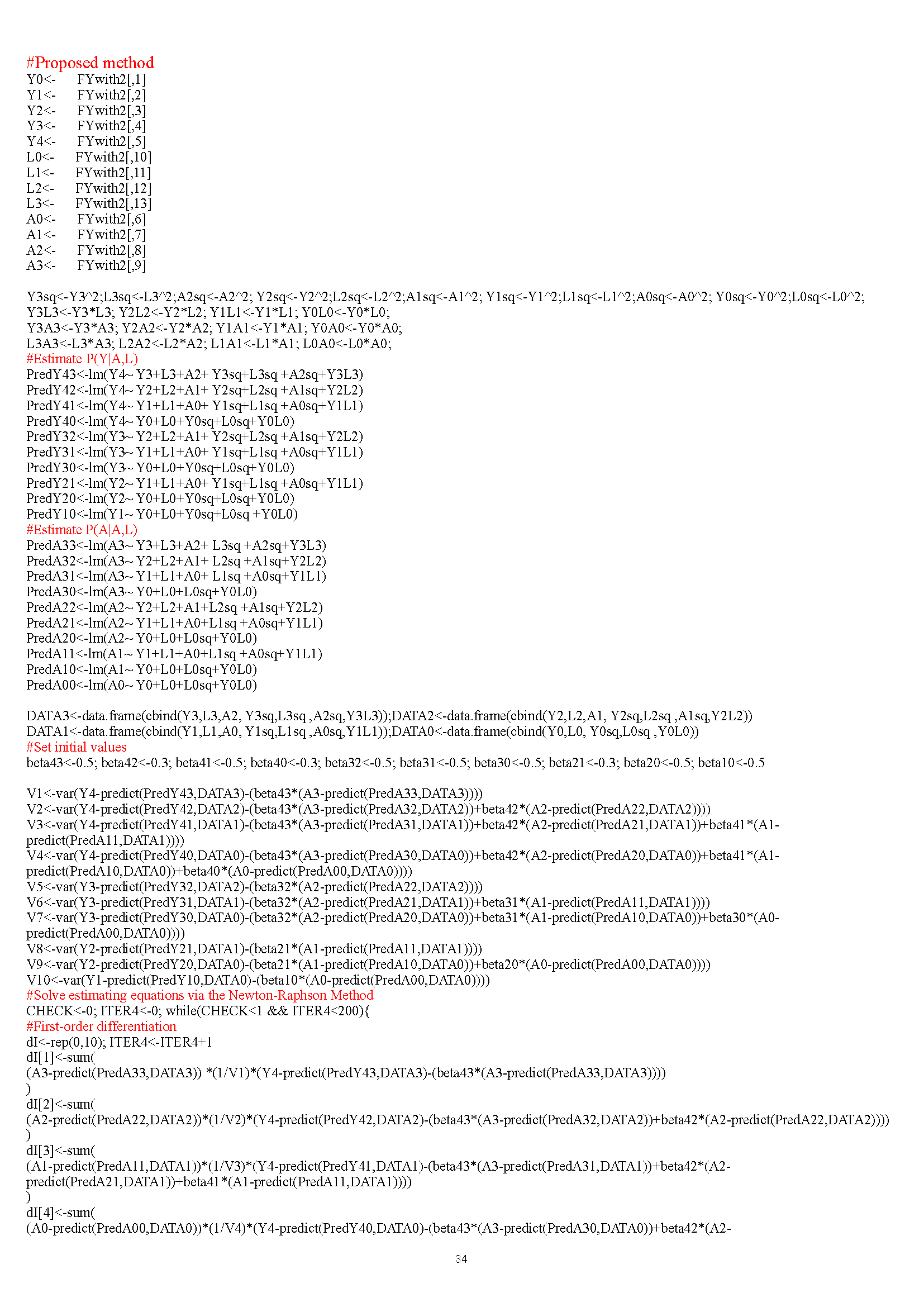}
\end{figure}
\begin{figure}[htbp]
\includegraphics[height=21cm,width=16cm,angle=0]{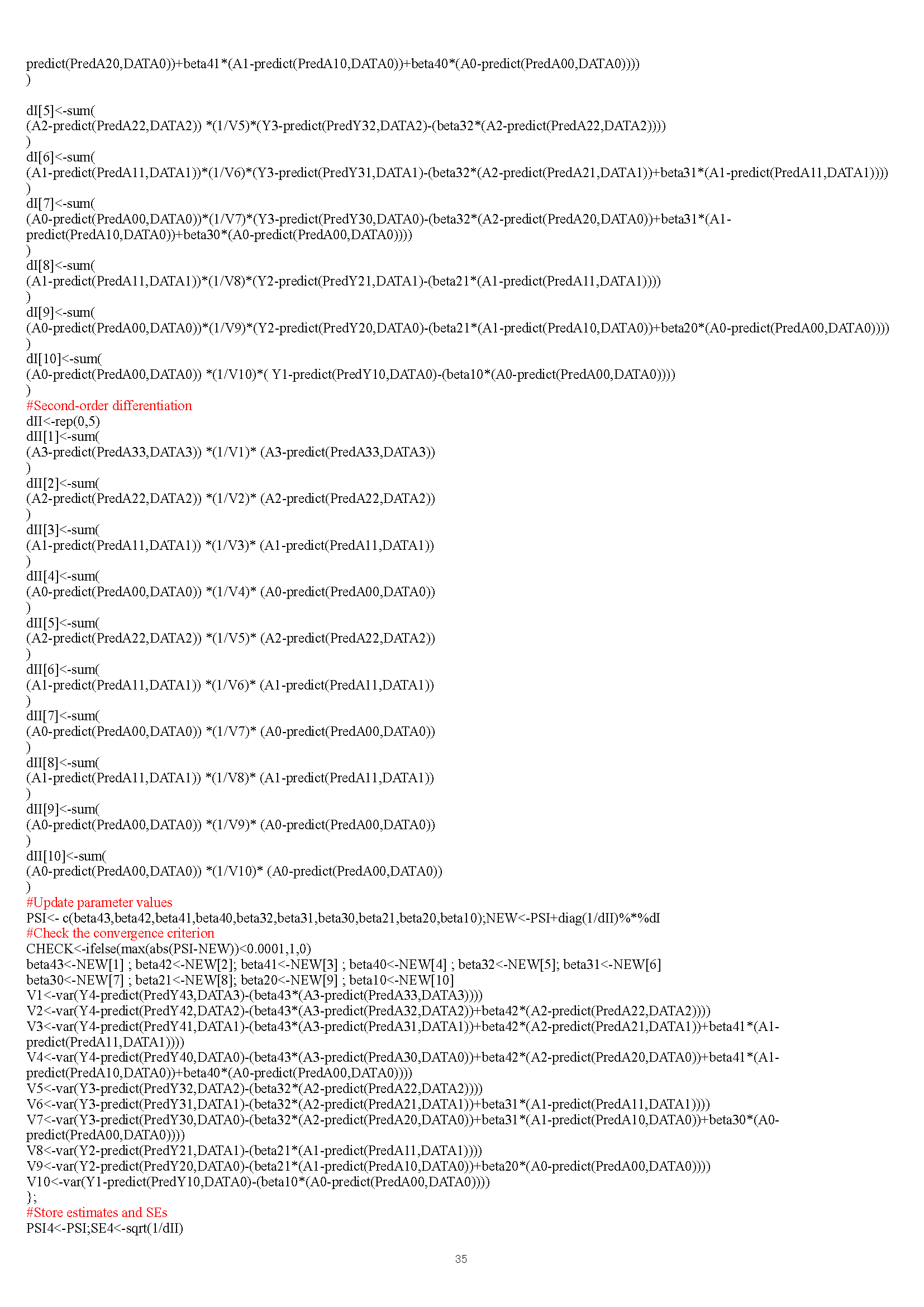}
\end{figure}
\begin{figure}[htbp]
\includegraphics[height=21cm,width=16cm,angle=0]{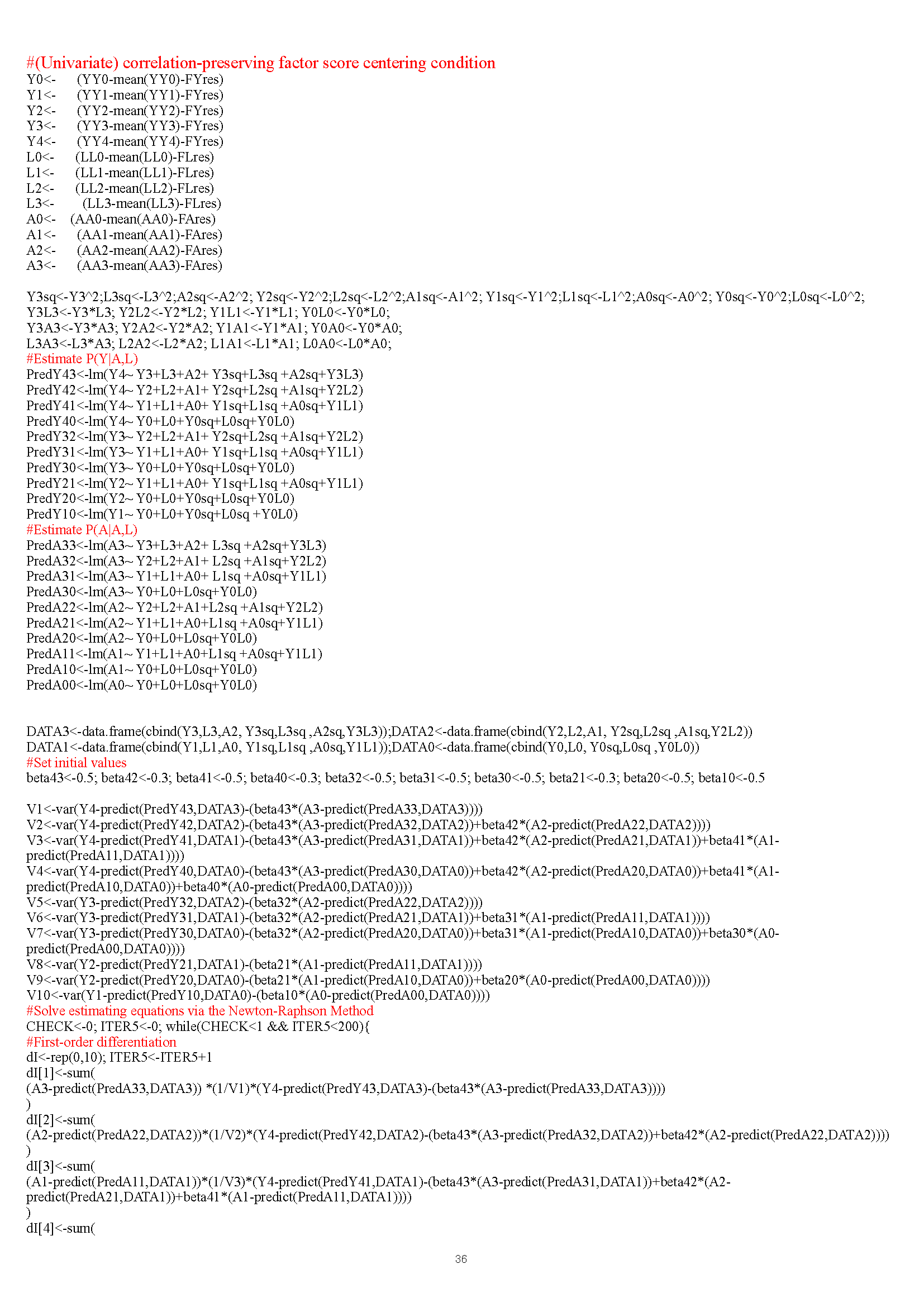}
\end{figure}
\begin{figure}[htbp]
\includegraphics[height=21cm,width=16cm,angle=0]{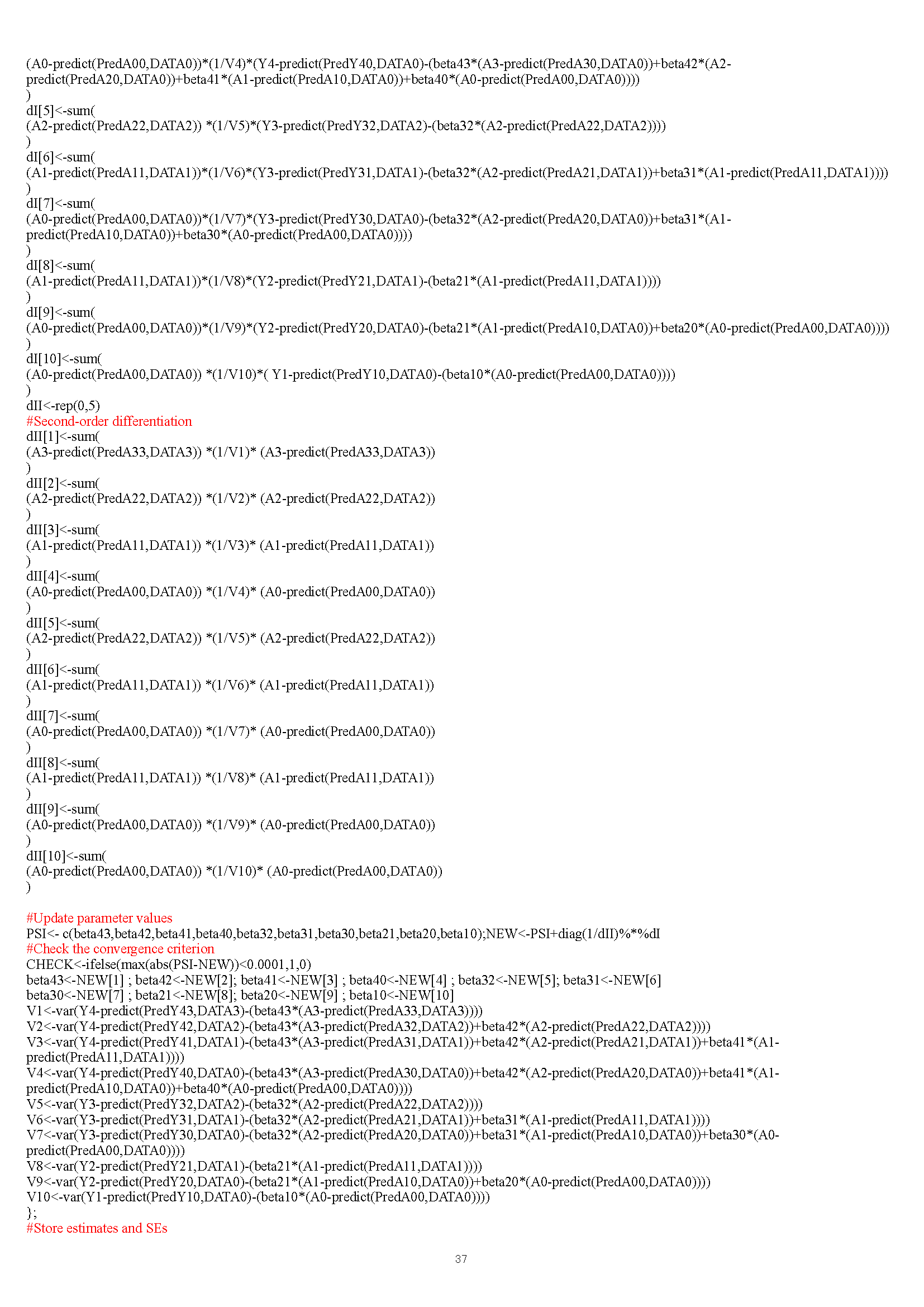}
\end{figure}
\begin{figure}[htbp]
\includegraphics[height=21cm,width=16cm,angle=0]{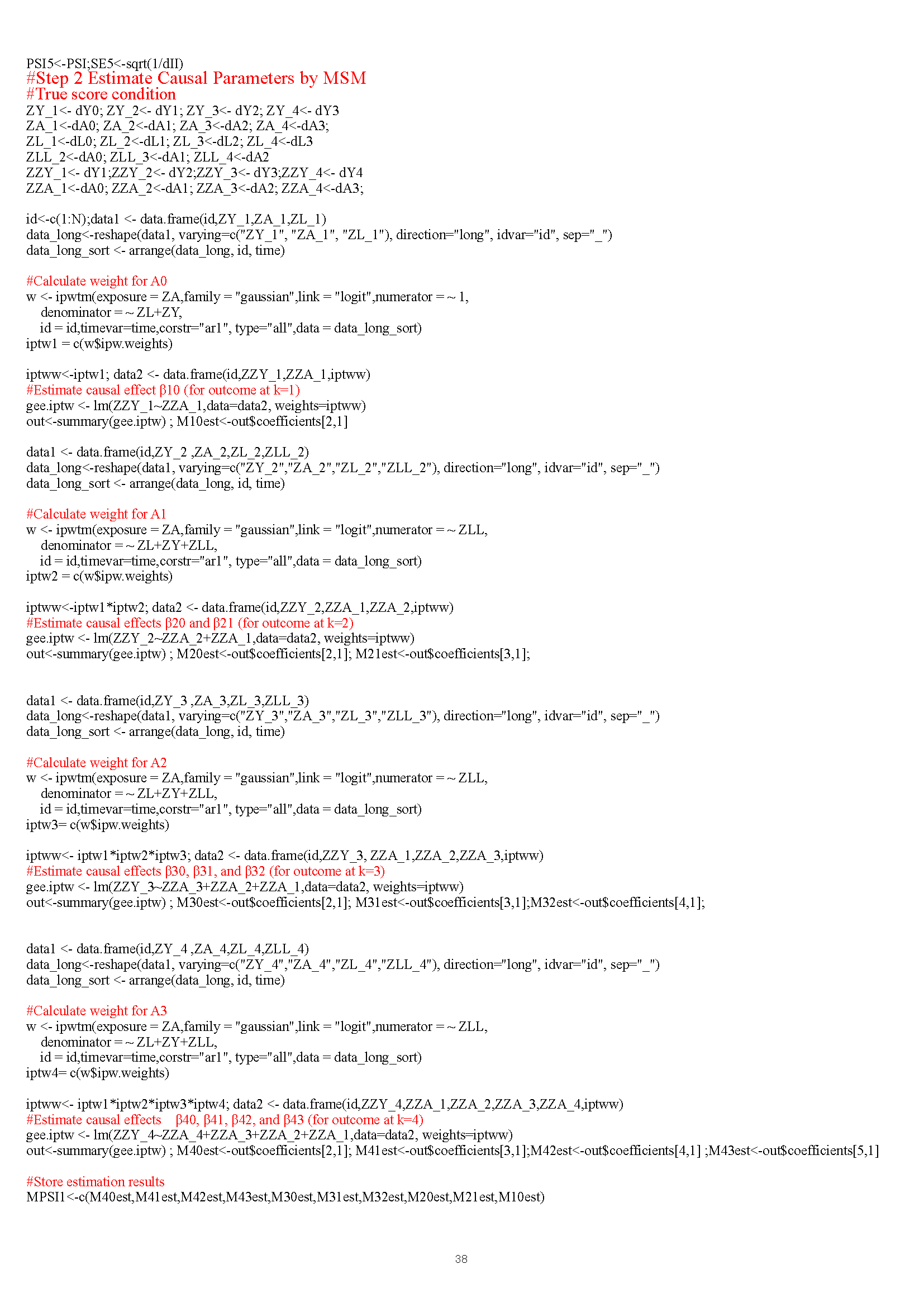}
\end{figure}
\begin{figure}[htbp]
\includegraphics[height=21cm,width=16cm,angle=0]{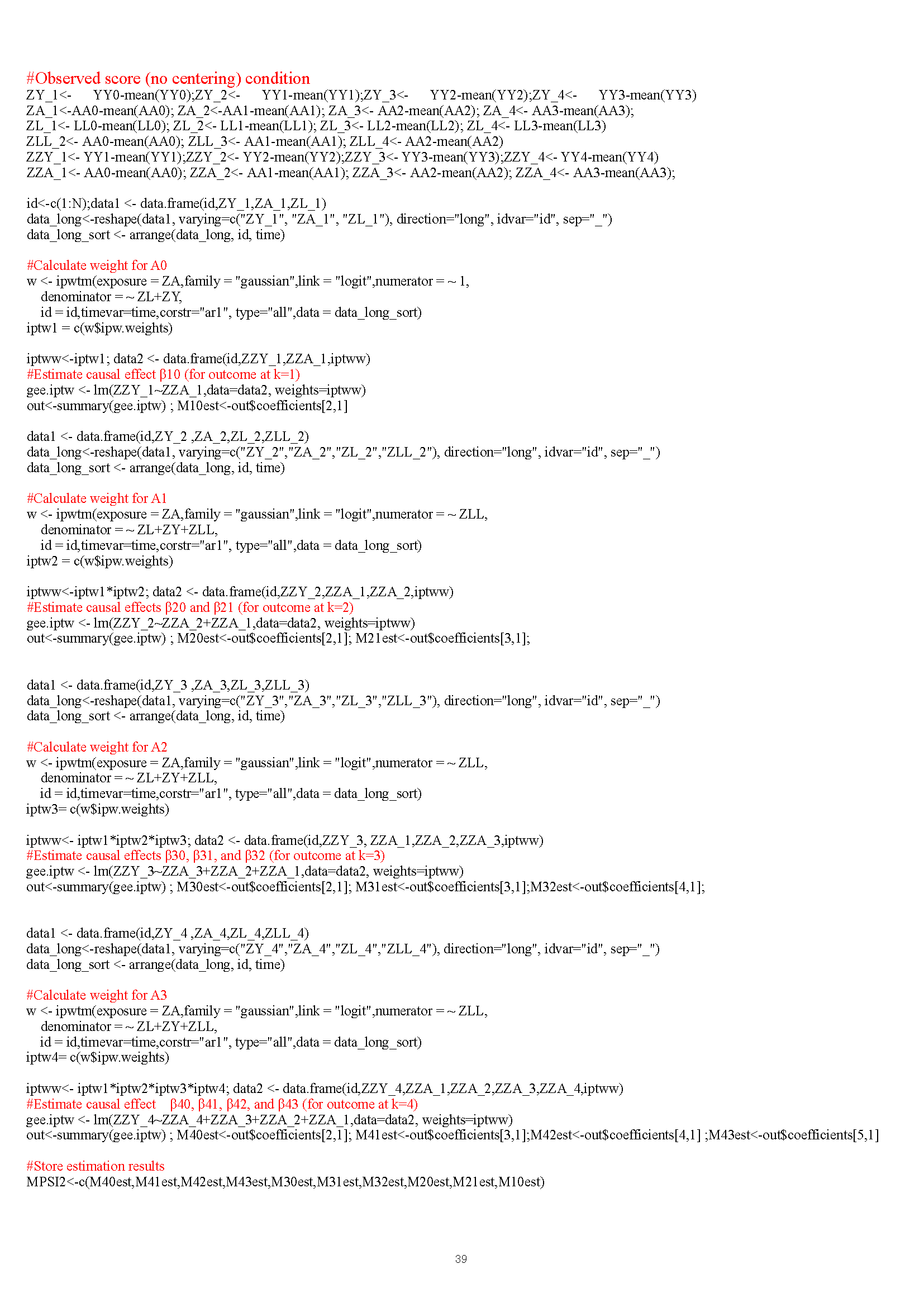}
\end{figure}
\begin{figure}[htbp]
\includegraphics[height=21cm,width=16cm,angle=0]{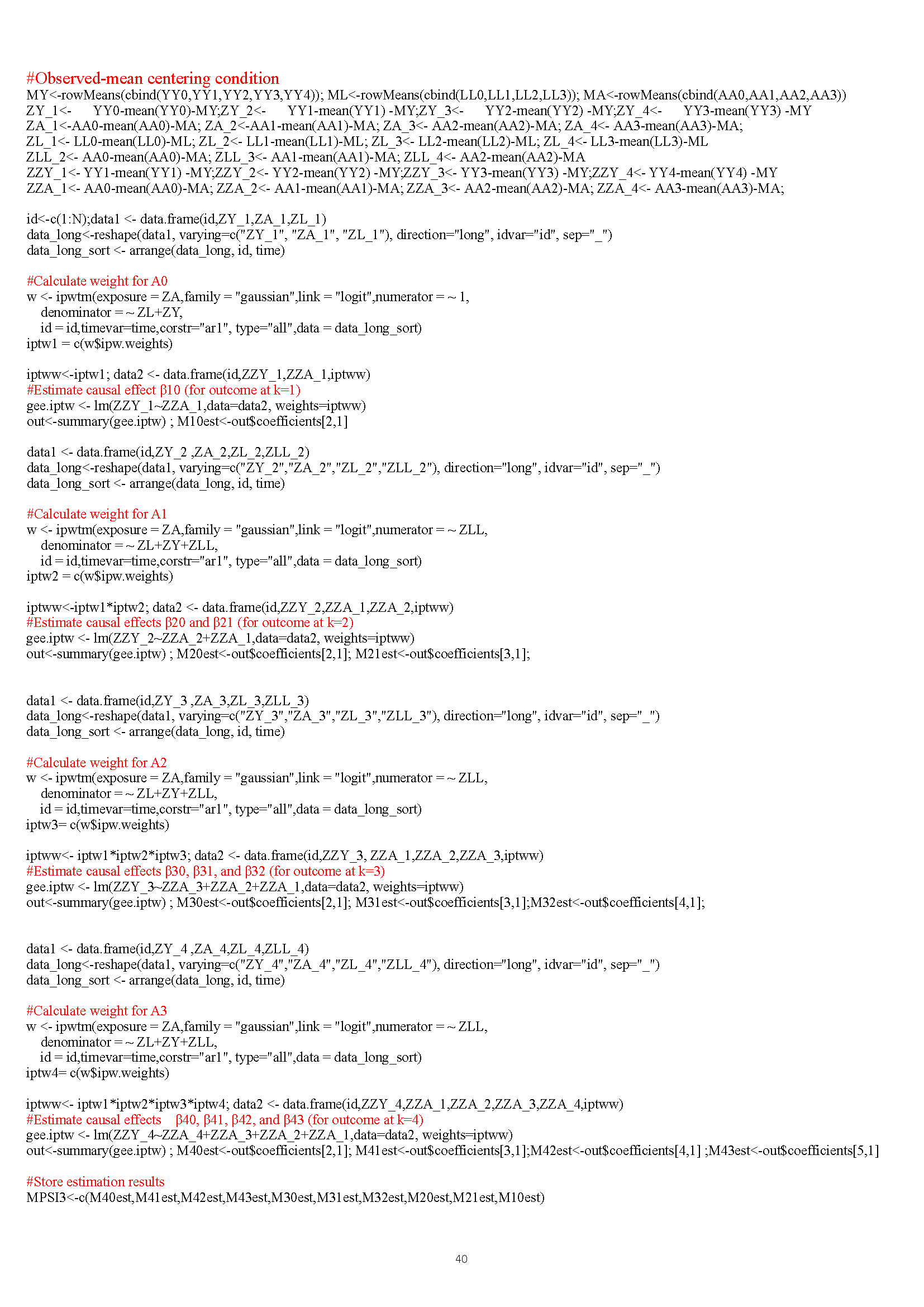}
\end{figure}
\begin{figure}[htbp]
\includegraphics[height=21cm,width=16cm,angle=0]{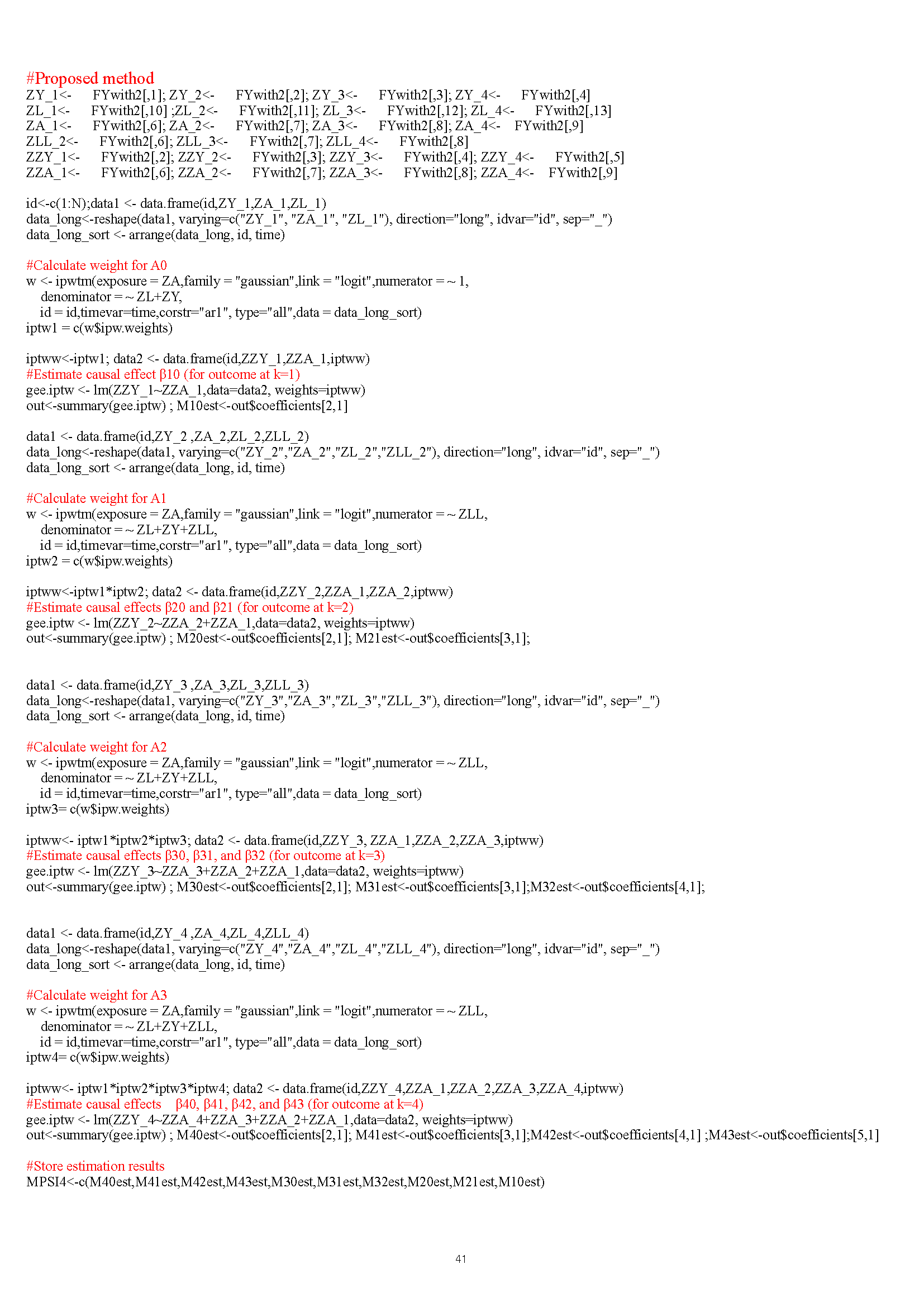}
\end{figure}
\begin{figure}[htbp]
\includegraphics[height=21cm,width=16cm,angle=0]{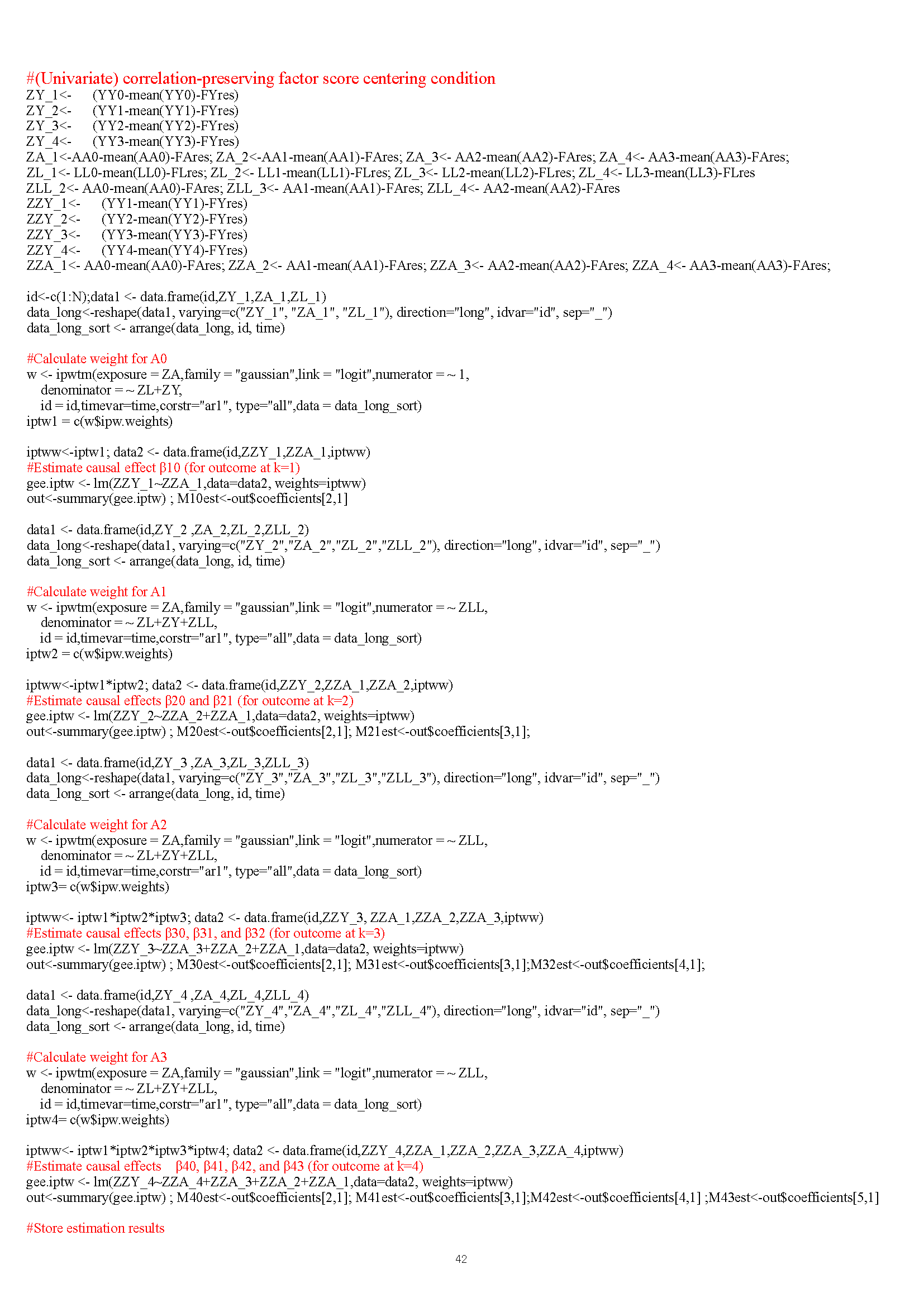}
\end{figure}
\begin{figure}[htbp]
\includegraphics[height=21cm,width=16cm,angle=0]{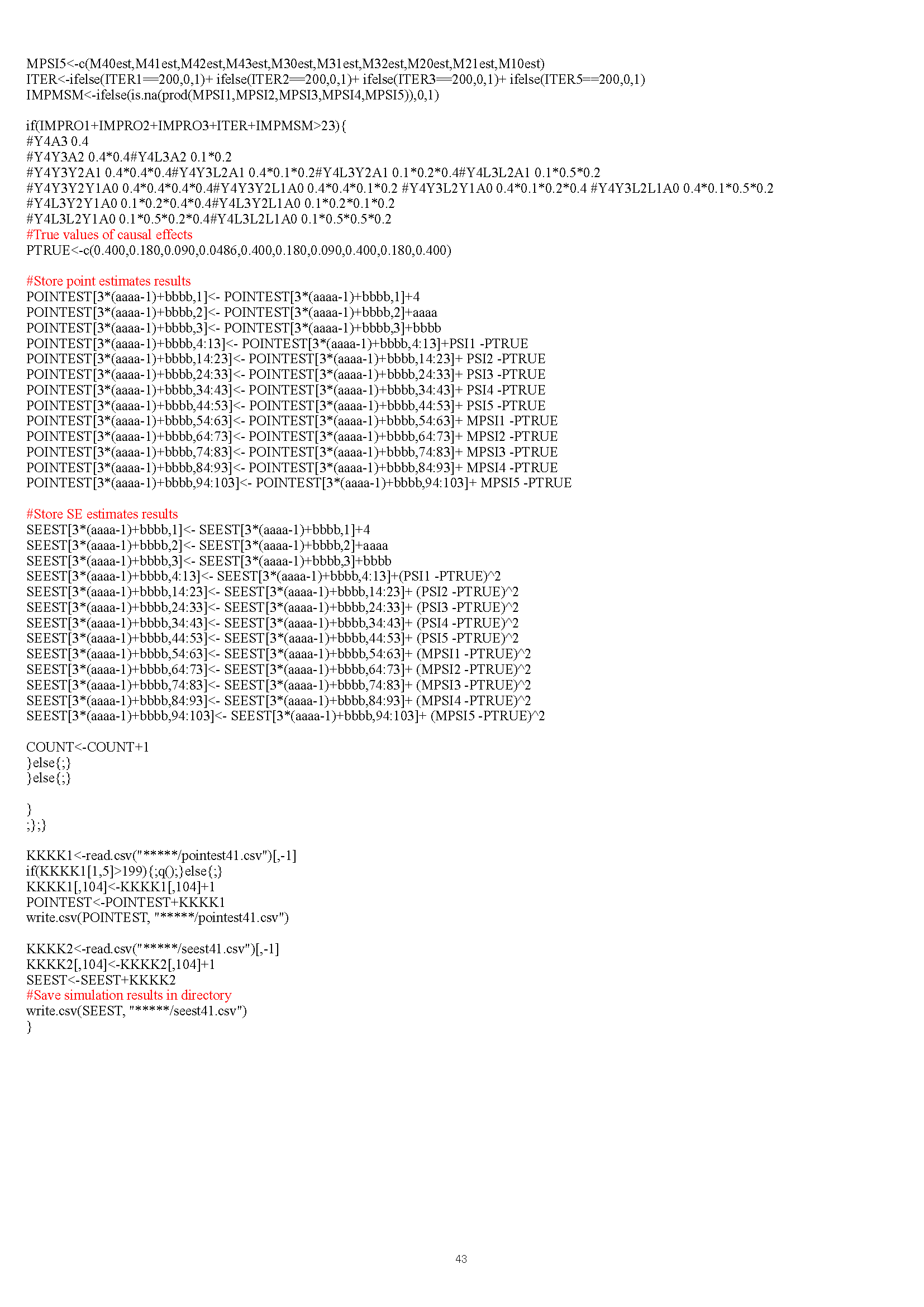}
\end{figure}\end{document}